\documentclass[iop,numberedappendix,twocolappendix]{emulateapj}   %use this later on
\usepackage{comment}
\usepackage{hyperref}
\usepackage[caption=false]{subfig}

\makeatletter
\newcommand*{\rom}[1]{\expandafter\@slowromancap\romannumeral #1@}
\makeatother

\usepackage{amsmath}
\usepackage{epstopdf}
\usepackage[flushleft]{threeparttable}
\shorttitle{}
\shortauthors{}

\begin{document}

\title{The CO Luminosity Density at High-z (COLDz) Survey: A Sensitive, Large Area Blind Search for Low-{\it J} CO Emission from Cold Gas in the Early Universe with the Karl G. Jansky Very Large Array}

\author{Riccardo Pavesi\altaffilmark{1}$^\dagger$, Chelsea E. Sharon\altaffilmark{1,2}, Dominik A. Riechers\altaffilmark{1}, Jacqueline A. Hodge\altaffilmark{3,4,5}, Roberto Decarli\altaffilmark{3,6}, Fabian Walter\altaffilmark{3}, Chris L. Carilli\altaffilmark{7,8}, Emanuele Daddi\altaffilmark{9}, Ian Smail\altaffilmark{10}, Mark Dickinson\altaffilmark{11}, Rob J. Ivison\altaffilmark{12,13},\\ Mark Sargent\altaffilmark{14}, Elisabete da Cunha\altaffilmark{15}, Manuel Aravena\altaffilmark{16}, Jeremy Darling\altaffilmark{17}, Vernesa Smol\v{c}i\'c\altaffilmark{18},\\ Nicholas Z. Scoville\altaffilmark{19}, Peter L. Capak\altaffilmark{20}, Jeff Wagg\altaffilmark{21}}

\affil{$^1$Department of Astronomy, Cornell University,  Space Sciences
Building, Ithaca, NY 14853, USA \\$^2$Department of Physics \& Astronomy, McMaster University,   1280 Main Street West, Hamilton, ON L85-4M1, Canada \\ 
$^3$Max-Planck Institute for Astronomy, K\"{o}nigstuhl 17, D-69117 Heidelberg, Germany\\
$^4$National Radio Astronomy Observatory, 520 Edgemont Road, Charlottesville, VA 22903, USA\\
$^5$Leiden Observatory, Leiden University, P.O. Box 9513, 2300 RA Leiden, The Netherlands\\
$^6$INAF - Osservatorio di Astrofisica e Scienza dello Spazio, via Gobetti 93/3, I-40129, Bologna, Italy\\
$^7$National Radio Astronomy Observatory, PO Box O, Socorro,
NM 87801, USA\\ 
$^8$Cavendish Astrophysics Group, University of Cambridge,
Cambridge, CB3 0HE, UK\\
$^9$Laboratoire AIM, CEA/DSM-CNRS-Universite Paris Diderot, Irfu/Service d'Astrophysique, CEA Saclay, Orme des Merisiers, F-91191 Gif-sur-Yvette cedex, France\\
$^{10}$Centre for Extragalactic Astronomy, Department of Physics, Durham University, South Road, Durham DH1 3LE, UK \\
$^{11}$Steward Observatory, University of Arizona, 933 N. Cherry Street, Tucson, AZ 85721, USA\\
$^{12}$European Southern Observatory, Karl-Schwarzschild-Strasse 2, D-85748, Garching, Germany\\
$^{13}$Institute for Astronomy, University of Edinburgh, Royal Observatory, Blackford Hill, Edinburgh EH9 3HJ, UK\\
$^{14}$Astronomy Centre, Department of Physics and Astronomy, University of Sussex, Brighton, BN1 9QH, UK\\
$^{15}$Research School of Astronomy and Astrophysics, Australian National University, Canberra, ACT 2611, Australia\\
$^{16}$N\'ucleo de Astronom\'ia, Facultad de Ingenier\'ia, Universidad Diego Portales, Av. Ej\'ercito 441, Santiago, Chile\\
$^{17}$Center for Astrophysics and Space Astronomy, Department of Astrophysical and Planetary Sciences, University of Colorado, 389 UCB, Boulder, CO 80309-0389, USA\\
$^{18}$University of Zagreb, Physics Department, Bijeni\v{c}ka cesta 32, 10002 Zagreb, Croatia\\ 
$^{19}$Astronomy Department, California Institute of Technology, MC 249-17, 1200 East California Boulevard, Pasadena, CA 91125, USA\\ 
$^{20}$Spitzer Science Center, California Institute of Technology, MC 220-6, 1200 East California Boulevard, Pasadena, CA 91125, USA \\
$^{21}$SKA Organization, Lower Withington, Macclesfield, Cheshire SK11 9DL, UK\\}
     \email{$^\dagger$rp462@cornell.edu}

%, Chris Carilli, Fabian Walter, Jacqueline Hodge, Roberto Decarli,Manuel Aravena, Elisabete da Cunha, Peter Capak, Nick Scoville; Ian R. Smail, R. J. Ivison, Emanuele Daddi, Mark Sargent, Vernesa Smol\v{c}i\'c, Jeff Wagg, Mark Dickinson, Jeremy Darling

%\affil{Astronomy Department, Cornell University,
  %  Ithaca, NY 14853}

%\altaffiltext{1}{Astronomy Department, Cornell University,
  % Ithaca, NY 14853, USA}
   %\altaffiltext{2}{Department of Physics \& Astronomy, McMaster University,
   %1280 Main Street West, Hamilton, ON L85-4M1, Canada}

\begin{abstract}
We describe the CO Luminosity Density at High-z  (COLDz) survey, the first spectral line deep field targeting CO(1--0) emission from galaxies at $z=1.95-2.85$ and CO(2--1) at $z=4.91-6.70$.  The main goal of COLDz is to constrain the cosmic density of molecular gas at the peak epoch of cosmic star formation. 
By targeting both a wide ($\sim$51 arcmin$^2$) and a deep area ($\sim$9 arcmin$^2$), the survey is designed to robustly constrain the bright end and the characteristic luminosity of the CO(1--0) luminosity function. An extensive analysis of the reliability of our line candidates, and  new techniques provide detailed completeness and statistical corrections as necessary to determine the best constraints to date on the CO luminosity function.
Our blind search for CO(1--0) uniformly selects  starbursts and massive  Main Sequence galaxies based on their cold molecular gas masses.
Our search also detects CO(2--1) line emission from optically dark, dusty star-forming galaxies at $z>5$.
We find a  range of spatial sizes for the CO-traced gas reservoirs up to $\sim40$ kpc, suggesting that spatially extended cold molecular gas reservoirs may be common in massive, gas-rich galaxies at $z\sim2$. Through CO line stacking, we constrain the gas mass fraction in previously known typical star-forming galaxies at $z=2$--3. The stacked CO detection suggests lower molecular gas mass fractions than expected for massive Main Sequence galaxies by a factor of $\sim3-6$. 
 We find  total CO line brightness at $\sim34\,$GHz  of $0.45\pm0.2\,\mu$K, which constrains  future line intensity mapping and CMB experiments.
\end{abstract}

\section{Introduction}
Although the process of galaxy assembly through star formation is believed to have reached a peak rate at redshifts of $z=2$--3 (i.e., $\sim$10--11 billion years ago), the fundamental driver of this evolution is still uncertain \citep{MadauDickinson}. In order to understand the physical origin of the cosmic star formation history (i.e., the rate of star formation taking place per unit comoving volume), we need to quantify the mass of cold, dense gas in galaxies as a function of cosmic time, because this gas phase controls star formation \citep{KennicuttEvans}. 
In particular, the evolution of the cold gas mass distribution can provide  strong constraints on models of galaxy formation by simultaneously measuring the gas availability and, through a comparison to the star formation distribution function, the global efficiency of the star formation process (see \citealt{CarilliWalter} for a review). In this work, we carry out the first fully ``blind" deep-field spectral line search for CO(1--0) line emission, arguably the best tracer of the total molecular gas mass at the peak epoch of cosmic star formation, by taking advantage of the greatly improved capabilities of NSF's Karl G. Jansky Very Large Array (VLA).

To date, observations of the immediate fuel for star formation, i.e., the cold molecular gas, have mostly been limited to follow-up studies of galaxies that were pre-selected from optical/near-infrared (NIR) deep surveys (and hence based on stellar light) or selected in the sub-millimeter based on dust-obscured star formation as sub-millimeter galaxies (SMGs; for reviews see, e.g., \citealt{Blain2002,Casey14}).  %The latter are mostly rare,  luminous infrared galaxies, the so-called submillimeter galaxies (SMGs), which are only a minor contributor to the cosmic star formation.
In particular, optical/NIR color-selection techniques (e.g., ``BzK", ``BM/BX"; \citealt{Daddi04,Steidel04}) have explored significant samples of massive, star forming galaxies at $z\sim$1.5 to 2.5 \citep{Daddi08,Daddi10a,Tacconi10,Tacconi13} and  the sub-mm selection has been particularly effective in identifying the most highly star-forming galaxies at this epoch for CO follow-up (e.g., \citealt{Bothwell13}). Although such targeted CO studies are fundamental to explore the properties of known galaxy populations, they need to be complemented by blind CO surveys that do not pre-select their targets, which may potentially reveal gas-dominated and/or systems with uncharacteristically low star formation rate missed by other selection techniques.

Targeted CO studies have found more massive gas reservoirs at $z\sim2$ compared to local galaxies. Cold molecular gas  is therefore believed to be the main driver for the high star formation rates of normal galaxies at these redshifts (e.g., \citealt{Greve2005,Daddi08, Daddi10a,Daddi10b,Tacconi10,Genzel10, Bothwell13}). 
Recent studies have claimed  tentative evidence for an elevated star formation efficiency, i.e.,  star formation rate generated per unit mass of molecular gas, at $z\sim2$ compared to local galaxies \citep[e.g.,][]{Genzel15, Scoville16, Schinnerer16, Scoville17, Tacconi17}. Such an elevated star formation efficiency could  be related to massive, gravitationally unstable gas reservoirs. The interstellar gas content of galaxies  therefore appears to be the main driver of the star formation history of the Universe, during the epoch when galaxies formed at least half of their stellar mass content (e.g., \citealt{MadauDickinson}). 
Although targeted molecular gas studies currently allow to observe larger galaxy samples more efficiently than blind searches, their pre-selection could potentially introduce an unknown systematic bias.  Critically, such studies may not uniformly sample the galaxy cold molecular gas mass function. The best way to address such potential biases, and thus, to complement targeted studies, is through deep field blind surveys, in which galaxies are directly selected based on their cold gas content. Although some targeted CO(1--0) deep studies have previously been attempted (most notably \citealt{Aravena12} and \citealt{Rudnick17}), these studies have typically targeted overdense (proto-)cluster environments. Hence, a blind search approach, to sample a representative cosmic volume is needed, in order to assess the statistical significance of such previous studies.

CO(1--0) line emission  is one of the most direct tracers of the cold, molecular inter-stellar medium (ISM) in galaxies\footnote{In this work, CO always refers to the most abundant isotopologue $^{12}$CO}. Its line luminosity can be used to estimate the cold molecular gas mass by means of a conversion factor ($\alpha_{\rm CO}$; see \citealt{Bolatto_review} for a review). Although other tracers of the cold ISM have been utilized to date, including mid-{\it J} CO lines and the dust continuum emission, these are less direct tracers because they require additional, uncertain conversion factors (e.g., CO excitation corrections and dust-to-gas ratios).
Specifically, while the ground state CO(1--0) transition traces the bulk gas reservoir, mid-{\it J} CO lines such as CO(3--2) and higher-{\it J} lines are likely to preferentially trace the fraction of actively star-forming gas.   Hence, their brightness requires additional assumptions about line excitation,  in order to provide a measurement of the total gas mass. Furthermore, different populations of galaxies may be characterized by significantly different CO excitation conditions (e.g., BzK, SMGs and quasar hosts; \citealt{Daddi10b,Riechers06,Riechers11a,Riechers11b,Ivison_GN19,Bothwell13,CarilliWalter,Narayanan14}), which also  show considerable individual scatter (e.g., \citealt{Sharon16}).

Long-wavelength dust continuum emission has been suggested to be a measure of the total gas mass, and is utilized to great extent in recent surveys with ALMA to investigate large samples of far-infrared (FIR)-selected galaxies \citep{Eales12,Bourne13,Groves15,Scoville16, Scoville17,ASPECS2}. Nonetheless, there remain substantial uncertainties in the accuracy of the calibration for this method at high redshift especially below the most luminous, most massive sources.\footnote{The dust continuum method to determine gas masses may be affected by the metallicity dependence of the dust-to-gas ratio \citep{Sandstrom13,Berta16},  by trends in dust temperature with redshift (e.g., \citealt{Magdis12}), or with galaxy population (e.g., \citealt{Faisst17}).} Another caveat to using FIR continuum emission instead of CO comes from the finding that the dust emission measured by ALMA may not always trace the bulk of the gas distribution. This is made clear by the small sizes of the dust-emitting regions compared to the star forming regions and the gas as traced by CO emission (e.g., \citealt{Riechers_GN19,Riechers14,Ivison_GN19,Simpson15,Hodge16,Miettinen17,Chen2017}).

%put this before blind searches
Disentangling the causes for the observed increased star formation activity at $z\sim2$ is not straightforward, since an increased availability of cold gas may be difficult to distinguish from increased star formation efficiency due to the uncertainty in deriving gas masses, for representative samples of galaxies.
%In particular, some previously employed samples may be partially affected by selection effects introduced by pre-selecting  samples based on stellar emission (e.g, \citealt{Tacconi13,Tacconi17}) or  dust-obscured star formation (e.g., \citealt{Scoville16,Scoville17}). 
Now, thanks to the unprecedented sensitivity and bandwidth of the VLA and the Atacama Large (sub-)Millimeter Array (ALMA), CO deep field studies can be carried out efficiently, and these are ideal to address such potential selection effects.
%we here report the first quantification of the total molecular gas mass through a deep field blind search of the most direct tracer of cold gas which is the CO(1--0) emission line.
%Specifically, the emission from the ground state rotational transition of the CO molecule, $^{12}$CO(1--0) , at 115.27 GHz can be related to the total molecular gas mass by means of a conversion factor ($\alpha_{CO}$; see \cite{Bolatto_review} for a review) and this line is redshifted into the VLA Ka-band at $z=2$--3 (see Fig.~\ref{fig:coverage}). Although other tracers of the cold inter-stellar medium (ISM) have been utilized to date and include mid-{\it J} CO lines and the dust continuum emission, they are less direct because they require additional, uncertain conversion factors (e.g., CO excitation and dust-to-gas ratios).
%Previous work to constrain the evolution of galaxies dense gas content have predominantly targeted pre-selected galaxies based on their optical/NIR emission tracing the stellar content \citep{Tacconi10, Tacconi13, Genzel15, Tacconi17} or their FIR emission tracing their dust emission \citep{Scoville16, Scoville17}. 
Previous deep field studies, with the Plateau de Bure Interferometer (PdBI; now the NOrthern Extended Millimeter Array, NOEMA) in the HDF-N \citep{Decarli14, Walter14} and ALMA in the HUDF (ALMA SPECtroscopic Survey in the Hubble Ultra-Deep Field Pilot or ASPECS-Pilot, \citealt{ASPECS1, ASPECS2}), have provided the first CO blind searches covering mid-{\it J}  transitions such as CO(3--2)\footnote{The ASPECS-Pilot survey simultaneously covered the CO(2--1)  line in the redshift range $z\sim$1.0--1.7, the CO(3--2) line at $z\sim$2.0--3.1, and higher-{\it J} CO transitions  at higher redshift.},  which are accessible at millimeter wavelengths. These studies have yielded crucial constraints on the molecular gas mass function at $z\sim1$--3, subject to assumptions on the excitation of the CO line ladder to infer the corresponding molecular gas content.\footnote{A key challenge in these studies is the uncertainty in assigning candidate emission lines to the correct CO transition, in cases where the redshift of the observed line candidates is not independently known.}  They  have found broad agreement with models of the CO luminosity evolution with redshift by finding an elevated molecular gas cosmic density at $z>1$ in comparison to $z\sim0$, but they may suggest a tension with luminosity function models at $z\gtrsim$1 by finding a larger number of CO line candidates than expected \citep{ASPECS2}.

In order to more statistically characterize the molecular gas mass function in galaxies at $z=2$--3 and 5--7 than previously possible, while avoiding some of the previous selection biases, we have carried out the COLDz survey\footnote{The COLDz survey data, together with complete candidate lists, and analysis routines may be found online at coldz.astro.cornell.edu.} a blind search for CO(1--0) and CO(2--1) line emission using the fully upgraded VLA\footnote{The recently expanded VLA, with its new Ka-band detectors, the new 3-bit samplers, the simultaneous 8 GHz bandwidth, and its improved sensitivity, for the first time, enables carrying out this survey study.}.
 The main objective of this survey is to constrain the CO(1--0) luminosity function at $z=2$--3, which provides the most direct census of the cold molecular gas at the peak epoch of cosmic star formation free from excitation bias, and based on a direct selection of the cold gas mass in galaxies.
As such, the COLDz survey is highly complementary to millimeter-wave surveys like ASPECS and targeted studies.
 The CO(1--0) intensity mapping technique explored by \cite{Keating15,Keating16} is complementary to our approach. Intensity mapping  offers sensitivity to the aggregate line emission signal from galaxies, but only measures the second raw moment of the luminosity function (therefore not distinguishing between the characteristic luminosity and volume density). While the intensity mapping technique allows to cover significantly larger areas of the sky, it does not directly measure gas properties of individual galaxies, and is therefore complementary to direct searches such as COLDz.

In a previous paper (\citealt{Lentati15}; Paper~0) we have described a first, interesting example of the galaxies  identified in this survey. 
In this work (Paper~\rom{1}), we describe the survey, present the blind search line catalog,  analyze the results of line stacking, and  outline the statistical methods employed to characterize our sample.  In Paper~\rom{2}, we present the analysis of the CO luminosity functions and our constraints on the cold gas density of the Universe at $z=2$--7, (Riechers et al., submitted). %In Paper~\rom{3}, we present the analysis of the continuum data in our observations and present a source catalog (Hodge et al., in prep.).  In Paper~\rom{4}, we present the detailed multi-wavelength properties of the CO(1--0) line candidates, and constraints on the star formation law, the gas dynamics and the gas excitation (Pavesi et al., in prep.). In Paper~\rom{5}, we study  $z>5$ dusty galaxies in our survey (Riechers et al., in prep.).

In Section 2 of this work, we describe the VLA COLDz observations, the calibration procedure and the methods  to mosaic and produce the signal-to-noise cubes. In Section 3, we describe our blind line search  through Matched Filtering in 3D. In Section 4, we present our ``secure" and ``candidate" CO line detections in both the deeper (in COSMOS) and wider (in GOODS-N) fields. In Section 5, we utilize stacking of galaxies with previously known spectroscopic redshifts, to provide strong constraints on their CO luminosity.   In Section 6, we derive constraints to the total CO line brightness at $\sim$34 GHz. In Section 7, we discuss the implications of our results in the context of previous surveys. We conclude with the implications for future surveys with current and planned instrumentation. A more detailed analysis of the line search methods, the statistical characterization of the candidate sample properties, and upper limits found for additional galaxy samples are presented in the Appendix.

In this work we adopt a flat, $\Lambda$CDM cosmology with $H_0=70\,$km s$^{-1}$ Mpc$^{-1}$ and $\Omega_{\rm M}=0.3$ and a Chabrier IMF.%Planck only changes lumi distances by 2\%.

\begin{table*}
\caption{COLDz Observations Summary.}
%\begin{center}
\centering
\begin{tabular}{ c c c c c c }
\hline
  Field & Pointing&D & D$\rightarrow$DnC  &DnC & DnC$\rightarrow$C  \\
   & &configuration &configuration & configuration& configuration\\
   \hline
   Baseline range (m) &&40--1000&40-2100&40--2100&40--3400\\
  \hline
   COSMOS&1--7&82 hr&{\textemdash}&11 hr&{\textemdash}\\
   GOODS-N&1--7&13 hr&{\textemdash}&{\textemdash}&{\textemdash}\\
   GOODS-N&6&{\textemdash}&{\textemdash}&3 hr&{\textemdash}\\
   GOODS-N&8--14&15 hr&{\textemdash}&{\textemdash}&{\textemdash}\\
   GOODS-N&15--21&15 hr&{\textemdash}&{\textemdash}&{\textemdash}\\
   GOODS-N&22--28&14 hr&1.4 hr&{\textemdash}&{\textemdash}\\
   GOODS-N&29--35&{\textemdash}&3 hr&12 hr&{\textemdash}\\
   GOODS-N&36--42&{\textemdash}&{\textemdash}&11 hr&3 hr\\
   GOODS-N&43--49&14 hr&{\textemdash}&1.3 hr&{\textemdash}\\
   GOODS-N&50--56&10.5 hr&{\textemdash}&3.5 hr&{\textemdash}\\
   GOODS-N&57&2 hr&{\textemdash}&{\textemdash}&{\textemdash}\\
    \hline
 
\end{tabular}
%  \end{center}
 \tablecomments{We list the total, on-source time in different array configurations, for all pointings in each group combined.}
 \label{obs_table}
\end{table*}

\section{Observations}
In order to constrain both the characteristic luminosity, $L^*_{\rm CO}$, or ``knee" of the CO(1--0) luminosity function and the bright end, we have optimized our observing strategy following the ``wedding cake" design, to cover a smaller deep area and a shallower, wide area.
We have used the wide-band capabilities of the upgraded VLA to obtain a continuous coverage of 8 GHz in the Ka band (PI: Riechers, IDs 13A-398; 14A-214) in a region of the COSMOS field (centered on the dusty starburst AzTEC-3 at $z=5.3$ as a line reference source, \citealt{Capak11,Riechers14}) and in the GOODS-N/CANDELS-Deep field, in order to take advantage of the availability  of excellent multi-wavelength data \citep{Grogin,Koekemoer,Giavalisco2004}.

The COSMOS data form a 7-pointing mosaic (center: R.A.=10h 0m 20.7s, Dec.=2$^{\circ}$35'17'') with continuous frequency coverage between 30.969 GHz and 39.033 GHz. The GOODS-N data form a 57-pointing mosaic (center: R.A.=12h 36m 59s, Dec.=62$^{\circ}$13'43.5'') with continuous coverage between 29.981 GHz and 38.001 GHz (Figs.~\ref{fig:coverage},\ref{fig:pointings}). The total on-source time was approximately 93 hrs in the COSMOS field and 122 hrs in the GOODS-N field. The frequency range targeted in this project covers CO(1--0) at $z=1.95$--2.85  and CO(2--1)  at $z=4.91$--6.70, such that the space density of CO(2--1)  line emitters is expected to be smaller than for CO(1--0) (Fig.~\ref{fig:coverage}; e.g., \citealt{Popping14,Popping16}). Both the large redshift spacing and the expected redshift evolution of the space density of CO emitters lessens the severity of the redshift ambiguity in our survey compared to previous studies. 

At 34 GHz the VLA primary beam can be described as a circular Gaussian with FWHM$\sim$80$^{\prime\prime}$, so our pointing centers were optimized to achieve a sensitivity that is approximately uniform in the central regions of the mosaics by choosing a spacing of 55$^{\prime\prime}$ ($<80^{\prime\prime}/\sqrt{2}$) in a standard hexagonally packed mosaic \citep{Condon98}.
During each observation, we targeted a set of 7 pointings in succession, alternating through phase calibration.  We performed pointing scans at the beginning of each observation, with additional pointing observations throughout for observations longer than 2 hours.
Most of the COSMOS and GOODS-N data were taken in the D configuration of the VLA. Some of the observations, especially for the GOODS-N pointings, were fully or partially carried out in DnC configuration, in re-configuration from D to DnC (D$\rightarrow$DnC) and in re-configuration from DnC to C (DnC$\rightarrow$C). Pointings are named sequentially from GN1 to GN57 (groups of GN1--7, GN8--14 etc. were observed together; Table~\ref{obs_table}).

The total area imaged, down to a sensitivity of $\sim30\%$ of the peak, in COSMOS is 8.9 arcmin$^2$ at 31 GHz and 7.0 arcmin$^2$ at 39 GHz. In GOODS-N the total area is 50.9 arcmin$^2$ at 30 GHz and 46.4 arcmin$^2$ at 38 GHz.
The correlator was set-up in 3-bit mode, at 2 MHz spectral resolution (corresponding to $\sim18\,$km s$^{-1}$ at 34 GHz), to simultaneously cover the full 8 GHz bandwidth for each polarization (Fig.~\ref{fig:coverage}). Tuning frequency shifts between tracks, and sometimes in the same track, were used to mitigate the edge channels noise increase in order to achieve a uniform depth across the frequency range (Table~2). 
\subsection{COSMOS observations}
The dataset in the COSMOS field consists of 46 dynamically scheduled observations between 2013 January 26 and 2013 May 14, each about 3 hours in duration.
Flux calibration was performed with reference to 3C286, and  J1041+0610 was observed for phase and amplitude calibration.
Three frequency tunings offset in steps of 12 MHz were adopted to cover the gaps between spectral windows and to obtain uninterrupted bandwidth.
\subsection{GOODS-N observations}
The GOODS-N dataset consists of 90 observations between 2013 January 27 and 2014 September 27, each about 2 hours in duration. Pointing 6, which covers the 3 mm PdBI pointing of the CO deep field in \cite{Decarli14} in the HDF-N, was observed both as part of the 1--7 pointing set, and in two additional, targeted observations to achieve better sensitivity.
%COSMOS data were taken in DnC configuration (5/46 tracks) and some GOODS-N pointings  In the following we list the pointing sets that were not fully observed in D-array configuration: pointings GN22--28 (D$\rightarrow$DnC in 1/11), GN29--35 (2/11 in D$\rightarrow$DnC and 9/11 in DnC), deep GN6 (2 in DnC and 10 tracks with the GN1--7 set in D), GN36--42 (8/10 in DnC and 2/10 in DnC$\rightarrow$C), GN43--49 (1/12 DnC), GN50--56 (3/12 in DnC).
%The total on source time was $\sim110$ min (in the 102--120 min range) for the GOODS-N pointings between GN1 and GN42 (294 min on GN6, the deeper pointing at the position of the HDF-N) and 120--135 min for the GN43--GN56 pointings, and 770--820 min for each COSMOS pointing. 
Pointing GN57 was observed for 3 hours (127 min on source) in D-array configuration on 18 December 2015, in order to follow up the most significant negative line feature in GN1--56 (see Appendix D for details).
 J1302+5748 was used for phase calibration, and the flux was calibrated by observing either 3C286 (in 7 observations) or in reference to the phase calibrator (in the remaining observations). An average phase calibrator flux at 34 GHz of  {\it S=}0.343 Jy and spectral index of $-0.2$ was assumed in the observations in 2013 and {\it S=}0.21 Jy and spectral index of $-0.6$ was assumed in 2014, as regularly measured in the tracks where a primary flux calibrator was observed. Based on  track-to-track variations of the calibrator flux, we estimate a $\sim$20\% total flux calibration uncertainty. The spectral setup employed uses two dithered sets of spectral windows, with a relative shift of 16 MHz, in order to fully cover the 8 GHz bandwidth available without gaps.

\begin{figure}[tbh]
 \includegraphics[width=.5\textwidth]{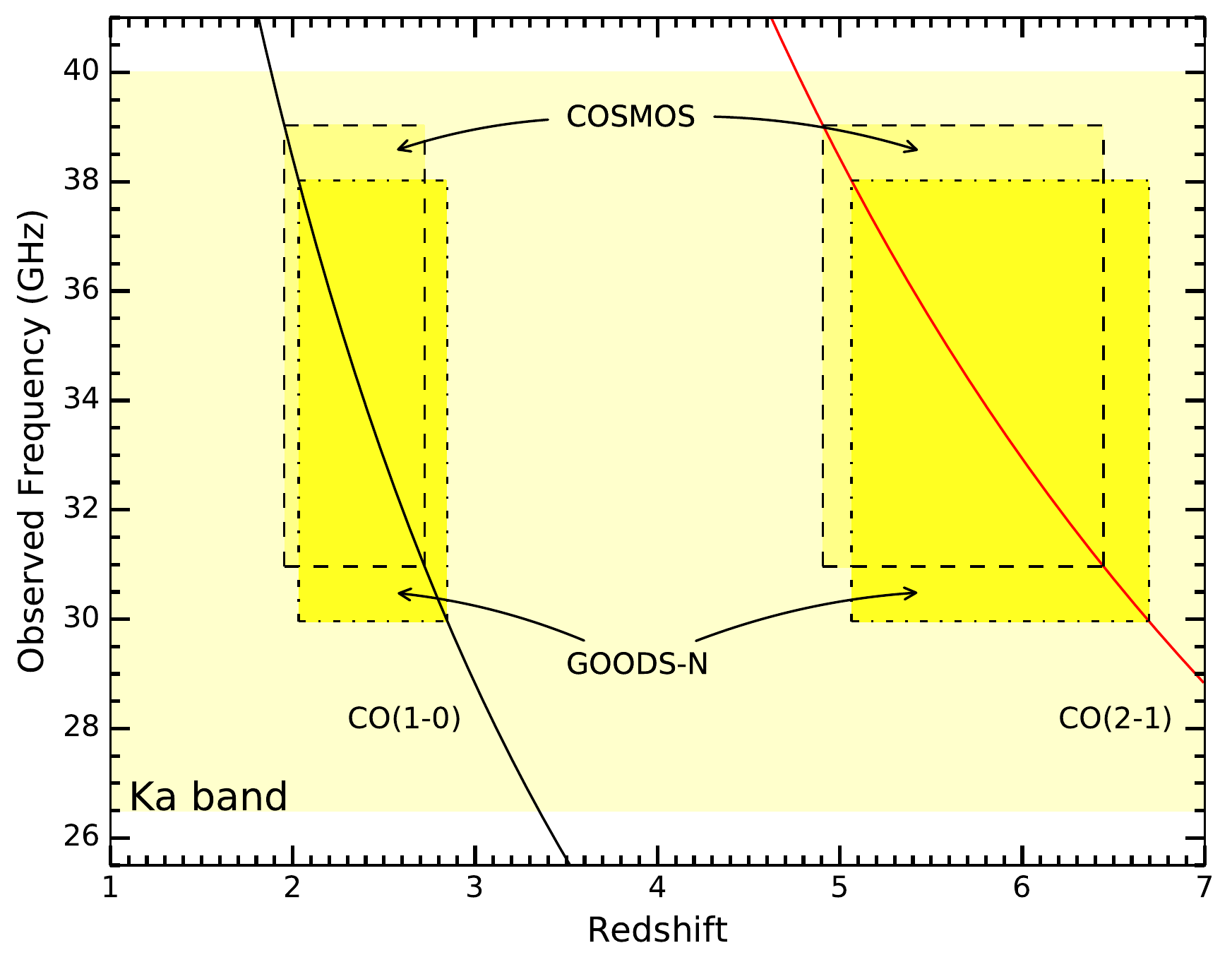}

\caption{Frequency coverage of the VLA COLDz survey, in the Ka band. The frequency range covers  CO(1--0) at $z=1.95$--2.85, and CO(2--1)  at z=4.91--6.70.}
\label{fig:coverage}
\end{figure}

\begin{figure*}[tbh]
\centering{
 \includegraphics[width=\textwidth]{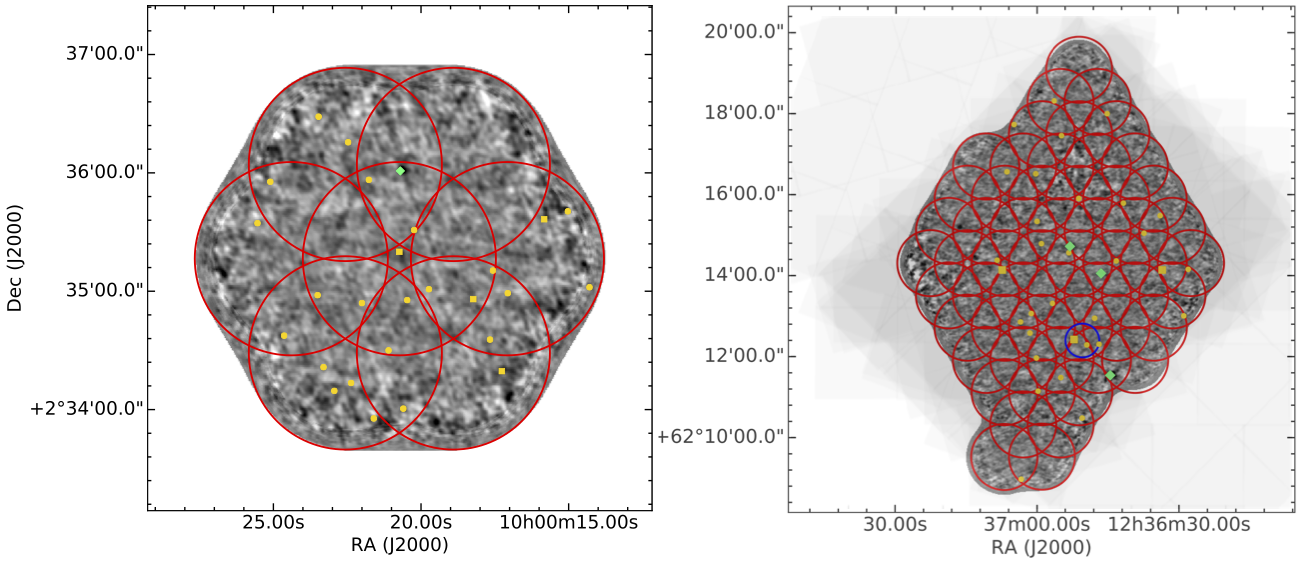}

}
\caption{CO deep field regions covered by COLDz in the COSMOS (left) and GOODS-N/CANDELS  fields (right). The mosaics are composed of 7 and 57 pointings, respectively (as shown by the red circles). The grayscale corresponds to the frequency-averaged signal data cube.  The positions of our line candidates from Tables~\ref{tab_lines} and \ref{tab_lines_appendix} are marked by yellow squares and circles, respectively. Green markers indicate the position of the most significant $\sim34$ GHz continuum sources. The field covered by the blind search from \cite{Decarli14} is shown by a blue circle. We covered the majority of the  CANDELS-Deep footprint in GOODS-N, shown as the background grayscale the {\em HST}/WFC3 F160W exposure map \citep{Grogin} for comparison.}
\label{fig:pointings}
\end{figure*}

\begin{figure*}[tbh]
\centering{
 \includegraphics[width=.8\textwidth]{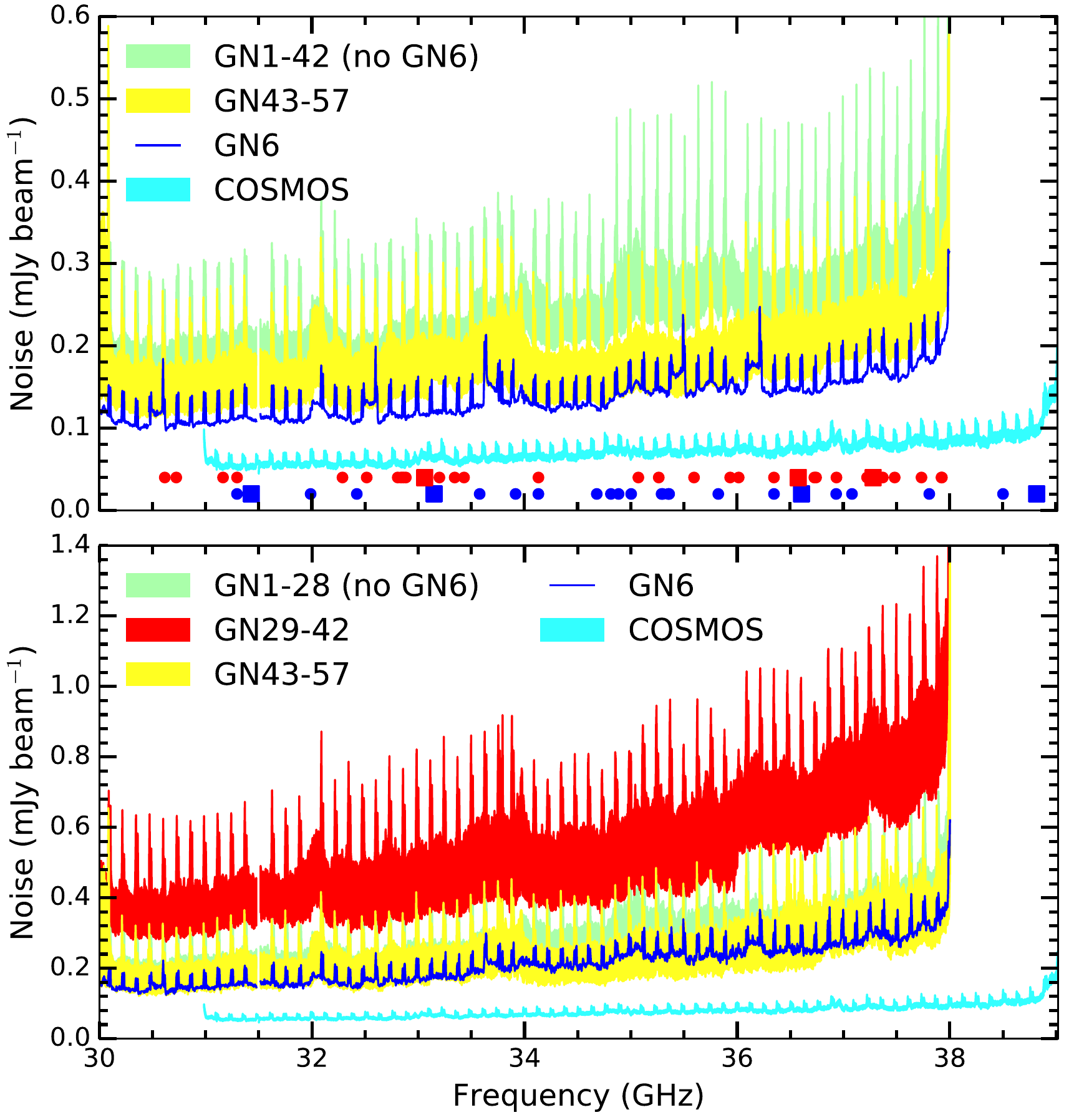}

}
\caption{Measured {\it rms} noise, per pointing, in 4 MHz channels as a function of frequency, at the native spatial resolution (top), and after smoothing to a common beam size (bottom). The bands are for groups of pointings that were observed in similar conditions and thus have similar noise characteristics. The GOODS-N pointings GN29--GN42, which were observed at higher resolution (predominantly in DnC configuration), suffer a significant noise increase in the bottom panel due to spatial smoothing. The frequencies of our line candidates (blue in COSMOS, red in GOODS-N) from Tables~\ref{tab_lines} and \ref{tab_lines_appendix}  are marked by squares and circles, respectively in the top panel.}
\label{fig:Noise}
\end{figure*}

\begin{figure}[tbh]
\centering{
 \includegraphics[width=.48\textwidth]{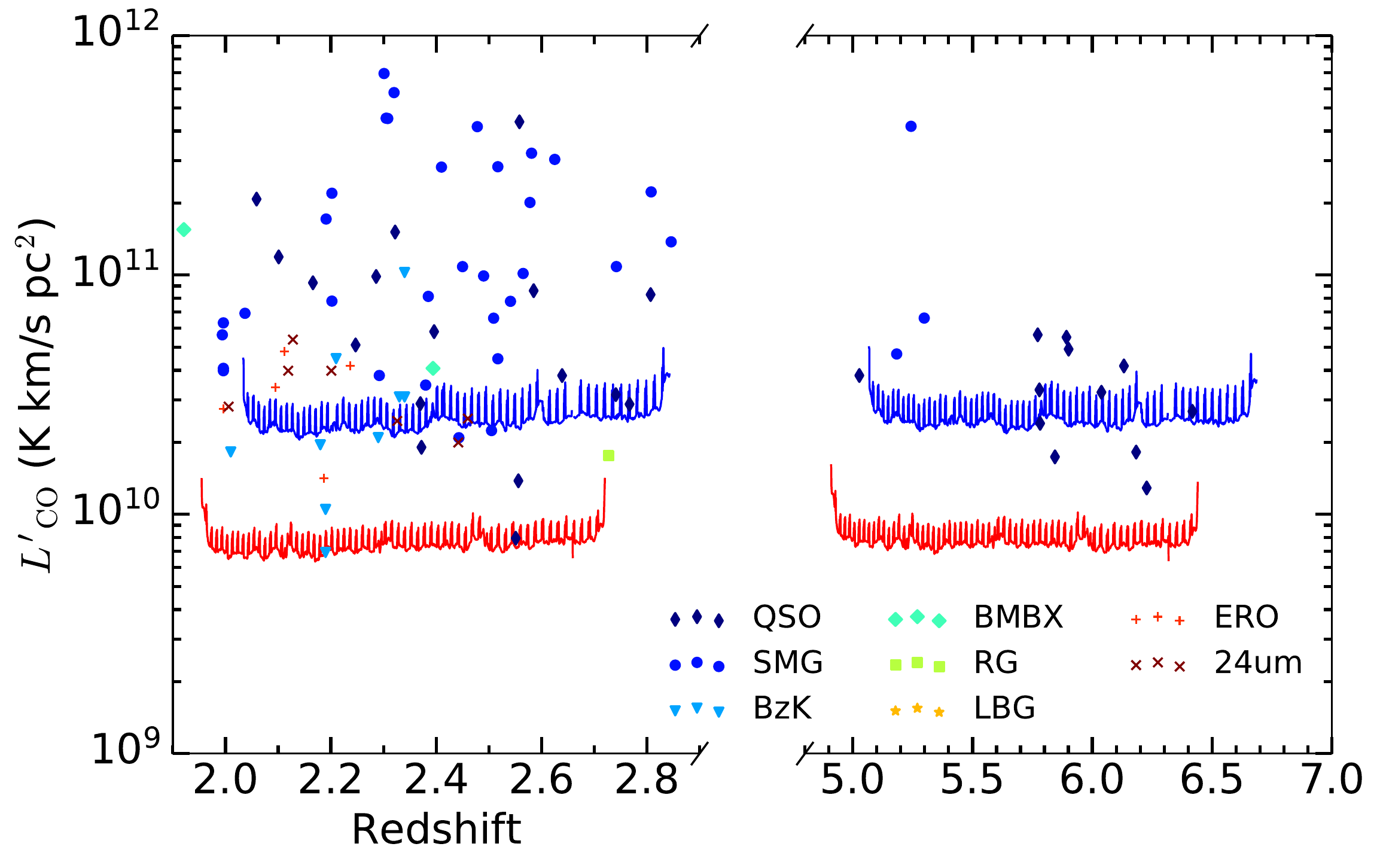}

}
\caption{Line detection sensitivity limit reached by our observations (GOODS-N in blue, COSMOS in red). We assume a line FWHM of 200 km s$^{-1}$, and a 5$\sigma$ limit on the average pre-smoothing noise limit for direct comparison to \cite{ASPECS2}. For comparison, we overplot all $z>1$ CO detections to date from the compilation by \cite{CarilliWalter}, as updated by \cite{Sharon16}. Colors mark different source types (quasars, submillimeter galaxies, 24$\,\mu$m-selected galaxies, Lyman-break galaxies, color-selected galaxies, radio galaxies). At the  sensitivity in the COSMOS field, we would be able
to detect all previously detected CO emitters at high redshift in this compilation.}
\label{fig:sensitivity}
\end{figure}

\begin{table}
\caption{Lines, Redshift Ranges and Volumes Covered by  COLDz.}
%\begin{center}
\centering
\begin{tabular}{ c c c c c c }
\hline
  Transition & $\nu_0$ & $z_{min}$ & $z_{max}$ & $\langle z \rangle$ & Volume \\
   & [GHz] & & & & [$\rm{Mpc}^3$] \\
   \hline
   \multicolumn{6}{c}{COSMOS}\\
   CO(1--0) & 115.271 & 1.953&2.723& 2.354&20,189\\
  CO(2--1) & 230.538 & 4.906& 6.445 &5.684 &30,398\\
  \hline
\multicolumn{6}{c}{GOODS-N}\\
   CO(1--0) & 115.271 & 2.032&2.847&2.443&131,042\\
   CO(2--1) &230.538 &  5.064&6.695 &5.861&193,286\\
  \hline
\end{tabular}
%  \end{center}
 \tablecomments{The comoving volume is calculated to the edges of the mosaic, and does not account for varying sensitivity across the mosaic, which is accounted for by the subsequent completeness correction. The average redshift is cosmic volume weighted.}
\end{table}

\subsection{Data Processing}
Data calibration was performed in {\sc {\sc casa}} version 4.1, using the VLA data reduction pipeline (v.1.2.0). {\sc casa} version 4.5 was used to re-calculate visibility weights using the improved version of \texttt{statwt} that excludes flagged channels when calculating weights, and for imaging and mosaicking \citep{CASA}.
The pipeline radio-frequency interference (RFI) flagging, which uses {\sc casa} \texttt{rflag} to identify transient lines, was switched off, as recommended by the developers, since it can potentially remove narrow spectral lines and because there is little RFI in the Ka-band (with the exception of the 31.487--31.489 GHz range, which we flag prior to running the {\sc casa} pipeline). The pipeline was further modified to only flag the first and last channel of each spectral window (instead of 3 channels), regardless of proximity to baseband edges, to minimize the gap between sub-bands. We find that the bandpass is sufficiently flat that this choice gives the best trade-off between sensitivity in the end channels and additional noise, although some noise increase at the band edges is visible in Figure~\ref{fig:Noise}. After executing the pipeline, we visually inspected the visibilities in the calibrator fields to identify any necessary additional flagging. We then re-executed the pipeline to obtain a final calibration. In addition, for most GOODS-N observations we modify the pipeline to flux-calibrate in reference to the gain calibrator (whenever a primary flux calibrator was not observed).

We identified a small number of noisy spectral channels in our observations that are not removed by the calibration pipeline.
The noisy channels were initially discovered as narrow spikes of a small number of channels in  amplitude vs. frequency plots of visibilities from the science target fields, and they are mostly associated with single antennas.
Being very narrow in frequency (one or two 2-MHz channels), the noise spikes are not significantly reduced by the statistical weights obtained from \texttt{statwt}, which minimizes the effects of all other noise features, since the weights are computed per spectral window.
Including one of these noisy channels for the affected antenna during the imaging of a
single pointing of the mosaic from a single observation track increases the rms noise by
 $\sim20\%$ in that frequency channel.
Selecting channels whose standard deviation exceeds the mean standard deviation in that spectral window for that antenna by $3\sigma$ is a sufficient criterion to exclude most of the problematic noise spikes (these are only of order $\sim0.2\%$ of all channel-antenna combinations). This method is partially redundant to the algorithms in \texttt{rflag} (which we did not execute as part of the pipeline), but it reduces the risk of removing real spectral lines since the noisy channels are selected for individual antennas.
We also found that many noisy channels in the same antenna repeat over time during an observation, and would therefore be more problematic if left in the data cube. We find a concentration of noise spikes in roughly four peaks over the frequency range, which correlate with peaks in the weighted calibrated amplitudes as a function of frequency. We consider this to be indicative of random electronic problems that manifest as increased noise and thus are more prevalent in certain hardware components of the correlator than others. The presence of four peaks is likely associated with the underlying basebands, since there appears to be one peak in each baseband, but no precise correlation of the noise peak frequencies to the baseband edges could be identified. The feature is stochastic and does not appear to preferentially affect any particular subset of antennae. These noise spikes are at least twice as narrow as the narrowest blindly selected line candidates (which are rare among all candidates) and therefore residual anomalous noise spikes are believed not to measurably affect our line search.

Calibrated data from each pointing were imaged separately without any \texttt{CLEAN} cycles, because the fields do not contain strong continuum or line sources (see Section 4). We imaged the total intensity (sum of the two polarizations) using natural weighting and choosing a pixel size of 0.5$^{\prime\prime}$ consistently in the two fields. The smallest adopted channel width is 4 MHz, equivalent to 35 km s$^{-1}$ mid-band, which is less than the typical line width from galaxies. With this choice, our data cubes have $\sim$2000 channels after averaging polarizations.
A crucial aspect of the imaging procedure, necessary for blind line searches, is to avoid any frequency regridding by interpolation, therefore we image using the nearest channel, rather than interpolating. Interpolation would introduce correlations between the noise of adjacent channels, undermining the statistical basis of the search for spectral lines, and producing a significant number of spurious noise lines.
Disabling frequency interpolation when imaging visibilities may introduce a very small, less than half a channel, frequency error that we consider negligible because 4 MHz channels (35 km s$^{-1}$) are smaller than the typical linewidth.

The geometric average synthesized beam size for COSMOS ranges between 2.2$^{\prime\prime}$ and 2.8$^{\prime\prime}$ as a function of frequency and the beam axes ratio is in the range of 0.8--1, while for GOODS-N the synthesized beam size differs more significantly from pointing to pointing. In particular, the main difference is between the subset of ``high resolution pointings" (GN20--GN42) and the rest  (GN1--GN19 and GN43--GN57). The geometric average beam size ranges between approximately 1.3$^{\prime\prime}$ to 2.0$^{\prime\prime}$ for the high resolution pointings, and between 2.1$^{\prime\prime}$ and 3.1$^{\prime\prime}$ for the rest. The beam axes ratio is in the range of 0.6--0.9.
The individual pointing cubes are subsequently smoothed to a common beam size of  3.38$^{\prime\prime}\times$ 2.91$^{\prime\prime}$ for COSMOS and 4.1$^{\prime\prime}\times$ 3.2$^{\prime\prime}$ for GOODS-N, using the {\sc casa} task \texttt{imsmooth}. This compromise in resolution and signal-to-noise is necessary in order to mosaic all pointings together, and to search for line emission in a uniform manner. This is called the Smoothed-mosaic. 
Separately, we have also mosaicked the pointings with their native resolution (after removing the beam information from the headers), and this Natural-mosaic (where the resolution is set by natural weighting) was used to exclusively search for spatially un-resolved sources, for which the spatial size information is not important. Fig.~\ref{fig:pointings} shows the spatial coverage provided by the individual pointings in our two mosaicked fields.

The {\sc casa} function \texttt{linearmosaic} was used to mosaic the images of our COSMOS data together. We wrote a custom script to optimally mosaic the images of the GOODS-N data, using Equation.~\ref{mosaic_formula}, which takes into account the different noise levels in different pointings, per channel, in order to compute optimal weights for mosaicking:
\begin{equation}
%I=\frac{\sum_p I_p A(x-x_p)}{\sum_p A^2(x-x_p)}
I=\frac{\sum_p I_p A({\bf x}-{\bf x_p})/\sigma_p^2}{\sum_p A({\bf x}-{\bf x_p})^2/\sigma_p^2},
\label{mosaic_formula}
\end{equation}
where $A$ is the primary beam function, ${\bf x_p}$ are the pointing center positions, $I_p$ represents the specific intensity data from pointing p and $\sigma_p$ is the noise level in pointing p (computed on a per channel basis).

All COSMOS pointings were always observed in every execution, and for comparable amounts of time. The GOODS-N pointings  were observed in blocks of 7, over the course of several months due to scheduling constraints. Therefore, they have slightly different noise levels, partly due to the upgraded 3-bit samplers in the later (2014) observations. Furthermore, some GOODS-N data were not taken in the D configuration but rather in a combination of the DnC configuration, in transition between the D and the DnC and in transition between the DnC and the C configuration.  Therefore, when smoothed to a common beam, these pointings have higher noise, because the information on the the longest baselines is effectively discarded.
For these reasons, the noise is significantly spatially varying in the smoothed version of the GOODS-N mosaic, which we take into account when analyzing the data (Fig.~\ref{fig:Noise} shows the noise before and after smoothing). 
All pointings suffer a noise increase due to smoothing, because the targeted beam for the smoothing process has to be larger than every beam in any pointing, at any frequency, and also includes those beams that have different position angles. 
%Optimal mosaicking in the case of unequal noise in the pointings would require weighting the linear combination by the inverse square of the noise in the individual maps, but the {\sc casa} function \texttt{linearmosaic} does not implement this possibility. We therefore do not correct for this imperfect weighting, which means paying a small noise penalty for the advantage of benefiting from the well tested {\sc casa} code.
In the case of the COSMOS mosaic, the {\sc casa} function \texttt{linearmosaic} produces the mosaic edge at the 30\% of peak level sensitivity (per-channel). For consistency, we therefore apply the same criterion in our GOODS-N mosaics. In order to do this, we define a mask which produces the mosaic edge at 30\% of peak sensitivity in the Natural-mosaic, and utilize the same mask for the Smoothed-mosaic for consistency.

\subsection{Constructing the Signal-to-Noise cubes}
In order to search for emission lines in our data, we produce a signal-to-noise ratio (SNR) cube by calculating a noise value for each pixel and in each frequency channel of the mosaics.
Spatial variations in the noise are introduced by mosaicking pointings with different noise levels and primary beam corrections. The noise in the resulting mosaic can be calculated assuming statistical independence of the noise in different pointings, and can therefore be calculated by summing their standard deviations in quadrature, with weights given by Eqn.~\ref{mosaic_formula}:
\begin{equation}
%\sigma(x)=\frac{\sqrt{\sum_p \sigma_p^2 A^2(x-x_p)}}{\sum_p A^2(x-x_p)}.
\sigma(x)=\frac{1}{\sqrt{\sum_p A({\bf x}-{\bf x_p})^2/\sigma_p^2}},
\label{mosaic_noise}
\end{equation}

\noindent where $\sigma_p$ is the measured noise in the individual pointing images, $A$ is the primary beam function and ${\bf x_p}$ are the pointing center positions. In the special case of pointings with approximately equal noise (as in our COSMOS data) we can use a simplified expression, where the denominator is simply the square root of the sensitivity map, output from {\sc casa}'s \texttt{linearmosaic} function:
\begin{equation}
\sigma(x)=\frac{\sigma}{\sqrt{\sum_p A({\bf x}-{\bf x_p})^2}}.
\end{equation}

The frequency variation of the noise is  accounted for by measuring the noise in each frequency channel, in the individual pointings.
In COSMOS, where the noise variations from pointing to pointing can be neglected, we calculate the signal-to-noise ratio by multiplying the signal cube by the square root of the sensitivity map (which gives spatially uniform noise), and then dividing each channel map by its standard deviation to normalize the pixel value distribution. In GOODS-N, we measure the noise in each pointing  and apply Eqn.~\ref{mosaic_noise} to compute noise and signal-to-noise ratio cubes.

\section{Line search methods} 
The main objective of this survey is to carry out a blind search for CO(1--0) and CO(2--1)  emission lines in the COLDz dataset. No other bright lines are expected to contaminate the 30--39 GHz frequency range (Fig.~\ref{fig:coverage}). The line brightness sensitivity is approximately equal for the low and high redshift bins (corresponding to CO(1--0)  and CO(2--1), respectively; Figure~\ref{fig:sensitivity}).  We therefore expect a higher source density in the low redshift bin due to the expected evolution of the cosmic gas density \citep{Popping14,Popping16}.
Hence, we will assume that all detected features correspond to CO(1--0) unless data at other wavelengths  suggest that they belong to the higher redshift bin. 
In order to detect emission lines in our data cubes, we have implemented a previously published method ({\sc spread}, \citealt{Decarli14}), and developed three new methods to explore the differences between different detection algorithms (see Appendix A for details).

The objective of a line search algorithm is to systematically assess the significance (expressed as Signal-to-Noise ratio, SNR) of candidate emission lines in the data. The relevant information available to us is the strength of the signal, the number of independent samples that make up the line, and the spatial and frequency structure (for which we have priors based on previous samples of CO detections at high redshift).
In particular, we expect most CO sources to be either unresolved or resolved over a few beams at most at the $\sim3^{\prime\prime}$ resolution of our mosaics (which corresponds to $\sim$25 kpc at $z\sim2.5$ and $\sim17$ kpc at $z\sim6$), and we expect the line FWHM to be in the range of 50 to 1000 km s$^{-1}$ \citep{CarilliWalter}.

Our line search method of choice, Matched Filtering in 3D (MF3D), expands on the commonly used spectral Matched Filtering (MF1D; e.g, {\sc aips} \texttt{serch}). Matched Filtering corresponds to convolving the data with a filter, or template, which is ``matched" to the sources of interest in order to attenuate the noise and concentrate the full signal-to-noise of the source in the peak pixel. A detailed description of the MF3D method is presented in Appendix A.

We also implement and test some of the previously used methods on our data, in particular {\sc spread} and Matched Filtering in the spectral domain, i.e., in 1D. The main limitation of {\sc spread} is that it does not employ the full spatial information available, but only utilizes signal strength. While Matched Filtering in 1D is arguably the optimal search method for completely unresolved sources (for which the spectrum at the peak spatial pixel contains the full information), it still requires a prescription for identifying pixels belonging to the same source, and it needs to be generalized to account for the possibility that some sources may be slightly extended. Besides accounting for extended sources, the 3D Matched Filtering also captures the spreading of the signal-to-noise over different spatial positions in different frequency channels, which is at least in part a consequence of moderate SNR. For this reason, it is natural to use the spatial information by using templates that include a spatial profile. Therefore we extended the method to Matched Filtering with 3D templates. A description of the detailed implementation of all line search methods and a more detailed comparison is presented in Appendix A.

%In the next section we will describe the details of what we think is the best solution to line searching our data-cube, then we will briefly elaborate on the comparison to the other methods we have implemented in the following sections.

\section{Results of The Line Search}
The 3D Matched-Filtering procedure provides an output including the maximal SNR for each line candidate, the position in the cube where that maximal SNR is achieved, the number of templates for which the candidate has $>4 \sigma$ significance, and the template size (spatial and frequency width) where the highest SNR is achieved.
We run the line search down to a low SNR threshold of $4\sigma$.  The number of identified features is very large, due to the large number of statistical elements in our data cubes. Specifically, we estimate approximately  $2.8\times 10^6$ and $1.7\times 10^7$ independent elements for the COSMOS and GOODS-N fields, respectively, by dividing the mosaic area by the beam area and dividing by a line FWHM of 200 km s$^{-1}$.  However, we caution that naively estimating the extent of the noise tails from these numbers does not provide a good estimate, as previously described by \cite{Vio16,Vio17} (also see Appendix F.2 for more details).

We mask  radio continuum  sources in our fields, which contaminate the line candidates: one in the COSMOS field at 10:00:20.67  +02:36:01.5 with a flux of 0.024 mJy beam$^{-1}$, and three in the GOODS-N field at 12:36:44.42 +62:11:33.5 with a flux of 0.3 mJy beam$^{-1}$, 12:36:52.92 +62:14:44.5 with a flux of 0.17 mJy beam$^{-1}$ and 12:36:46.34 +62:14:04.46 with a flux of 0.07 mJy beam$^{-1}$ (Hodge et al., in prep.).  Even though the continuum fluxes of these sources only have low significance in the individual channels ($<0.3\sigma$ and $<2\sigma$ per 4 MHz channel for the brightest source in COSMOS and GOODS-N, respectively), we remove any candidate within 2.5$^{\prime\prime}$ of the spatial positions of these sources, because they are likely spurious and caused by noise superposed to the continuum signal. Specifically, once we remove the continuum flux from their spectra, the significance of those line candidates becomes lower than $\sim4.5\sigma$, indicating that they likely correspond to noise peaks. 

%In Table~\ref{tab_lines}, we present the list of our highest significance candidate lines in COSMOS and in GOODS-N. While we are confident that our highest SNR ($>6.4\sigma$) candidates correspond to real CO emission lines because they all have identified multi-wavelength counterparts, we here present the longer lists of line candidates which have significantly lower purity defining a statistical sample. Although only a fraction of the tabulated sources are real emission lines, they provide statistical information once we account for their fractional purity, and therefore they may be used to constrain the CO luminosity function.
In Table~\ref{tab_lines}, we present the list of the secure line emitters in COSMOS and in GOODS-N which were independently, spectroscopically confirmed. While we are confident that our highest SNR ($>6.4\sigma$) candidates correspond to real CO emission lines because they all have identified multi-wavelength counterparts, we also define a longer lists of line candidates which have significantly lower purity ($\sim$5\%-40\%) as a statistical sample in Table~\ref{tab_lines_appendix} in Appendix E, as described below. Although only a fraction of those tabulated sources are real emission lines, they provide statistical information once we account for their fractional purity, and therefore they may be used to constrain the CO luminosity function. 
While a fraction of these lower significance candidates may be expected to correspond to real CO emission, we advise caution in interpreting these lower significance candidates on a per-source basis until they are independently confirmed.

In order to determine the reliability of the line candidates presented in Table~\ref{tab_lines_appendix}, we compare the SNR distribution to that for ``negative" line candidates, following the standard practice (e.g., \citealt{Decarli14, ASPECS1}) which relies on the symmetry of interferometric noise.
We provide a detailed description of our candidate purity estimation in Appendix F, but we point out that an excess of positive candidates over the negatives, for signal-to-noise ratios above a threshold is an indication that at least a fraction of those positive candidates may correspond to real sources, rather than due to noise.
By adopting this criterion, we determined the SNR thresholds for our candidate lists consistently for both fields by cutting at the SNR level that includes as many negative line candidates as unconfirmed positive candidates. Thus, we exclude from the count the high signal-to-noise, confirmed sources (4 in COSMOS and 2 in GOODS-N, Table~\ref{tab_lines}), and we require that the number of unconfirmed sources is greater than the number of negative lines down to the same SNR threshold, thereby constituting an excess.
This procedure determines SNR thresholds on the candidates catalog of 5.25$\sigma$ for the smaller COSMOS field and 5.5$\sigma$ for the wider GOODS-N field (Table~\ref{tab_lines_appendix}). The threshold is chosen to be higher in the wide, GOODS-N mosaic  because the larger number of statistical elements produces more pronounced noise tails.

\begin{table*}
\tiny
%\begin{deluxetable}
\caption{{\rm Catalog of the secure line candidates identified in our analysis, which have been independently confirmed (see Table~\ref{tab_lines_appendix} for the remainder of the full statistical sample). Columns are: (1) Line ID. (2-3) Right ascension and Declination (J2000).  (4) Central line frequency and uncertainty, based on Gaussian fitting. (5) CO(1--0) redshift and uncertainity, unless otherwise noted. (6) Velocity integrated line flux and uncertainty. (7) Line Full Width at Half Maximum (FWHM), as derived from a Gaussian fit. (8) SNR measured by MF3D. (9) Presence of a spatially coincident optical/NIR counterpart (10) Comments.}} \label{tab_lines}
\begin{center}
\begin{tabular}{cccccccccl}
\hline
ID  & RA	  & Dec 	& Frequency	 & Redshift & Flux	     & FWHM     & $S/N$ & Opt/NIR & Comments \\  
    & (J2000.0)   & (J2000.0)	& [GHz] 	&  & [Jy\,km\,s$^{-1}$] & [km\,s$^{-1}$] &		   & c.part? &          \\  
 (1) & (2)	  & (3) 	& (4)		  & (5) 	     & (6)	& (7)		   &  (8)    & (9)  &(10)    \\  
\hline
\multicolumn{9}{l}{\bf COSMOS} \\
COLDz.COS.0  & 10:00:20.70 & +02:35:20.5 & $ 36.609 \pm 0.002$ &$5.2974\pm 0.0003^a$& $0.17\pm0.02$ & $390\pm40$  & 14.7     & Y  &  AzTEC-3 \\
COLDz.COS.1  & 10:00:15.80 & +02:35:37.0 & $ 31.430\pm0.003$ &$2.6675\pm0.0004$& $0.11\pm0.03$ & $430\pm80$  & 10.6     & Y & \\
COLDz.COS.2  & 10:00:18.20 & +02:34:56.5 & $ 33.151 \pm0.006$ &$2.4771\pm0.0006$& $0.13\pm0.03$ & $830\pm130$  & 9.6    & Y & source reported by \cite{Lentati15}\\
COLDz.COS.3  & 10:00:17.23& +02:34:19.5 & $ 38.822\pm0.003$ & $1.9692\pm0.0002$&$0.37\pm0.10$ & $ 240\pm50$  &     9.2  & Y & extended\\
 \hline
 \hline
 \multicolumn{9}{l}{\bf GOODS-N} \\
 COLDz.GN.0 & 12:36:33.45 & +62:14:08.85 & $ 36.578\pm0.005$ & $ 5.3026\pm0.0009^a$ & $ 0.344\pm0.074$ & $ 610\pm100$ & 8.56 & Y  &  GN10 \\
COLDz.GN.3 & 12:37:07.37 & +62:14:08.98 & $ 33.051\pm0.006$ & $ 2.4877\pm0.0006$ & $ 0.34\pm0.12$ & $ 580\pm120$ & 6.14 & Y & GN19 \\
{\it COLDz.GN.31$^b$} & 12:36:52.07 & +62:12:26.49 & $ 37.283\pm0.007$ & $ 5.1833\pm0.0008^a$ & $ 0.148\pm0.057$ & $ 490\pm140$ & 5.33 & Y  &  HDF850.1 \\
\hline
\end{tabular}
\tablecomments{$^a$ CO(2--1)  redshift.
$^b$ Source is below the formal catalog threshold adopted here, and therefore, not part of the statistical sample.}
\end{center}
%\end{deluxetable}
\end{table*}

\subsection{Measuring line candidate properties}

After selecting the blind search line candidates, we separately measure their line properties using a standard method described in the following. The statistical corrections  were computed adopting identical methods in the artificial source analysis (Appendix F.3).

In order to extract the spectrum of the line candidates, we fit a 2D-Gaussian to the velocity-integrated line maps and extract the flux in elliptical apertures with sizes equal to the FWHM of the fitted Gaussians. For the integrated line maps, we use a velocity range equal to the FWHM of the template that maximizes the SNR. This procedure is expected to provide the highest SNR of the extracted flux. In the infinite SNR case, this aperture choice includes half of the total flux, and we therefore correct the extracted flux scale of the spectrum by a factor of two.
We then fit a Gaussian line profile to the aperture spectrum and measure its peak flux and velocity width, from which we derive the integrated fluxes reported in Tables~\ref{tab_lines} and \ref{tab_lines_appendix}.
We also measure peak fluxes for the candidates, which are expected to best represent the correct flux for unresolved sources. For the peak fluxes, we extract the spectrum at the highest pixel in the integrated line map. 
We find that the peak fluxes are compatible with aperture fluxes for point-like sources, and so we choose to adopt the aperture fluxes because they measure the full flux of extended sources at the expense of slightly larger uncertainties.
We calculate the positional and size uncertainty of the 2D Gaussian fitting using the {\sc casa} task \texttt{imfit}, applied to the same integrated line maps described above. The positional uncertainty is relevant when establishing  counterpart associations  (as detailed in Appendix E for the full candidate list). It is dominated by the detection SNR and the spatial size of the synthesized beam or extended emission. 
%, our artificial sources analysis suggests that the flux recovered from the single-pixel analysis may be more biased than for aperture fluxes, due to the moderate SNR of our line detections and the positional uncertainty. 

In the COSMOS field, we can measure aperture fluxes in the Natural-mosaic, to make full use of the highest SNR (the fluxes are typically within 20\% of the values measured in the Smoothed-mosaic). Specifically, the 7 pointings of the mosaic have an approximately equal beam size.  This allows us to calculate an average beam size for each channel and hence, to correctly measure aperture fluxes. These are the fluxes we report in Tables~\ref{tab_lines} and \ref{tab_lines_appendix} for the most significant candidates,  which we also use for the luminosity function.

In the GOODS-N field, on the other hand, we are limited to measuring aperture fluxes  for resolved objects in the Smoothed-mosaic, because of the strong beam size variations across the mosaic which make it impossible to precisely define a beam in the Natural-mosaic. Nonetheless, since most of the candidates are unresolved in the original data (show highest SNR in the Natural-mosaic), in those cases we report the peak fluxes, measured in the Natural-mosaic, without concern for missing any flux, and without being affected by the beam size variations.

In the GOODS-N field, there is another beam size effect that needs to be taken into account even in the Smoothed-mosaic. The measured beam size is actually larger than the formal $4.1^{\prime\prime} \times 3.2^{\prime\prime}$ size which was targeted with the {\sc casa} task \texttt{imsmooth}, and is slightly pointing-dependent, as explained in Appendix C. The measured beam area is $\sim1.4$ times larger in the D-array only pointings, and $\sim1.7$ times larger in the higher resolution pointings than the target size for the smoothing procedure, because of the precise $uv$-plane coverage and the effect of tapering. Therefore, we measure the correct beam size after smoothing, by Gaussian-fitting to the smoothed dirty beam, in each pointing, for each channel. 
We correct the aperture flux for each candidate line detection in the Smoothed-mosaic by calculating an effective beam area given by a weighted average of the beams of the overlapping pointings, weighted by the square of the primary beams (the same weighted average that determines the flux in the mosaic).
We calculate aperture fluxes in this way in the Smoothed-mosaic, and confirmed that the peak pixel flux in unresolved sources matches this corrected aperture flux, within the uncertainties.

The measured CO line fluxes are affected by the effect of a warmer cosmic microwave background (CMB) at the redshift of our sources, which is a uniform background (hence invisible to an interferometer) at the small scales  of galaxy sizes \citep{daCunha13}. While we do not expect corrections for our $z=2$--3 sources to be significant ($\sim20-25$\%)  a larger correction (up to a factor $\sim2$) may be required if the gas kinetic temperature were lower than expected.
On the other hand, the CO(2--1)  line luminosity from the $z>5$ sources may be underestimated by up to a factor of  $\sim2-5$ \citep{daCunha13}. We do not apply any of these corrections to the measured line flux values reported here. These effects will be further discussed in Paper~\rom{2}, in the context of the CO luminosity function.

\subsection{Individual candidates}
\begin{figure*}[tbh]
\includegraphics[scale=1.7]{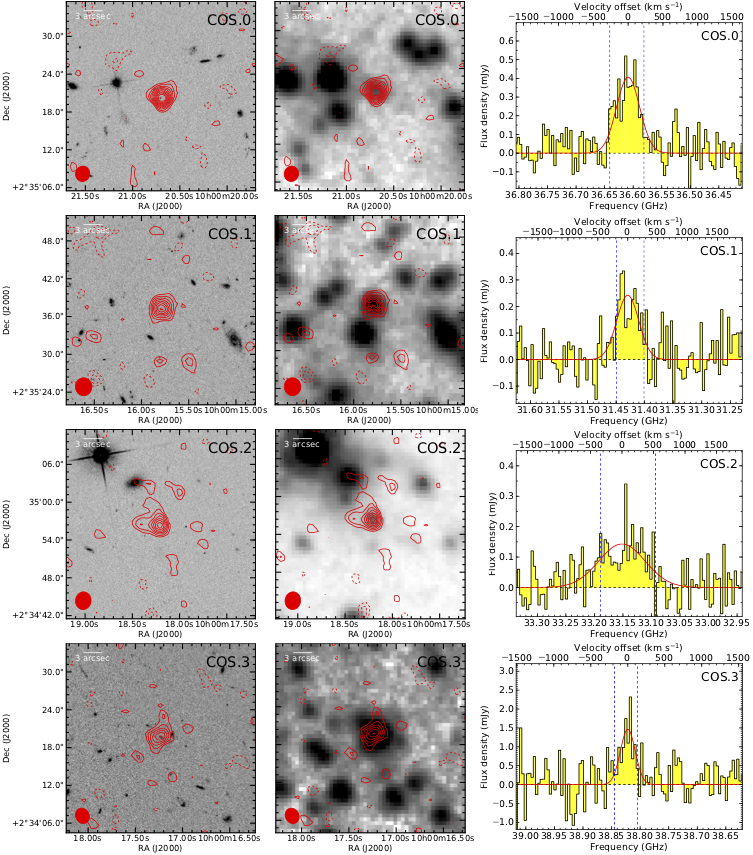} 

\caption{Independently confirmed  candidates  from our blind line search in the COSMOS field. CLEANed integrated line emission (contours) is shown overlaid on {\em HST} I-band (left) and IRAC $3.6\,\mu$m images (middle) from SPLASH (grayscale; \citealt{Steinhardt14}).  Contours are shown in steps of $1\sigma$, starting at $\pm2\sigma$. COS.0 corresponds to CO(2--1)  emission from AzTEC-3.
Right: Extracted line candidate aperture spectra (``histograms") and Gaussian fits (red curves) to the line features. The observer-frame frequency resolution of 4 MHz corresponds to $\sim35$ km s$^{-1}$ mid-band.  
The velocity range that was used for the overlays is indicated by the dashed blue lines.}
\label{fig:spectra_COS1}
\end{figure*}

We have identified 26 line candidates in the COSMOS field down to a SNR threshold of 5.25, and 31 candidates in the GOODS-N field down to a SNR threshold of 5.5 (Tables~\ref{tab_lines} and \ref{tab_lines_appendix}).
The top four sources in COSMOS and two among the highest SNR sources in GOODS-N have been independently confirmed through additional CO transitions (\citealt{Daddi09,Riechers10a,Riechers_GN19}; and in prep.; \citealt{Ivison_GN19}; Pavesi et al., in prep.).
Furthermore, we include COLDz.GN.31 in this set of independently confirmed sources (Table~\ref{tab_lines}),  although it is slightly below the formal 5.5$\sigma$ cutoff, because it corresponds to CO(2--1)  line emission from HDF850.1 \citep{Walter12}. This line source does not contribute to our evaluation of the CO(2--1)  luminosity function because it does not satisfy the significance threshold to be included in the statistical sample (Paper~\rom{2}).
 For reference, we here briefly describe these individual secure candidates and we show their CO line maps and spectra in Figs.~\ref{fig:spectra_COS1} and \ref{fig:spectra_GN_main}.  Line maps and spectra of the complete statistical sample are presented in Appendix E for reference.

We detect four previously known dust-obscured massive starbursting galaxies, and three  secure sources in the COSMOS field that lie within the scatter of the high-mass end of the Main Sequence at $z\sim2$ (\citealt{Lentati15}; Pavesi et al., in prep.) These galaxies may be representative of a galaxy population that has not been well studied to date, due to our novel selection technique.

\textbf{COLDz.COS.0}: We identify the brightest candidate in the COSMOS field with CO(2--1)  from the z=5.3 sub-millimeter galaxy AzTEC-3, detected at a SNR of 15, and which was chosen to be near the center of our survey region. This galaxy is known to reside in a massive proto-cluster \citep{Riechers10a,Riechers14, Capak11}. The line flux is compatible with the previously measured value of $0.23\pm0.03$ Jy km s$^{-1}$  \citep{Riechers10a} within the relative flux calibration uncertainty. This source is also detected at 3 GHz, with a flux of $20\pm3\,\mu$Jy  \citep{Smolcic17}, and by SCUBA-2 at $850\mu$m as part of the S2COSMOS survey with a significance of $9.3\sigma$ and a flux of $8.1^{+1.1}_{-1.3}$ mJy (J.M. Simpson, et al. in prep).

\textbf{COLDz.COS.1}: This high signal-to-noise detection is matched in position (offset $0.3^{\prime\prime}\pm0.3^{\prime\prime}$) and CO(1--0) redshift to a source with photometric redshift ($z_{phot}$=2.6--2.9) in the COSMOS2015 catalog \citep{Laigle16}. We have confirmed its redshift with ALMA through a detection of the CO(3--2) line (Pavesi et al., in prep.). This source is also detected at 3 GHz, with a flux of $15\pm2\,\mu$Jy \citep{Smolcic17}, and  at $850\mu$m with a significance of $6.0\sigma$ and a flux of $4.9_{-1.2}^{+1.1}$ mJy (J.M. Simpson, et al. in prep).

\textbf{COLDz.COS.2}: This high signal-to-noise detection is matched in position (offset $0.3^{\prime\prime}\pm0.3^{\prime\prime}$) to a source in the COSMOS2015 catalog \citep{Laigle16}.  We have confirmed its redshift with ALMA through a detection of the CO(3--2) line (Pavesi et al., in prep.), and some of its properties were previously presented in \citep{Lentati15}. The photometric redshift in the COSMOS2015 catalog is highly uncertain, and not compatible with the CO redshift of 2.477 within $1\sigma$ ($z_{phot}$=2.9--4.4). This source is also detected at 3 GHz, with a flux of $19\pm3\,\mu$Jy \citep{Smolcic17}, and  at $850\mu$m with a significance of $5.9\sigma$ and a flux of $4.0_{-1.0}^{+0.9}$ mJy (J.M. Simpson, et al. in prep).

\textbf{COLDz.COS.3}: This high signal-to-noise detection, is a significantly spatially extended CO source with a deconvolved size of $(4.0^{\prime\prime}\pm1.1^{\prime\prime})\times(1.8^{\prime\prime}\pm1.2^{\prime\prime})$. It is matched in position to two galaxies in the COSMOS2015 catalog (\citealt{Laigle16}; offsets of $0.14^{\prime\prime}\pm0.3^{\prime\prime}$ and $1.8^{\prime\prime}\pm0.3^{\prime\prime}$). We have confirmed its CO(1--0) redshift with ALMA through a detection of the CO(4--3) line (Pavesi et al., in prep.). The cataloged photo-$z$ for both galaxies ($z_{phot}$=1.8--1.9) is not  compatible with the CO redshift of 1.97 within $1\sigma$. This source is also detected at 3 GHz, with a flux of $27\pm3\,\mu$Jy \citep{Smolcic17}. The S2COSMOS survey shows a weak signal at $850\mu$m  with a significance of $3.7\sigma$. The formal $4\sigma$ limit on the deboosted flux is $<4.0$ mJy, and the tentative detection suggests a potential source at a flux level of $\sim2-3$ mJy (J.M. Simpson, et al. in prep).

\textbf{COLDz.GN.0}: We identify the brightest candidate in the GOODS-N field with CO(2--1)  line emission from GN10, a massive, bright dust-obscured starbursting galaxy \citep{Pope06, Dannerbauer08, Daddi09}. We find a CO redshift of $z=$5.3, showing that the previous redshift determination ($z$=4.04) was incorrect.  Its properties are described in Riechers et al., in prep. This source is also detected at 1.4 GHz with a flux of $36\pm4\,\mu$Jy \citep{Morrison10}, and by SCUBA-2 at $850\mu$m in the SCUBA-2 Cosmology Legacy Survey (S2CLS)  with a significance of $9.2\sigma$ and a flux of $7.5\pm1.5$ mJy \citep{Geach17}.

\textbf{COLDz.GN.3}: We identify this source with CO(1--0) line emission from GN19, a merger of two massive, bright dust-obscured starbursting galaxies at $z$=2.49 found by \cite{Pope06} and characterized in detail by \cite{Tacconi06,Tacconi08}, \cite{Riechers_GN19}, and \cite{Ivison_GN19}. 
It is detected by the 5.5 GHz eMERGE survey, with a flux of $9.6\pm1.7\,\mu$Jy \citep{eMerge}. Its line flux is compatible with the previously measured total flux of $0.33\pm0.04$ Jy km s$^{-1}$ from \cite{Riechers_GN19}. This source is also detected at 1.4 GHz, with a flux of $28\pm4\,\mu$Jy and $33\pm4\,\mu$Jy for the W and E components, respectively  \citep{Morrison10}, and at $850\mu$m with a significance of $7.9\sigma$ and a flux of $6.5\pm1.1$ mJy \citep{Geach17}.

\textbf{COLDz.GN.31}: We also detect CO(2--1)  line emission from the bright, dust-obscured starbursting galaxy HDF850.1 ($z$=5.183), with a moderate significance of SNR=5.3. We include this line detection here given the known match, but we do not include it in the statistical analysis because it does not reach the significance threshold for detection by the blind line search.
The measured flux is compatible with the previously reported flux of $0.17\pm0.04$ Jy km s$^{-1}$ \citep{Walter12}. It is detected by the 5.5 GHz eMERGE survey, with a flux of $14\pm3\,\mu$Jy \citep{eMerge}, but is not detected at 1.4 GHz \citep{Morrison10}. This source is also detected  at $850\mu$m with a significance of $7.1\sigma$  and a flux of $5.9\pm1.3$ mJy \citep{Geach17}.

The other line candidates identified by our blind line search with moderate significance are to date not independently confirmed (Table~\ref{tab_lines_appendix}). Thus, we only use their properties in a statistical sense in the following, to place more detailed constraints on the CO luminosity function. We point out that three out of the seven secure, confirmed sources in our blind search belong to the high redshift bin, and therefore suggest caution in interpreting the indicated CO(1--0) redshift, especially for those line candidates without strong counterparts. We describe the complete candidate sample in Appendix E, where we also discuss   potential counterpart associations. In Appendix F, we develop novel statistical techniques to evaluate the purity and completeness of this statistical sample, which yield the best constraints to the CO(1--0) luminosity function at $z\sim$2--3 to date (Paper~\rom{2}).

\begin{figure*}[tbh]
\includegraphics[scale=1.7]{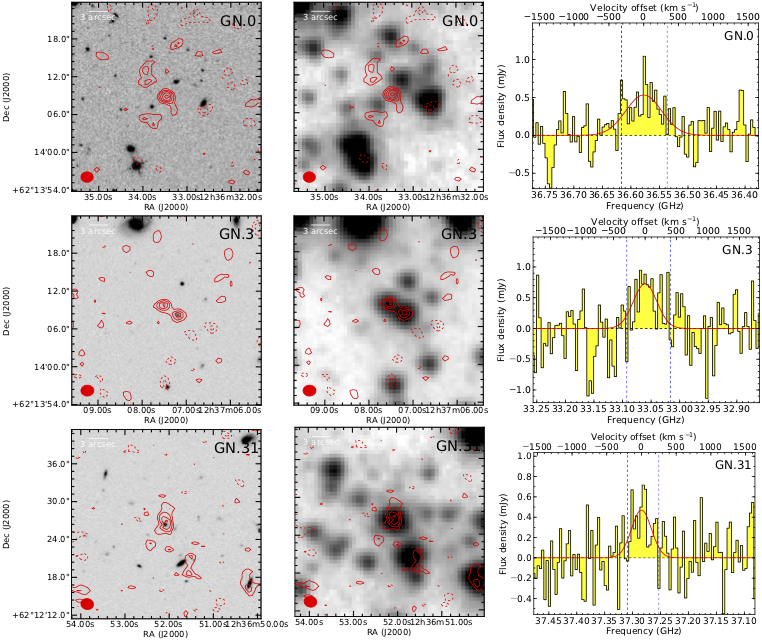} 
\caption{Independently confirmed  candidates  from our blind line search in the GOODS-N field. Integrated line emission (contours) is shown overlaid on {\em HST} H-band (left) and IRAC $3.6\,\mu$m images (middle; grayscale). The {\em HST} and {\em Spitzer} images were obtained from the CANDELS database. The contours are shown in steps of $1\sigma$, starting at $\pm2\sigma$. COLDz.GN.0 (GN10), and COLDz.GN.3 (GN19) were separately CLEANed because the high SNR allows to meaningfully deconvolve the emission. GN.31 corresponds to CO(2--1)  emission from HDF850.1.
Right: Extracted line candidate single-pixel/aperture spectra (``histograms"; for unresolved/resolved emission) and Gaussian fits to the line features (red curves). The observer-frame frequency resolution of 4 MHz corresponds to $\sim35$ km s$^{-1}$ mid-band.  The velocity range  used for the overlays is indicated by the dashed blue lines.}
\label{fig:spectra_GN_main}
\end{figure*}

\subsection{Statistical counterpart matching}

All  SNR$>$6.4 candidates  in COSMOS and GOODS-N  have optical, NIR and/or radio/(sub)-mm counterparts (in addition to GN19 and HDF850.1). At lower SNR, it becomes more difficult to establish definitive counterparts due to the modest precision of photometric redshifts and potential (apparent or real) spatial offsets of the emission.
Our purity analysis (Appendix F) suggests that the contamination from noise is considerable. As an example, for the candidates shown below SNR=6, we may expect only 1 or 2 out of 10 to be real CO line emitters due to the large sizes of the data cubes. Therefore, we consider the lack of counterparts as a possible indication that a line candidate  may be due to noise.
On the other hand, the very objective of a blind search for CO emitting galaxies is to address a potential bias against optical/NIR-faint galaxies. Possible explanations for the lack of counterparts are:
1) the stellar light could be too dust-obscured to be visible in the rest-frame optical/NIR; 2) the CO line may correspond to the {\it J}=2--1 transition; placing the galaxy at $z>5$, such that counterparts may only exist below the detection limit; 3) a CO-bright emitter may be gas-rich but have low star formation rate and/or stellar mass, which would make it optically ``dark".

\subsubsection{Optical-NIR counterparts}

We here consider the uncertain line candidates near and below the SNR threshold only. If we match all $5<SNR<6$ candidates in COSMOS (60 in total) to the COSMOS2015 photometric catalog \citep{Laigle16}, by requiring a spatial separation of $<2^{\prime \prime}$ and a $z_{\rm CO10}$ or $z_{\rm CO21}$ within the 68th percentile range of the photometric redshifts, we find 10 matches. This is $\sim2.7\sigma$  higher than the number of matches found for random displacements of the positions of our candidates (randomly expecting $\sim4.7\pm2.0$   associations). We therefore conclude that some ($\sim3-7$) of the 10 associations (out of these top 60  candidates) are likely to be real physical counterparts to real CO line emitters, in agreement with our typical purity estimate of order $\sim10\%$ for the statistical sample in this SNR interval (Appendix F).
Consistently, we also find a $1.8\sigma$ excess of positional matches within $<2^{\prime \prime}$ for this extended candidates sample, 20 matches with a $13.8\pm 3.4$ false positive rate, by spatially  associating to the {\it Spitzer}/IRAC-based catalog by the deep  SEDS survey  \citep{Ashby2013}. This confirms that at least a fraction of our line candidates in the COSMOS field at these lower SNR levels may have real counterpart associations, to be confirmed by future spectroscopic observations.

%We also find a $1.3\sigma$ excess of {\em Herschel} position matches within $5^{\prime \prime}$ of our candidates. Specifically, COLDz.COS.1, COS.2, COS.3 all have {\em Herschel} matches, and 11 more of the SNR$>5$ candidates (randomly expecting $\sim7.5\pm2.8$). All of these {\em Herschel} matches but one are different than those found  in the COSMOS2015 catalog.

We repeat the same procedure in GOODS-N, for the  candidates with SNR$>$5.4, excluding the independently confirmed ones (51 in total). We employ the best redshifts available from \cite{Skelton14} and \cite{Momcheva16}, using the same selection criteria with a separation requirement of $<2^{\prime \prime}$. The grism spectroscopy does not significantly impact our matched counts, as almost all of the potential counterparts are too faint and only have photometric redshifts.  
We only find a slight excess relative to chance associations ($\sim1.1\sigma$), by finding 9 associations at an expected chance rate of $6.3\pm2.5$.
The latest ``super-deblended" GOODS-N catalog from \cite{superdeblended} does not yield any additional associations besides the secure sources corresponding to GN10 and GN19.
In addition, we search for positional matches within $<2^{\prime \prime}$ for this extended candidate sample by searching for spatial  associations in the \cite{Ashby2013} {\it Spitzer}/IRAC-based catalog from the deep  SEDS survey. We do not find any excess of matches over the expected false positive rate  .

%The COSMOS2015 catalog has a total source surface density of 117 arcmin$^{-2}$. If we restrict to the photo-$z$ range of $z=2$--3, the COSMOS catalog has a source surface density of $\sim15$ arcmin$^{-2}$. On the other hand, the GOODS-N/CANDELS catalog has a total source surface density of 260 arcmin$^{-2}$, and if we restrict to the best-$z$ range of $z=2$--3, the catalog has a source surface density of $\sim33$ arcmin$^{-2}$. Therefore, the higher rate of chance associations may dilute a potential counterpart matching signal at the limited statistical significance of our study.
%The lower sensitivity of our GOODS-N observations may be responsible for the lower rate of candidate association. Alternatively, the deeper archival observations available in GOODS-N, relative to our field within COSMOS, may be responsible for a higher false positive rate, which may be diluting the candidate matching signal.
 The counterpart association signal in GOODS-N does not constitute a significant excess, perhaps due to contamination by chance associations with low redshift galaxies. However, at least approximately $\sim6-10$ line candidates out of the top $\sim200$ have a very close {\it Spitzer}/IRAC counterpart ($<1^{\prime\prime}$) and a photometric redshift estimate which is compatible with the CO(1--0) line candidate, as would be expected for real counterpart matches.

%We also fit the SED of the potential counterparts from the photometric catalog, and extract the spectra in our MUSE data of the objects that were covered by the MUSE observations.
In the following, we evaluate the implications of a lack of $3.6\,\mu$m counterparts for some of our lower SNR CO line candidates.
The deep {\em Spitzer}/IRAC images in Fig.~\ref{fig:spectra_COS1} and Appendix E are derived from the {\sc splash} observations \citep{Steinhardt14}, while the {\em Spitzer}/IRAC images in GOODS-N were obtained as part of the legacy  GOODS program \citep{Giavalisco2004}. Due to the moderate resolution of {\em Spitzer} observations, these images are sometimes contaminated by lower redshift galaxies or stars, reducing our ability to detect counterparts at higher redshift and hence, in those cases the following limits may not apply. In order to asses the implications of a counterpart non-detection in the IRAC $3.6\mu$m images, we use template spectral energy distributions (SEDs) for star forming galaxies from \cite{BruzualCharlot}, redshift them to $z\sim2.3$ and convolve them with the IRAC $3.6\mu$m filter curve, using {\sc magphys} to estimate the stellar mass limits placed by a lack of detection in COSMOS or GOODS-N \citep{magphys08,magphys15}. The expected mass-to-light ratio at this wavelength depends on the stellar population ages and star formation histories, as well as on the degree of dust extinction.  The following estimates are thus only indicative.  We estimate that the lack of IRAC $3.6\,\mu$m counterparts at the $\sim 0.2\,\mu$Jy  and $\sim 0.06\,\mu$Jy limits ($\sim 3\sigma$; \citealt{Ashby2013}) of the COSMOS and GOODS-N data correspond to approximate stellar mass upper limits of $\sim6\times 10^{9}$ and $\sim2\times 10^{9}\,{\rm M}_\odot$ respectively at $z\sim2.3$ for a representative $A_V\sim2.5$\footnote{For reference, the limits would be $\sim1.6\times 10^9$ and $5\times 10^8\,{\rm M}_\odot$ for $A_V<0.5$, and $\sim 2\times10^{10}$ and $7\times10^{9}\,{\rm M}_\odot$ at $A_V\sim5$}. These limits suggest that a lack of infrared counterparts implies either a very low stellar mass, or a high degree of dust obscuration. The stellar mass limits would be significantly higher for a line candidate  associated with CO(2--1)  emission at $z>5$. Indeed, repeating the same calculations for $z\sim5.8$ we obtain significantly less constraining stellar mass limits of $\sim1.3\times 10^{11}$ and $\sim4\times 10^{10}\,{\rm M}_\odot$ for a representative $A_V\sim2.5$, in COSMOS and GOODS-N, respectively.

 %deeper IR observations are required in order to conclusively identify or rule out  counterpart matches, and that a moderate amount of dust obscuration may be responsible for the lack of unambiguous counterparts in at least a fraction of the candidates.

\begin{figure*}[tbh]
\centering{
 \includegraphics[width=\textwidth]{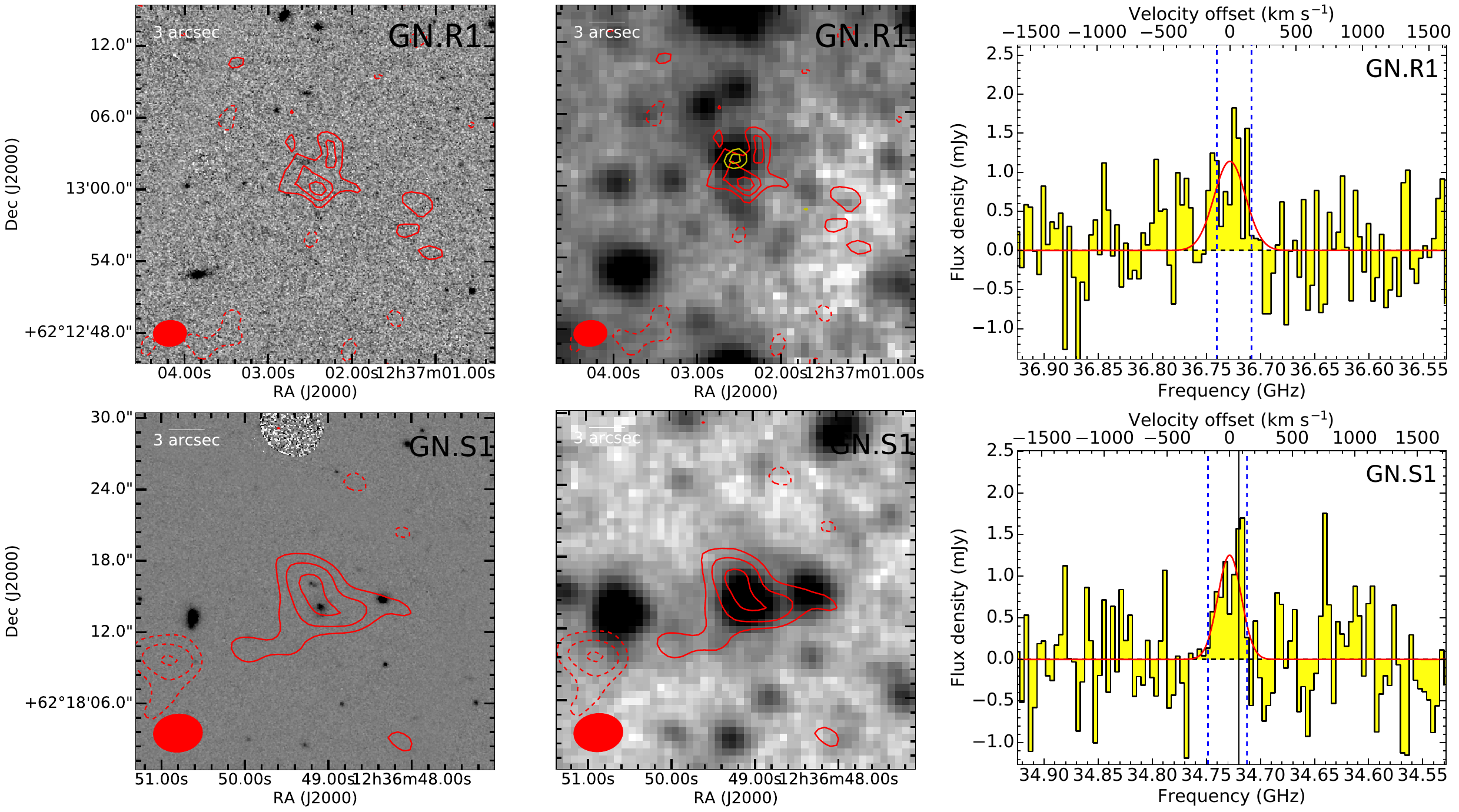}
}
\caption{Candidates with radio continuum counterpart, and with optical spectroscopic redshift below the catalog threshold (i.e., not part of our statistical sample). Left: Line map overlays (red contours) over {\em HST} H-band (left) and IRAC $3.6\,\mu$m images (middle; grayscale).   Red line contours show the CO line in steps of $1\sigma$, starting at $\pm2\sigma$. Yellow contours show the radio continuum  emission in steps of $2\sigma$, starting at $\pm3\sigma$  \citep{ Morrison10}.
Right: Extracted line candidate aperture spectra (``histograms") and Gaussian fits (red curves) to the line features. The frequency resolution is the same as in Fig.~\ref{fig:spectra_COS1}.
The velocity range  used for the overlays is indicated by the dashed blue lines. The  solid black line shows the CO(1--0)  frequency corresponding to the optical spectroscopic redshift of $z_{\rm spec}=2.320$.}
\label{fig:radio_cand_figs}
\end{figure*}

\subsubsection{Radio counterparts}
We also searched for counterpart matches in the deep COSMOS 3 GHz continuum catalog \citep{Smolcic17}, only finding associations for COLDz.COS0, COS1, COS2 and COS3 by using a 3$^{\prime\prime}$ search radius. Of the 18 sources from the \cite{Smolcic17} catalog  located within the boundaries of our mosaic, our secure sources represent the only ones with a redshift estimate (photometric or spectroscopic when available) falling within our survey volume. All the remaining sources from the \cite{Smolcic17} catalog within our survey area lie in the range $z=$0.1--1.6.
We performed an equivalent search in the catalog from the eMERGE 5.5 GHz survey of the GOODS-N field \citep{eMerge}, finding a single association for GN19 by using a 3$^{\prime\prime}$ search radius. We also searched the VLA 1.4 GHz catalog of the GOODS-N field from \cite{Morrison10} with the same criteria, finding two matches for GN19 %(for the two individual components of this source, with fluxes of $28\pm4$ $\mu$Jy and $33\pm4$ $\mu$Jy for the W and E components respectively) 
and a radio counterpart for GN10.
% with a flux of $36\pm4$ $\mu$Jy. 
We also used these radio catalogs to search for counterpart associations with CO candidates to a lower significance of $5\sigma$. We found one candidate at a SNR=5.13 which satisfies the requirement of close association with a radio source (within $3^{\prime\prime}$), and with a CO redshift which is compatible with the $1\sigma$ interval for the photometric redshift listed by the optical-NIR photometric catalogs. This candidate, named COLDz.GN.R1 in the following, is at J2000 12:37:02.53 +62:13:02.1, and has an offset of $1^{\prime\prime}.7\pm0^{\prime\prime}.6$ from the radio source. We show this candidate in Fig.~\ref{fig:radio_cand_figs}. The photometric redshift estimate is 4.73--5.30, which is compatible with the CO(2--1)  redshift of $z_{\rm 21}=5.277\pm0.001$ implied by the COLDz data. We measure a CO(2--1)  line luminosity of $(8\pm3)\times10^{10}\,$K km s$^{-1}$pc$^2$,  which implies a gas mass of $(2.9\pm1.1)\times10^{11}\,{\rm M}_\odot$ for a standard $\alpha_{\rm CO}$=3.6 M$_\odot$ (K km s$^{-1}$ pc$^2$)$^{-1}$ and $r_{\rm 21}=1$. The \cite{Skelton14} catalog reports a stellar mass of $2.8\times10^{11}\,$M$_\odot$, which suggests a molecular gas mass fraction of $\sim1$. The radio continuum  fluxes are S$_{\rm 1.4\,GHz}=(26\pm4)\,\mu$Jy and S$_{\rm 5.5\,GHz}=(20 \pm4)\,\mu$Jy  (\citealt{Morrison10,eMerge}, respectively). This  suggests a star formation rate of $\sim200-400\,$M$_\odot$ yr$^{-1}$ when applying the radio-FIR correlation \citep{Delhaize17}.
Although this line candidate has a higher probability of corresponding to real emission than implied by its SNR, we do not include it in the statistical analysis to preserve the un-biased (i.e., CO SNR-limited) nature of our selection.

The deep radio catalogs by \cite{Smolcic17} in COSMOS and by \cite{Morrison10} in GOODS-N have a $5\sigma$ sensitivity limit of $\sim11\,\mu$Jy at 3 GHz and 20 $\mu$Jy at 1.4 GHz, respectively which can be converted to  upper limits on the $L_{\rm FIR}$ for radio counterparts to our line candidates through the radio-IR correlation. By adopting the relationship from \cite{Delhaize17}, we deduce a detection limit of $L_{\rm FIR}<4-7\times10^{11}\,L_{\odot}$ in the $z=1.953$--2.847 redshift range. On the other hand, recent results have suggested that the radio-FIR correlation in disk-dominated star-forming galaxies may not show a redshift evolution as used by \cite{Delhaize17} \citep{Molnar}. If true, this would suggest less constraining limits of $L_{\rm FIR}<1-3\times10^{12}\,L_{\odot}$.
 The $5\sigma$ sensitivity limit of the eMERGE catalog at 5.5 GHz is approximately 15 $\mu$Jy, and corresponds to  limits of $L_{\rm FIR}<8-14\times10^{11}\,L_{\odot}$ according to \cite{Delhaize17}, and to $L_{\rm FIR}<2.5-6\times10^{12}\,L_{\odot}$ according to \cite{Molnar}. These limits may be constraining, because  our measured $L'_{\rm CO}$ would imply median $L_{\rm FIR}\sim10^{12}\,L_{\odot}$ and $\sim4\times10^{12}\,L_{\odot}$ based on the star-formation law \citep{Daddi10b, Genzel10} for our unconfirmed line candidates in the COSMOS and GOODS-N fields, respectively. Possible reasons for the lack of radio counterparts may be due to  fainter radio fluxes in our sample than expected from the  radio-IR correlation, lower star formation rates  than expected based on the gas masses, or that candidates may correspond to CO(2--1)  emission at $z>5$. Alternatively, line candidates may not be real and be due to noise. The  possibility of gas-rich, low star-formation rate galaxies would be particularly interesting, because surveys like the one reported here may be the only way to uncover such a hidden population.

\section{Identification and stacking of galaxies with previous spectroscopic redshifts}

\begin{table*}[htb]
 \caption{Low-{\it J} CO Counterpart Search for HDF-N PdBI Blind Mid-{\it J} CO Candidates.}
\tiny
\hspace{-10pt}
\centering
\begin{tabular}{ccccccc}
\hline
  ID &Preferred Redshift&PdBI preferred  &PdBI covered  & This survey  & Low-{\it J} Flux & $L'_{\rm CO}$ constraint \\
  &&Mid-{\it J} line&Mid-{\it J} line & Low-{\it J} line & or 3$\sigma$ limit & \\
 & &  & & &(Jy km s$^{-1}$) & \\
   \hline
   ID.01&1.88&2--1&5--4&2--1&$<$0.05 &r$_{52}>$1.6\\
   ID.02&1.81&2--1&5--4&2--1&$<$0.02 &r$_{52}>$2.6\\
    ID.03& 1.78 (secure)&2--1&5--4&2--1 & $<$0.05 &r$_{52}>$1.8\\
   ID.04&1.71&2--1&5--4&2--1&$<$0.03 &r$_{52}>$0.9\\
    ID.05&2.85&3--2&5--4&2--1&$<$0.06 &r$_{52}>$1.2\\
   ID.08 (HDF850.1)&5.19 (secure)&5--4&5--4*&2--1&$0.17\pm0.06$ &r$_{52}=0.40\pm0.16$\\
   ID.10 &2.33& 3--2&3--2*/6--5&1--0/2--1 &$<$0.03 & r$_{31}\&r_{62}>$0.7\\
    ID.11&2.19& 3--2& 3--2*/6--5&1--0/2--1& $<$0.04& r$_{31}\&r_{62}>$0.9\\
   & & 3--2&7--6&2--1& $<$0.03 &r$_{72}>$1.0\\
   ID.12&2.19& 3--2&3--2*/6--5&1--0/2--1& $<$0.05 & r$_{31}\&r_{62}>$0.6\\
   && 3--2&7--6&2--1 & $<$0.03 & r$_{72}>$0.7\\
    ID.13&2.18& 3--2&3--2*/6--5&1--0/2--1& $<$0.04 & r$_{31}\&r_{62}>$0.6\\
      && 3--2&7--6&2--1 & $<$0.03 & r$_{72}>$0.7\\
   ID.14 &2.15& 3--2&3--2*/6--5&1--0/2--1&$<$0.04  &r$_{31}$or r$_{62}>$0.8\\
   && 3--2&7--6&2--1 &$<$0.03 & r$_{72}>$0.7\\
    ID.15&2.15& 3--2& 3--2*/6--5&1--0/2--1& $<$0.04 & r$_{31}\&r_{62}>$1.2\\
   & & 3--2&7--6&2--1 &$<$0.03 & r$_{72}>$1.2\\
    ID.17 (HDF850.1)&5.19 (secure)& 6--5&6--5*&2--1&$0.17\pm0.06$&r$_{62}=0.24\pm0.10$\\
   ID.18&2.07& 3--2&3--2*/6--5&1--0/2--1& $<$0.05 & r$_{31}\& r_{62}>$1.3\\
      && 3--2&7--6&2--1 &$<$0.04 & r$_{72}>$1.3\\
    ID.19&2.05 (secure)& 3--2&  3--2*/6--5& 1--0/2--1&  $<$0.07 &r$_{31}$ \& r$_{62}>$0.7\\
    && 3--2&7--6&2--1 &$<$0.04 & r$_{72}>$0.9\\
 ID.20&2.05& 3--2& 3--2*/6--5&1--0/2--1& $<$0.05 & r$_{31}\& r_{62}>$0.8\\
  & & 3--2 &7--6&2--1 &$<$0.03 & r$_{72}>$1.1\\   
      ID.21&3.04&4--3&7--6&2--1 &$<$0.04& r$_{72}>$0.8\\
      \hline
\end{tabular}
\label{Decarli_counterpart}

 \begin{tablenotes}
 \small
 \item 
\textbf{Note} Preferred redshifts are quoted from \cite{Decarli14}. Although we systematically constrain every possible J line assignment that would place a CO line in our data, the asterisk marks those cases where the assignment is preferred by \cite{Decarli14}. ID.03 has a secure redshift identification which places it outside our redshift coverage.
ID.19 has a secure redshift which places it within our coverage.
%For each mid-{\it J} CO line candidate from \cite{Decarli14} we search for potential low-{\it J} counterparts in our dataset, while remaining agnostic regarding the identification of which mid-{\it J} line may have been detected. We only detect CO(2--1)  line emission in the confirmed candidates corresponding to CO(5--4) and CO(6--5) from HDF850.1. For each remaining candidate we place 3$\sigma$ flux upper limits assuming the same line FWHM as measured in the mid-{\it J} lines. We use these flux upper limits to constrain the line luminosity $L'_{\rm CO}$ ratio for the allowed J-line combinations. A line luminosity ratio larger than one is unlikely to be physical (see text for details), suggesting that the mid-{\it J} line candidate may not correspond to real emission.
 \end{tablenotes}
\end{table*}

\subsection{Identification and stacking of previous mid-{\it J} blind  CO surveys}

We searched the GOODS-N dataset for low-{\it J} CO counterparts to the candidate mid-{\it J} CO detections from our previous  CO blind survey in the HDF-N with the PdBI \citep{Decarli14}. We find a single match in our candidate list, which corresponds to  CO(2--1)  line emission from HDF850.1. We systematically searched for every possible mid-{\it J}/low-{\it J} CO line combination that would place a low-{\it J} CO line in our surveyed volume (Table~\ref{Decarli_counterpart}).
Several of these possible mid-{\it J}/low-{\it J} CO line combinations are not the preferred line identifications by \cite{Decarli14}. Therefore, our non-detections are consistent with their preferred redshift in those cases constraining or ruling out several alternative redshift solutions allowed by the PdBI data alone.
In order to search for lower significance candidate lines,  we extract spectra at the mid-{\it J} candidate positions, and evaluate the significance of any features or place $3\sigma$ upper limits to the line fluxes. By assuming the same line FWHM as the candidate mid-{\it J} CO lines, we then derive limits on the line brightness temperature ratios (Table~\ref{Decarli_counterpart}). We evaluate the signal-to-noise by spectral (1D) match-filtering of individual spectra, extracted in the central pixel and within 5 frequency channels around the expected position of the lines. We do not find any significant detections above a $3\sigma$ threshold.
%We point out that our constraining upper limits often imply un-physical line ratios; in particular, although it is possible that a lower-{\it J} level may be less populated than a higher-{\it J} level or that low optical depth may cause these line ratios, this is unlikely to occur and therefore brightness line ratios that are significantly larger than unity likely  rule out the viability of the mid-{\it J} candidate as a line detection.
Some of our upper limits imply super-unity line brightness temperature ratios for the mid-{\it J} CO candidates. While a lower-{\it J} level can be less populated than a higher-{\it J} level, or  low optical depths can cause such high line ratios, the physical conditions that give rise to such ratios are rare. %Instead, it is more likely that high line ratios indicate that the mid-{\it J} detection is a false positive.
In cases where super-unity line ratios are found for redshifts  (i.e., mid-{\it J}/low-{\it J} CO line combinations) disfavored by \cite{Decarli14}, our data provide supporting evidence for the preferred redshifts identified by \cite{Decarli14}, under the assumption that those line candidates are real.
%Our constraining limits for the mid-{\it J}/low-{\it J} CO line combinations which are not preferred by the analysis in \cite{Decarli14} therefore reinforce the likelihood that those redshift assignments may be correct.
As an example, in the case of  ID.03, multiple line detections have determined a secure  redshift. Since this redshift does not lie within our surveyed volume, our non-detection is consistent  with the redshift identification by \cite{Decarli14}.
On the other hand, candidate ID.19 was confirmed to lie at $z=$2.0474 based on optical grism spectroscopy  \citep{Decarli14}.  Therefore, the candidates lies in our survey volume and our line ratio limit ($r_{31}>0.7$) is significant. This suggests moderately elevated CO excitation compared to the average ratio found for a sample of Main Sequence galaxies at $z=$1.5 ($r_{31}=0.42$, \citealt{Daddi15}), and even compared to the average ratio ($r_{31}=0.52\pm0.09$) found for a sample of sub-millimeter galaxies \citep{Bothwell13}.

For the mid-{\it J} line candidates  ID.15 and ID.18, our constraints on the line ratios are higher than unity. This suggests that an alternative lower redshift mid-{\it J} line assignment of CO(2--1)  in the PdBI data may be more likely (since it would imply a redshift outside our survey volume), if the line candidate were confirmed to correspond to real emission.

 %We evaluate the rate of false positives at random positions, and verify that the adopted $4\sigma$ cut corresponds to a $\sim$2.7\%  false positive rate. We adopt a $4\sigma$ cut rather than the more conventional $3\sigma$ in this regard because we allow 
%We note that although one of the \cite{Decarli14} line candidates, ID.03, shows a weak signal ($\sim4.4 \sigma$), we do not consider it significant for a detection, especially in light of the fact that ID.03 was already identified to be at $z=1.784$, by multiple CO line detection, and by showing an unphysically high putative $R_{52}$. For the other line candidates we provide constraining  upper limits, which often already imply un-physical line ratios, therefore ruling out their viability as line detections.

We also stack the extracted spectra to obtain more sensitive limits. In particular, we select random subsets of candidates for stacking, to take into account a possible mis-identification of the correct {\it J} value for some CO lines. To search for lines in the stacked spectra, we match-filter using the same set of spectral templates as the main line search (Table~\ref{template_sizes}). We find no   line signal in the stacks above $3\sigma$ significance. Assuming an average line FWHM of 300 km s$^{-1}$, we therefore obtain a sensitive $3\sigma$ upper limit of $0.014\pm0.002$ Jy km s$^{-1}$ to the line fluxes, where the quoted uncertainties on the limit depend on the number of stacked spectra.
From the same sample, we also separately stack the nine candidates for which the lines were identified as likely CO(3--2) emission by \cite{Decarli14}, and whose redshift would place their CO(1--0) line in our data cube. We obtain a $3\sigma$ limit of $\sim0.019$ Jy km s$^{-1}$, which implies a constraining limit of $r_{31}>2.0$ when using the mean CO(3--2) flux in the limit (0.34 Jy km s$^{-1}$, using the same weights as in our stack). We estimate how many of these stacked spectra need to be removed in order for the line luminosity ratio to become smaller than unity. We find that at least six of them may not correspond to real emission, subject to the stated assumptions.
%In summary, we have shown that many of the line candidates from the blind line search for mid-{\it J} CO emitters carried out by \cite{Decarli14} are either spurious and not corresponding to real CO emission, or correspond to galaxies with particularly high CO excitation, as might be expected from a mid-{\it J} CO selection. This is in contrast to our $^{12}$CO(1--0) -based selection, which therefore provides the only unbiased approach to a molecular gas deep field survey.
In summary, we find some tentative evidence suggesting that mid-{\it J} blind CO searches may preferentially select galaxies with relatively high CO excitation. An alternative interpretation may be that some of the candidate mid-{\it J} CO emitters considered here, may be spurious and do not correspond to real CO emission. In order to more strongly differentiate between these possibilities, more sensitive 3 mm observations need to be carried out.

\begin{figure*}[htb]
\centering{
 \includegraphics[width=\textwidth]{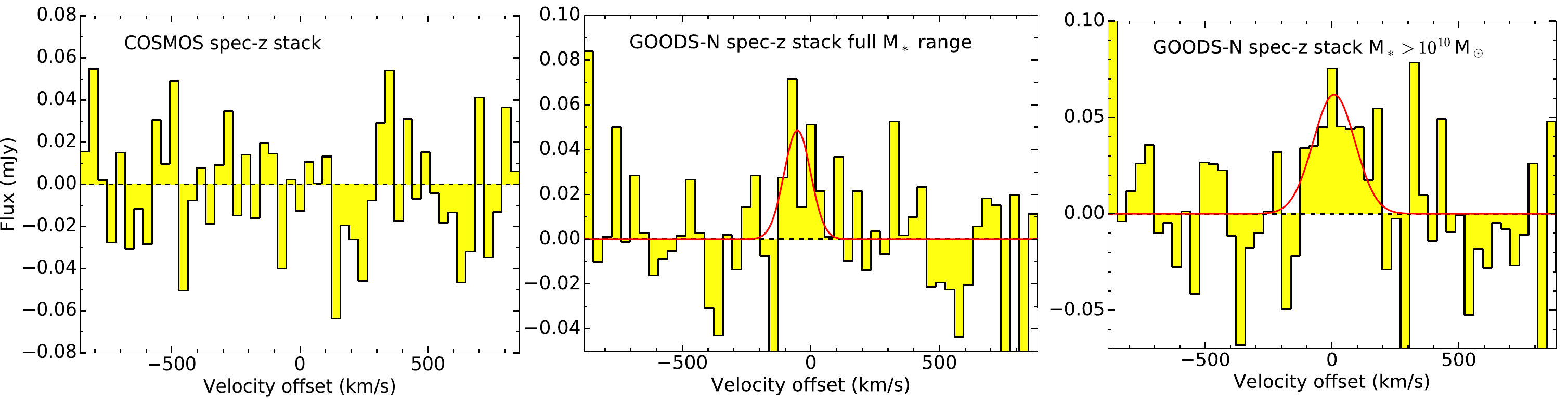}
}
\caption{Spectral stacks of sets of galaxies  with  spectroscopic redshifts. Stack of galaxies with spectroscopic redshifts in COSMOS (left,  7 galaxies) and GOODS-N (right). The  GOODS-N spec-z stack including the full  stellar mass range  (78 galaxies) displays a tentative $3.4\sigma$ detection while the stack of massive galaxies (M$_*>10^{10}\,$M$_\odot$,  34 galaxies) shows a 3.5$\sigma$ detection. The best-fitting Gaussian line profiles are shown in red. %Middle: Stacks of subsets of galaxies with accurate grism spectra, with redshift uncertainty $<$500 (left) and $<$700 km s$^{-1}$ (right), respectively.  Bottom: Stacks for  samples of potential CO(2--1) , $z>5$ galaxies  belonging to  previously identified over-densities in our fields, i.e. the AzTEC-3 proto-cluster at z=5.3 (left) and the overdensity around HDF850.1 (right) \citep{Capak11,Walter12}. 
The spectral resolution is the same as in Fig.~\ref{fig:spectra_COS1}.}
\label{fig:stacked_spectra}
\end{figure*}

\subsection{Identification and stacking of galaxies with optical redshifts}

In order to obtain additional constraints on the CO luminosity of galaxies that remain individually undetected in the volume covered by our survey, we utilize the available optical/NIR spectroscopic redshift information for galaxies in our well studied target fields for stacking. 
We extract  single pixel spectra of the sources described in the following, and we stack them with a weighted average. As weights, we used the inverse of the variance of the local noise following \cite{ASPECS2}. We present additional, less constraining, stacks of galaxies in Appendix G, where we consider galaxies with grism redshifts and galaxies at higher redshifts, for which CO(2--1)  may lie within our data.

\subsubsection{Spectroscopic redshifts in the COSMOS field}
Only seven galaxies have known ground-based optical spectroscopic redshifts that place them within our COSMOS data cube, all of which were obtained as part of the zCOSMOS-deep survey (Lilly et al., in prep.). These galaxies have relatively low stellar masses ($\lesssim10^{10}\,M_\odot$). Therefore, we do not expect to detect their CO emission individually.
We also do not detect their averaged CO line emission down to a deep $3\sigma$ limit of $<$0.008 Jy km s$^{-1}$ (assuming a  line FWHM of 300 km s$^{-1}$), after stacking spectra extracted at their positions (stacked spectrum shown in Fig.~\ref{fig:stacked_spectra}). This limit implies $L^\prime_{\rm CO}<1.7-2.7\times 10^{9}\,$K km s$^{-1}$ pc$^2$ for different redshifts within our surveyed range.
In order to determine the implications of this limit, we perform SED fitting of the same galaxies with {\sc magphys} \citep{magphys08,magphys15} to estimate their stellar masses, finding that they are compatible with the tabulated values in COSMOS2015 whenever the photo-$z$ is similar to the spectroscopic redshift (only 3/7 cases).  These stellar masses are in the range $10^9\,{\rm M}_\odot-10^{10}\,{\rm M}_\odot$. Assuming that these galaxies lie on the Main Sequence, we use the fitting functions from \cite{Speagle} to determine star formation rates (SFR) $\sim2-25\,{\rm M}_\odot \, {\rm yr}^{-1}$. These values are consistent with the lack of detections by the 3 GHz survey by \cite{Smolcic17}, which implies SFR$<40-70\,{\rm M}_\odot \, {\rm yr}^{-1}$ based on the radio-FIR correlation\footnote{Recent work by \cite{Molnar} suggests that  radio-FIR correlation for disk-dominated galaxies may not show redshift evolution, and would imply a less constraing limit of  SFR$<100-300\,{\rm M}_\odot \, {\rm yr}^{-1}$} estimated by \cite{Delhaize17}.  The SFRs estimated by {\sc magphys} for these galaxies (with great uncertainty due to the lack of FIR detections) span the 6--150$\,{\rm M}_\odot \, {\rm yr}^{-1}$ range, with a mean of 55$\,{\rm M}_\odot \, {\rm yr}^{-1}$.  Assuming the star formation law found for Main Sequence galaxies at high redshift\footnote{These star formation law estimates were mostly based on CO(3--2) observations, therefore variations in the r$_{31}$ line ratio may contribute additional uncertainty.} \citep{Daddi10b,Genzel10}, we can use the star-formation rate estimates to infer an expected $L^\prime_{\rm CO}$  $\sim8\times 10^9 \,$K km s$^{-1}$ pc$^2$, which is higher than our measured limit, and therefore not fully consistent. The adopted chain of scaling relations, including SED fitting and the star formation law have large scatter, and therefore introduce  large uncertainties in the $L^\prime_{\rm CO}$ estimate.  The apparent tension with our upper limit would disappear if the average SFR were $\sim10-15\,{\rm M}_\odot \, {\rm yr}^{-1}$.  The non-detection of the stacked CO(1--0) emission therefore  provides a valuable constraint on the CO luminosity of faint, modestly massive $z=2$--3 galaxies, as our best estimates for their SFR appear to be in tension with the expectated $L^\prime_{\rm CO}$  based on the star formation law by \cite{Daddi10b,Genzel10}. %In Figure~\ref{fig:gas_frac} we also show the implication of these upper limits for the gas fraction in these galaxies.

\subsubsection{Spectroscopic redshifts in the GOODS-N field}
The 3D-{\em HST} catalog \citep{Skelton14, Momcheva16} provides 67 galaxies in the region in GOODS-N covered in our survey area  with ground-based optical spectroscopic redshift, whose CO(1--0) line is covered by our data. We also include 13 more galaxies with more recent spectroscopic redshifts from the catalog by \cite{superdeblended} in our analysis. One of the galaxies in this combined sample corresponds to GN19 and is individually detected. Therefore, we exclude it from further investigation here. %Five more galaxies are potentially associated with blind-search line candidates at $4.6\sigma$, $4.4\sigma$, $4.4\sigma$, $4.1\sigma$ and $4.1\sigma$ significance respectively (matching within $<2^{\prime\prime}$ and $<500\,$km s$^{-1}$).

One more galaxy with a spectroscopic redshift of $z_{\rm spec}=2.320$ is perfectly matched, in position and redshift with a SNR=4.6 CO line candidate at coordinates J2000 12:36:49.10 +62:18:14.0 and $z_{\rm 10}=2.3192\pm0.0003$, which we name COLDz.GN.S1 (Figure~\ref{fig:radio_cand_figs}). This CO line candidate is significantly extended (FWHM of $\sim9.0^{\prime\prime}\pm0.5^{\prime\prime}\sim74\pm4$ kpc) and appears to be associated with two, potentially interacting, galaxies with a projected separation of $\sim20$ kpc. The galaxy to the south  is associated with the spectroscopic redshift measurement, while the galaxy to the North has a compatible photometric redshift estimate. The total aperture CO flux of GN.S1 is $(0.22\pm0.11)$ Jy km s$^{-1}$,  corresponding to $L'_{\rm CO}=(5\pm3)\times10^{10}\,$ K km s$^{-1}$ pc$^2$. We find a molecular gas mass of $(2.0\pm1.1)\times10^{11}\,$M$_\odot$ when assuming $\alpha_{\rm CO}=3.6\,{\rm M}_\odot$ (K km s$^{-1}$ pc$^2$)$^{-1}$. The stellar mass of the southern (brightest) component is $\sim1.2\times10^{11}\,$M$_\odot$, and that of the northern component $\sim3\times10^{9}\,$M$_\odot$, suggesting a high gas mass fraction $\sim1.7$. This gas fraction may be elevated due to the galaxy interaction, although the star formation rate reported by \cite{superdeblended} of $\sim160$ M$_\odot$ yr$^{-1}$ is approximately $\sim2-3\times$ lower than what may be expected from the total CO luminosity  based on the star formation law  \citep{Daddi10b,Genzel10}. Therefore, we may be witnessing a gas-rich early phase of the merger, which may precede a starburst. This source is also tentatively detected in the S2CLS map, with a significance of $\sim3.3\sigma$ and a $850\,\mu$m flux of $3.2\pm1.0$ mJy \citep{Geach17}, which is compatible with the moderate star formation rate estimate.
%3.3sigma in SCUBA2 and (3.2\pm1.0)mJy

Allowing for an offset of $<2^{\prime\prime}$ and $<500\,$km s$^{-1}$ results in four more potential candidate association, but they are not likely to be real due to apparent offsets in the emission  and because they are not associated with the most massive, or most star-forming  galaxies in the sample.
%due to the low stellar masses and star formation rates of these galaxies relative to the candidate CO luminosities. %Figure~\ref{fig:gas_frac} shows the gas mass fractions in GN19 and GN.S1, and calculated upper limits for all the other galaxies in this spectroscopic sample. 
%In order to search for lower significance detections, we also extract single pixel spectra at the expected CO line positions based on the optical position prior. We find  three more galaxies showing tentative $>3\sigma$ lines (Table~\ref{GN_specz_counterparts}).
In the set of galaxies with spectroscopic redshifts covered by our data, nine galaxies have stellar mass estimates of $>5\times10^{10}\,{\rm M}_\odot$, corresponding to a gas fraction ${\rm M}_{\rm gas}/{\rm M}_*\sim1$ at our approximate  $3\sigma$ sensitivity limit of $L'_{\rm CO}\sim1.5\times10^{10}\,$ K km s$^{-1}$ pc$^2$. These galaxies may therefore be expected to be individually detectable. Excluding GN19 and GN.S1, the remaining seven galaxies remain undetected, implying  M$_{\rm gas}/{\rm M}_*<1$. Previous samples of Main Sequence galaxies at $z=2-3$ have shown typical molecular gas mass fractions of order M$_{\rm gas}/{\rm M}_*\sim1-1.5$ in this stellar mass range \citep{Genzel15,Scoville17}, i.e., higher than those limits. We note that adopting the same CO conversion factor as utilized by \cite{Genzel15}, including the correction for a stellar-mass dependent metallicity, would result in approximately 50\% higher molecular gas mass estimates. Overall, the observed limits may be consistent with previous observations of the molecular gas fractions in Main Sequence galaxies, although they appear to be at the low end of the expected scatter of the relation.

Stacking all 78 spectra, i.e, excluding GN19 and COLDz.GN.S1, yields a tentative ($\sim3.4\sigma$) CO line detection in the deep stack ($6\pm3\times 10^{-3}$ Jy km s$^{-1}$; Fig.~\ref{fig:stacked_spectra}). The noise in this stacked spectrum is $\sim23\,\mu$Jy beam$^{-1}$ in 35 km s$^{-1}$-wide channels.  The stacked galaxies in this sample have a wide range of stellar masses, and are therefore expected to show a range of CO luminosities. While constraining the average CO luminosity for this sample, we note that such an average does not represent common properties of the stacked galaxies.
%The line FWHM in  the stack is narrow, 70--170 km s$^{-1}$.  However,  a set of simple simulations which also include the effect of the spec-$z$ uncertainty in the stacking, shows that at these limited signal-to-noise ratios, the line FWHM may be under-estimated by up to $\sim50\%$ purely due to the low SNR.
The measured flux in the stacked spectrum corresponds to  an average CO luminosity for this galaxy sample of $L'_{\rm CO}=(1.5\pm0.8)\times10^9$ K km s$^{-1}$ pc$^2$ at an average $z\sim2.4$. 
%Assuming Main Sequence star formation law this would imply star formation rates $\sim3-13\,M_\odot$ yr$^{-1}$ \citep{Daddi10b,Genzel10}.
These 78 galaxies have a mean stellar mass of $2.1\times10^{10}\,M_\odot$ and an average star formation rate of $\sim45\,{\rm M}_\odot$ yr$^{-1}$ (with quartiles of 3.6, 22 and 66 M$_\odot$ yr$^{-1}$, respectively), according to \cite{superdeblended} when available, and to \cite{Skelton14} otherwise. %The main sequence estimate from \cite{Speagle} for the star formation rate in this stellar mass range at this redshift is $\sim20-40\,{\rm M}_\odot$ yr$^{-1}$, which is compatible with the direct result of SED fitting. %The lower values of star formation rate obtained by SED fitting in \cite{Skelton14} are consistent with this estimate, as the available SEDs do not include FIR continuum measurements, and therefore are lower limits to the total star formation rates.
We also stack the subset of 34 spectra corresponding to the most massive galaxies with M$_*>10^{10}\,$M$_\odot$, expecting them to contribute the strongest CO signal. We detect emission in this sub-stack with a  significance of 3.5$\sigma$, corresponding to a line flux of $1.2\pm0.7\times 10^{-2}$ Jy km s$^{-1}$ and a line FWHM of $200\pm80$ km s$^{-1}$ (Fig.~\ref{fig:stacked_spectra}). The line flux in the stacked spectrum corresponds to $L'_{\rm CO}=(3.2\pm1.8)\times10^9$ K km s$^{-1}$ pc$^2$ at an average $z\sim2.4$, which corresponds to a molecular gas mass of M$_{\rm gas}=(1.2\pm0.6)\times 10^{10}\,$M$_\odot$ according to our earlier choice of $\alpha_{\rm CO}$.
This gas mass should be compared to the mean stellar mass $\sim3.6\times10^{10}\,$M$_\odot$ of this sub-sample (with a median of $\sim2.8\times10^{10}\,$M$_\odot$) and the average star formation rate of $\sim66\,{\rm M}_\odot$ yr$^{-1}$ (median of $\sim52\,{\rm M}_\odot$ yr$^{-1}$) \citep{superdeblended}. To further investigate these star formation rate estimates, we also stack SCUBA-2 S2CLS 850 $\mu$m images and derive a 3$\sigma$ upper limit of $<0.7$ mJy at the positions of the sample galaxies \citep{Geach17}. Utilizing the average FIR SED from the ALESS sample, at approximately the same redshift as these galaxies, the sub-mm flux upper limit implies a star formation rate constraint of $\lesssim60-100$ M$_\odot$ yr$^{-1}$ \citep{ALESS_daCunha}. A modified black body model with a dust temperature of $T_d=35$K also implies comparable limits.
We find that the star formation rate of this set of galaxies is  compatible with random scatter around the star-forming Main Sequence reported by \cite{Speagle}. The average expected CO luminosity based on the star formation law \citep{Daddi10b,Genzel10} may be  higher than our measurement by about a factor of 3 ($L'_{\rm CO}\sim9.6\times10^9$ K km s$^{-1}$ pc$^2$).

%The low $L'_{\rm CO}$ in the stacks is not compatible with the  estimates based on the previous calibration of the star formation law-CO relationship, or with the molecular gas fraction expectations (Fig.~\ref{fig:gas_frac}). This may suggest somewhat weaker than expected CO emission in these galaxies but this finding is to be considered tentative at the current uncertainties in the estimates based on scaling relations.

\begin{figure*}[bht]
\centering{
 \includegraphics[width=\textwidth]{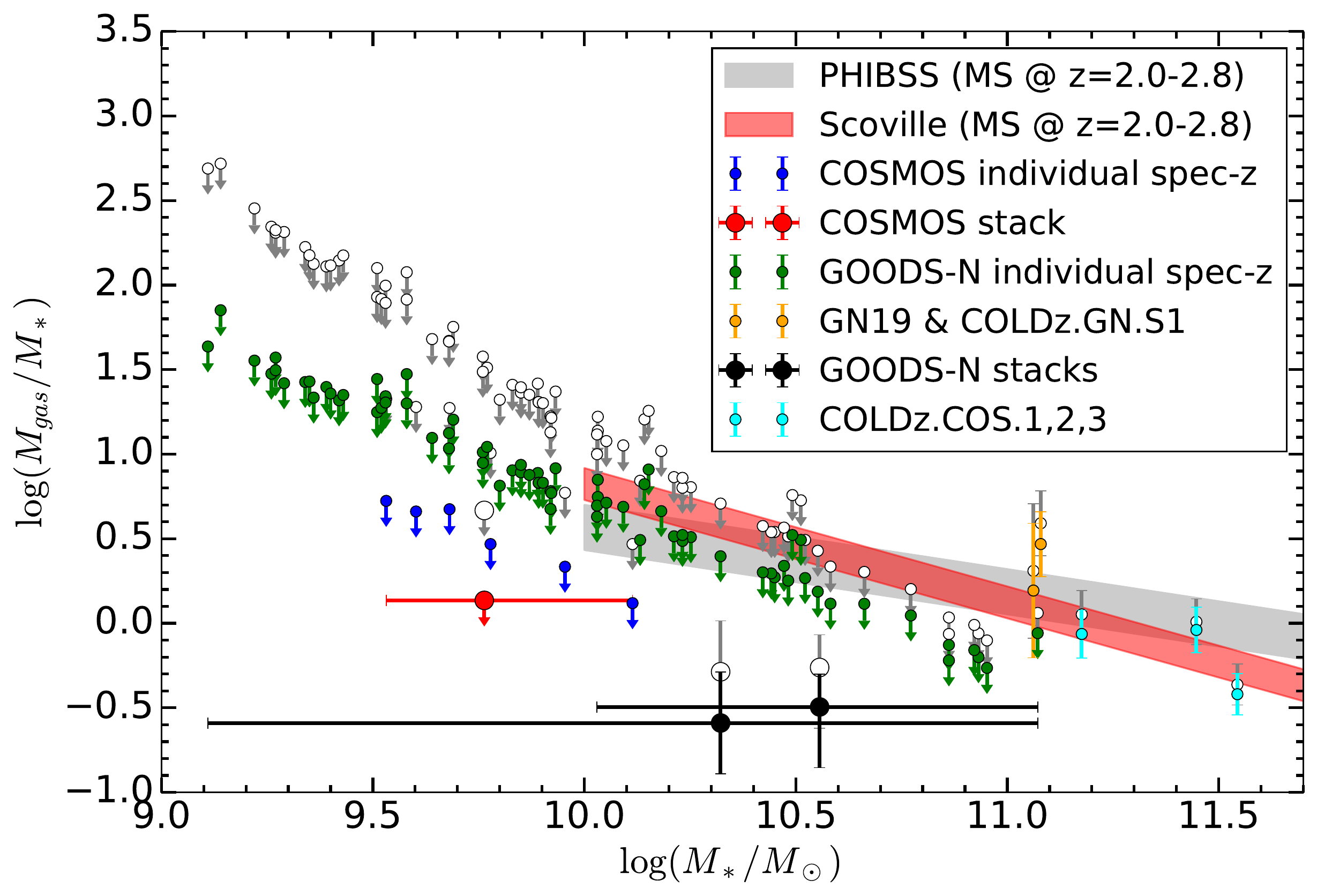}
}
\caption{Molecular gas mass fraction constraints for galaxies with known spectroscopic redshifts for which the CO(1--0) emission can be constrained by the COLDz data. The points in color assume $\alpha_{\rm CO}$=3.6 M$_\odot$ (K km s$^{-1}$ pc$^2$)$^{-1}$, while the gray points correspond to adopting the metallicity-dependent $\alpha_{\rm CO}$ from \cite{Genzel15}. All upper limits correspond to $3\sigma$ limits, and assume a line FWHM of 300 km s$^{-1}$. The sets of individual galaxies, the origin of their stellar masses and the constraints from stacking are described in Section~6. The grey and red shaded regions correspond to the reported average expected gas mass fraction in the range $z\sim2.0-2.8$ for Main Sequence galaxies according to \cite{Genzel15} and \cite{Scoville17}, respectively. These regions do not show the expected scatter, but rather represent the evolution within the covered redshift range. The shown scaling relations were measured over the stellar mass range ${\rm M}_*\gtrsim10^{10}{\rm M}_\odot$.}
\label{fig:gas_frac}
\end{figure*}

\subsubsection{Implications}
Figure~\ref{fig:gas_frac} summarizes our constraints on the molecular gas mass fraction in the analyzed samples of galaxies. In order to convert CO luminosity to molecular gas mass, we consider  two different assumptions for $\alpha_{\rm CO}$. First, we assume a constant  value of $\alpha_{\rm CO}=3.6\,{\rm M}_\odot$ (K km s$^{-1}$ pc$^2$)$^{-1}$ adopted by some previous studies (e.g., \citealt{Daddi10b, Decarli14, ASPECS1}). We then also consider  a metallicity-dependent conversion factor, evaluated by assuming a redshift and stellar mass-metallicity relation \citep{Genzel15}.  
Previous studies investigating optically and FIR-selected galaxy samples have estimated the relationship between gas mass fraction, stellar mass, redshift and SFR-offset from the Main Sequence (e.g., \citealt{Genzel15,Scoville17}). While the PHIBSS project estimated molecular gas masses by measuring the CO(3--2) line emission \citep{Tacconi13,Genzel15}, \cite{Scoville16,Scoville17} have used the flux on the Rayleigh-Jeans tail of the dust continuum emission to estimate the total gas masses. 
We here assume that the samples of galaxies plotted, although not complete to any degree due to their pre-selection for having a spectroscopic redshift, may be somewhat representative of the star-forming Main Sequence (Figure~\ref{fig:gas_frac}). Their  star formation rates are consistent with scatter around the Main Sequence  and appear to include as many galaxies above and below the Main Sequence estimated by \cite{Speagle}. % Therefore, Figure~\ref{fig:gas_frac} shows the gas fraction estimates from \cite{Genzel15} and \cite{Scoville17} for reference, under the assumption of Main Sequence star formation rates, which is expected to apply on average across our galaxy samples.
Although our CO detections (both blind and with previous spectroscopic redshifts) are indicative of gas fractions compatible  with or above (GN19) expectations for Main Sequence galaxies,  the individual CO non-detections and the stacked signal appear to be systematically lower than the  predicted averages, suggesting lower gas mass fractions than might be expected (Figure~\ref{fig:gas_frac}).
  We can quantify the apparent deficit in stacked signal relative to expectations for the ${\rm M}_*\gtrsim10^{10}\,{\rm M}_\odot$ sample, by  calculating expected gas masses for the individual stacked galaxies, predicted as a function of their redshift, stellar masses and star formation rates. The expected sample average molecular gas mass  is $5.8\times10^{10}\,{\rm M}_\odot$ adopting the best fit relation by \cite{Genzel15} and $7.5\times10^{10}\,{\rm M}_\odot$ according to the relation by \cite{Scoville17}.  The constant CO luminosity conversion factor above would imply a ratio between expected and observed stacked CO luminosity of $4.8\pm2.4$ and $6.3\pm3.1$ according to the relations by \cite{Genzel15} and by \cite{Scoville17}, respectively.  Applying instead the metallicity dependent CO conversion factor suggested by \cite{Genzel15} to individual galaxies would somewhat reduce the tension, implying ratios of $3.0\pm1.7$ and $3.8\pm2.1$ according to the relations by \cite{Genzel15} and by \cite{Scoville17}, respectively.
While the constraints for low stellar mass galaxies may be compatible with an evolving CO conversion factor due to low metallicity, this is unlikely to resolve the apparent conflict at the high mass  end, and may point to lower than expected gas masses.

\section{Total CO line brightness at 34 GHz}
One additional key measurement that becomes possible with the COLDz survey is to determine  the total CO line brightness at 30--39 GHz in the survey volume. We follow the simple procedure outlined by \cite{Carilli16}, and as a first conservative estimate, include only the independently confirmed line candidates, for each field in order to derive secure lower bounds. We derive lower brightness temperature limits (since we only include the securely detected sources, without any completeness correction) at the average frequency of 34 GHz  of $T_B\gtrsim0.4\pm0.2\,\mu$K for the COSMOS field and $T_B\gtrsim0.05\pm0.04\,\mu$K for the GOODS-N field, respectively.   The uncertainties are dominated by Poisson relative uncertainties, due to the limited number of sources considered. Sources near the knee of the CO luminosity function (Paper~\rom{2}) dominate the total surface brightness, as expected. Since the two measurements are sensitive to different parts of the CO luminosity function, we add the two values to obtain our best estimate for a lower limit on the average surface brightness of $T_B\sim0.45\pm0.2\,\mu$K.
Next, we attempt to include a longer list of candidates, down-weighted by their purities (evaluated in Appendix F), to estimate a plausible uncertainty range. In the COSMOS field, also including all  moderate SNR candidates presented in Table~\ref{tab_lines_appendix}, we obtain $T_B\sim0.48\,\mu$K and $T_B\sim0.57\,\mu$K without and with the completeness corrections evaluated in Appendix F, respectively.
In the GOODS-N field, also including all  candidates in Table~\ref{tab_lines_appendix}, we obtain $T_B\sim0.18\,\mu$K and $T_B\sim0.3\,\mu$K without and with the completeness corrections, respectively.
Because the complete candidate list in GOODS-N overlaps in flux ranges with the candidates in COSMOS, it is not clear that the best estimate for this case may simply be derived by adding the two contributions; the plausible range of values from our data should therefore be considered to be the full range $T_B\sim0.2-0.6\,\mu$K, with a likely lower limit of $T_B\sim0.45\pm0.2\,\mu$K.
These measurements are consistent with that of $T_B\sim0.94\pm0.09\,\mu$K at 99 GHz by \cite{Carilli16} within the expectation that the total (all CO lines) average surface brightness may  slightly increase between 34 GHz and 99 GHz due to adding more CO transitions together (e.g., \citealt{Righi08}).
Our measurement of the average surface brightness is in agreement with theoretical predictions (e.g, \citealt{Righi08,Pullen13}) which suggest a range of $T_B=0.3-1\,\mu$K. 
Our constraints on the total CO brightness at 34 GHz suggest that the CO signal will be an important  contribution to CMB spectral distortion at these frequencies, which is relevant for upcoming experimental efforts. In particular, as shown in Figure~2 by \cite{Carilli16}, our constraints at 34 GHz are already higher than the PIXIE sensitivity limit \citep{Kogut11,Kogut14} and, while lower than the low-redshift Compton distortion component, it is higher than the relativistic correction to the low-redshift signal, the primordial Silk damping distortion, and the imprint of primordial hydrogen and helium recombination radiation contributions. A measurement of these important cosmological probes will therefore necessarily require a subtraction scheme that will remove the CO line signal (also see \citealt{Carilli16}). %{\bf The average CO line brightness, proportional to the first moment of the CO luminosity function, cannot be directly inferred from the power spectrum measured by intensity mapping, because that would require distinguishing the characteristic CO luminosity from the volume density.}

\section{Discussion and Conclusions}

In this work, we have carried out the first blind search deep field targeting CO(1--0) line emission at the peak epoch of cosmic star formation at $z=2$--3. This allowed us to provide the least biased measurement of the molecular gas content in a representative  sample of galaxies at this epoch. One of our main findings is the absence of a population of massive, gas-rich galaxies with suppressed star formation in our high signal-to-noise sample, which would have been missed by previous selection techniques. The lower signal-to-noise, and hence lower purity, CO line candidate sample includes several candidates without clear multi-wavelength counterparts, which are therefore possible candidates for such a population of ``dark", gas-rich galaxies. Nonetheless, the low purity of such candidate lines requires further, independent confirmation as the absence of a counterpart is more likely to indicate that the line feature may be spurious.

Interestingly, the CO line sources detected with confidence in this study include a mix of different galaxy populations. In particular, our sub-sample of independently confirmed CO emitters contains previously known starbursts like AzTEC-3 (by design), GN10, GN19, and HDF850.1 but also COLDz.COS.1, 2 and 3 and COLDz.GN.S1 which belong to the massive end of the Main Sequence at $z\sim2$ (Pavesi et al., in prep.). This highlights the CO(1--0)-based selection, which does not preferentially select outliers in star formation such as starbursts as preferentially selected by sub-millimeter continuum selected samples. Also, the total gas mass is accurately traced by these measurements, without the extinction biases that affect optical/NIR selected samples.

Most studies of molecular gas in galaxies at high redshift to-date have targeted mid-{\it J} CO lines. Although these lines have higher fluxes than the ground state, {\it J}=1--0 transition, and therefore are typically easier to detect, their higher critical densities and level energies imply that they do not always faithfully trace the bulk of the gas mass, but that they can be biased towards  the dense and warm fraction of the gas reservoir. Therefore, in order to derive gas masses from those mid-{\it J} CO lines, an excitation correction needs to be assumed (i.e. a ratio of those lines to the CO(1--0) line brightness), which introduces a source of  uncertainty. 
Previous blind CO searches have  targeted mid-{\it J} CO lines \citep{Decarli14,ASPECS1}, and therefore relied on similar excitation correction assumptions in order to derive constraints to the total molecular gas mass. In this study, we have shown that blind CO(1--0) searches, selecting galaxies uniquely through their total gas masses, find a varied sample of galaxies belonging to a mix of different populations, which may be characterized by significant differences in CO excitation (e.g., starbursting and  Main Sequence galaxies; \citealt{Daddi10b, CarilliWalter,Riechers06,Riechers11a,Riechers11b,Ivison_GN19,Bothwell13}). We also find significant excitation differences among the individual sources  (to be described in detail by Pavesi et al., in prep.).
Furthermore, our limits on the CO(1--0) line luminosities in the candidates previously selected by \cite{Decarli14} indicate that the corresponding galaxies either have substantially elevated CO excitation, or that a large fraction of them may not correspond to real line emission.

The so called ``wedding cake" design of the COLDz survey, targeting a shallow wide field and a deep narrower field,  allows us to provide valuable, independent constraints to different parts of the CO luminosity function (Paper~\rom{2}) which would not have been possible with a single field due to the limited accessible volume and depth. While the sensitivity of our deeper field (in COSMOS) is within a factor of two of the sensitivity that was previously achieved by ASPECS through ALMA in a comparable redshift bin (after correcting for CO excitation), the volume that we could sample in that field is six times larger. Furthermore, the  volume covered in both fields combined  is $>$50 times as large as that covered by ASPECS-Pilot and $>60$ times as large as that carried out in the HDF-N with the PdBI, given the $>60$ times larger survey area ($\sim$60 vs $\sim$1 arcmin$^2$).

In this study, we have also significantly further developed the methods utilized to carry out blind searches for emission lines in interferometric datasets. In particular, we have generalized the Matched Filtering technique that is commonly used in the spectral dimension to identify spectral lines, to the regime of interferometric data cubes where sources may be spatially extended. By taking advantage of this new source selection method, we have blindly detected significantly extended CO(1--0) line sources like COLDz.COS.3, which hosts a very large cold gas reservoir ($\sim30-40$ kpc). Furthermore, one of our highest SNR line emitters in the GOODS-N field (GN19) and a galaxy with optical spectroscopic redshift (COLDz.GN.S1), also appear extended ($\sim40-70$ kpc) in CO observations due to a major gas-rich merger in this galaxy (see also \citealt{Riechers_GN19,Ivison_GN19}). The high incidence, two out of the eight most significant CO(1--0) sources, suggests that extended CO(1--0) sources may in fact be  prevalent, in agreement with previous findings  \citep{Riechers_GN19,Ivison_GN19}.  Indeed, through the blind search we have selected other CO candidates which may be significantly extended, some of which might have been missed by previous blind line search techniques searching only for unresolved sources \citep{Decarli14,ASPECS1,ASPECS2}. % (with typical CO sizes of 5 kpc--15 kpc; \citealt{Daddi10a, Riechers_GN19,Ivison_GN19,Hodge12,Tacconi13}).

The candidate CO lines span a large range in line FWHM, from $\sim60$ km s$^{-1}$ to $\sim800$ km s$^{-1}$, although the narrowest ones have not yet been independently confirmed. This demonstrates the need for the inclusion of a broad range in line width  templates in order not to miss a significant fraction of the signal. The spread and distribution in the line FWHM  we find is comparable to those previously measured (e.g., \citealt{Tacconi13}), although the occurrence of a particularly broad line (in COLDz.COS.2) in our limited, highest quality sample, suggest that there may be a larger incidence of broad lines in blindly selected CO sources, compared to optical/NIR selections. Nonetheless, due to the limited number of sources, this finding requires further, independent study.

Although the molecular gas fraction estimates for the CO detections are comparable to expectations, the lack of CO detections for a number of massive galaxies with good quality spectroscopic redshifts and the detection of their  stacked CO signal suggests that molecular gas mass fractions for typical Main Sequence galaxies may be somewhat lower than expected (Figure~\ref{fig:gas_frac}; \citealt{Genzel15,Scoville17}). 
A possible caveat to this interpretations may come from  higher systematic uncertainty than expected of the optical spectroscopic redshifts used in the stacking, which may lead to missing a fraction of the CO flux in the stack. While this analysis could only be carried out on  samples of galaxies with previous spectroscopic redshift measurements, its conclusion is in agreement with the finding from the blind search. In particular, if the gas mass fraction and hence the CO luminosity of the known galaxies (in addition to all other galaxies in the observed cosmic volume without spectroscopic redshifts) were closer to expectations,   the number of blind CO detections would have been  higher.  Empirical predictions based on SED fitting and scaling relations suggest an expected number of CO emitters in the range 10--20 for our field in COSMOS and 5--15 for our field in GOODS-N \citep{daCunha_predict}. In addition,  the high mass end of the galaxy distribution in our cosmic volume provides the strongest result for lower than expected gas masses, but also the full sample over the shown stellar mass range provides  important constraints, compatible with a metallicity evolution in the CO conversion factor (Figure~\ref{fig:gas_frac}).
The significant detection of CO emission from  galaxies in the stack suggests that our dataset is rich in additional signal, that is too faint to be reliably blindly identified at the current depth, but which can be mined through spectroscopic observations at other wavelengths.  We have therefore demonstrated the power of stacking the CO signal from galaxies with spectroscopic redshifts, in order to fully take advantage of the information in CO deep field  data.
%This measurement of the CO luminosity density at the faint end ($\sim2\times10^9$ K km s$^{-1}$ pc$^2$) of the luminosity function suggests a fairly flat faint end slope, in agreement with the models by \cite{Vallini16}. The non-detections in the other stacked spectra we have explored, which also reach comparable depth, also suggests that the faint end slope may be very shallow. This may also be due to a rapidly increasing $\alpha_{CO}$ for low stellar mass galaxies, as suggested for example by \cite{Bouwens16} and \cite{Tacconi17}.

We have also developed statistical methods (presented in Appendix F) to evaluate the purity, completeness and recovered candidate properties with higher accuracy than previous techniques.  This enables us to infer the best constraints to date on the CO(1--0) luminosity function at $z\sim$2--3 (Paper~\rom{2}).

With this CO deep field study, we also further demonstrate that  blind CO searches are sensitive to ``optically dark", dust-obscured galaxies at very high redshift, such as GN10 and HDF850.1. In particular, the massive molecular gas reservoirs of these galaxies are among the largest in our field (Riechers et al., in prep.).
Our sample of new, high-SNR CO(1--0) spectra for COLDz.COS1, 2 and 3 and for GN.S1 provides a significant contribution to the state of current CO(1--0) measurements of Main Sequence galaxies at $z>2$\footnote{The number of high SNR CO(1--0) detections in non-quasar hosts or sub-millimeter selected galaxies is very limited to-date \citep{Riechers10b, Aravena10, Aravena12, Aravena14, Bolatto15, Spilker16}.} (see Pavesi et al., in prep., for details).

Finally, the Next Generation VLA (ngVLA)  is necessary to significantly improve the constraints  presented here (e.g., \citealt{Casey_ngVLA,Carilli_ngVLA, ngVLA_McKinnon, ngVLA_memo}). In particular, an equivalent survey in the 30--38 GHz range with five to ten-fold sensitivity improvement for point-source detection as provided by the ngVLA will allow reaching the depth of these observations in a small fraction of the time ($\sim1/50$),  therefore routinely reaching depths of $\log(L'_{\rm CO}/L_\odot)\sim9.5$ in one to two hours of observation. The high survey speed of the ngVLA will uniquely enable the deep, wide area surveys which are necessary to build large statistical samples, currently inaccessible to the VLA. These future surveys will constrain the luminosity function to well below the knee, with percent precision, for a comparable observing effort as the present survey. A significant benefit of the ngVLA will also come from the planned smaller antennas, which increase the field of view for a fixed total collecting area, therefore enhancing the survey speed.
In addition, the vast bandwidth of the ngVLA  will allow us to simultaneously cover CO(1--0) emission over a large fraction of the age of the Universe, and will therefore allow us to probe CO(1--0) over the almost complete redshift range up to $z\sim$10.

\smallskip
\acknowledgments 
The National Radio Astronomy Observatory is a facility of the National Science Foundation operated under cooperative agreement by Associated Universities, Inc.
DR and RP acknowledge support from the National Science Foundation under grant number
AST-1614213 to Cornell University.
RP acknowledges support through award SOSPA3-008 from the NRAO. We thank Tom Loredo for helpful discussion.
IRS acknowledges support from the ERC Advanced Grant DUSTYGAL (321334), STFC (ST/P000541/1) and a Royal Society/Wolfson Merit Award.
VS acknowledges support from the European Union's Seventh Frame-work program under grant agreement 337595 (ERC Starting Grant, CoSMass)
RJI acknowledges ERC funding through Advanced Grant 321302 COSMICISM.
This research has made use of the NASA/ IPAC Infrared Science Archive, which is operated
by the Jet Propulsion Laboratory, California Institute of Technology, under contract with the National Aeronautics and Space Administration.
This work is based on observations taken by the CANDELS Multi-Cycle and 3D-{\em HST} Treasury Program (GO 12177 and 12328) with the NASA/ESA {\em HST}, which is operated by the Association of Universities for Research in Astronomy, Inc., under NASA contract NAS5-26555.
We acknowledge funding towards the 3-bit samplers used in this work from ERC Advanced Grant 321302, COSMICISM.
JAH acknowledges support of the VIDI research program with project number 639.042.611, which is (partly) financed by the Netherlands Organisation for Scientific Research (NWO).
MTS acknowledges support from the Science and Technology Facilities Council (grant number ST/P000252/1).

\bibliographystyle{aasjournal}

\bibliography{biblio_CODF}

\appendix

\section{Additional details on the line search methods}
In this section we provide additional details of our line search methods in interferometric data cubes, which were used to carry out the blind line search presented in Section~4 of the main text.
We first provide a more complete description of our method of choice, Matched Filtering in 3D\footnote{We provide a Python implementation for this algorithm at https://github.com/pavesiriccardo/MF3D} (MF3D, which extends Matched Filtering in the spectral dimension; i.e., MF1D), and then we compare its performance to three alternative methods that we have also investigated. 

\subsection{Matched filtering in 3-D interferometric data cubes}
Since we do not expect the CO line emission in $z=2$--3 galaxies to be resolved over more than a few beams at most, we expect our sources to be spatially well described  by a family of 2-D Gaussian templates.
Therefore, under the prior of source shape, Matched Filtering is theoretically an optimal detection method. 

The Matched Filtering method can be thought of as concentrating all of the extended (spatially and in frequency) signal to a peak pixel that captures both the overall strength of the original signal and how closely this matches the template shape. At the same time, the smoothing of the noise in  regions without signal allows us to reliably measure the noise level on the scale probed by the template size. In this way, the problem of finding emission lines with structure is effectively reduced to the problem of just examining peak heights to assess their significance.

We compute templates that are Gaussians in frequency and circular 2D Gaussians spatially (sizes given in Table~\ref{template_sizes}).  We then convolve the signal-to-noise cube with these templates by multiplication in Fourier space to produce multiple Matched-Filtered cubes, one for each template. %We refer to this convolving as Matched-Filtering because the templates match what we expect the signal to look like, we also occasionally refer to it as smoothing in the following. 
%We include a template with a spatial size smaller than a beam to better characterize the SNR of point sources (beam-sized templates give higher SNR to point sources that are surrounded by some positive noise, while the smallest template is not sensitive to the noise surrounding the peak beam). This smallest template ensures that our method represents a generalization of the single-pixel-based (1D) matched-filtering including its contribution as a subset, this is important because MF1D is optimal for unresolved sources (which have the full SNR in the single peak pixel)

The main difference between the traditional application of Matched Filtering in astronomical images and our application to interferometric data comes from the spatially correlated nature of the noise in interferometric images. In the case of un-correlated (i.e., white) noise, the matched filter simply corresponds to the expected source shape and size, but correlated noise introduces deviations from this matching, as described in the following.

The frequency width of a template approximately matches the line width that it selects, because the noise in different channels is uncorrelated.
On the other hand, spatially the noise has a non-zero correlation length, as determined by the synthesized beam.  Therefore, the ``matching" to a template is not the intuitive relation for which spatial template size matches the size that it  selects. As an example, for un-resolved sources, i.e. sources whose image is beam-sized, the maximal SNR is realized at the peak pixel rather than over an extended area. 
To calculate the relationship between template size and selected size, we therefore considered the idealized problem of circular Gaussian beams and Gaussian sources, which can be treated analytically.  We calculate the correspondence between template size which maximizes the SNR and source size (see Appendix B).
To carry out the calculation, we have to make the approximation of  source positions being known a priori, evaluating the signal-to-noise at this position. This is not what is done in practice, since  positions are unknown.  The pixel with the locally highest SNR is  utilized instead. We briefly discuss the effects of this approximation on the recovered SNR in  Appendix B.

The results of this calculation show that template ``matching" (i.e. providing the maximum SNR) takes place approximately when:
\begin{equation}
\sigma_A^2=\sigma_h^2+2\sigma_b^2
\end{equation}
where $\sigma_A$ is the size (radial standard deviation) of the source in the image (which is  given by the sum in quadrature of the real source size and the beam), $\sigma_h$ is the size of the template Gaussian, and $\sigma_b$ is the beam size.
For source sizes smaller than $\sqrt{2}\sigma_b$, the template size that maximizes the SNR is  an infinitely narrow source template. Therefore, we include a single-pixel template (we call this ``0-size", being the limit where the radius of the spatial Gaussian tends to 0), which implements a single-pixel search (i.e., MF1D as a subset of MF3D) and therefore selects un-resolved and very slightly resolved sources. The analytical expression above only provides an indication of the matching dependence, but we do not make use of it in the following. The main simplification comes from measuring the SNR at the (in practice unknown) real position of the source rather than at the local maximum. In Appendix F, we explore the analysis of simulated artificial sources through our MF3D algorithm, and we use those simulations to numerically estimate a probabilistic connection between template sizes and injected source sizes.

\begin{figure}[htb]
\centering{
 \includegraphics[width=.45\textwidth]{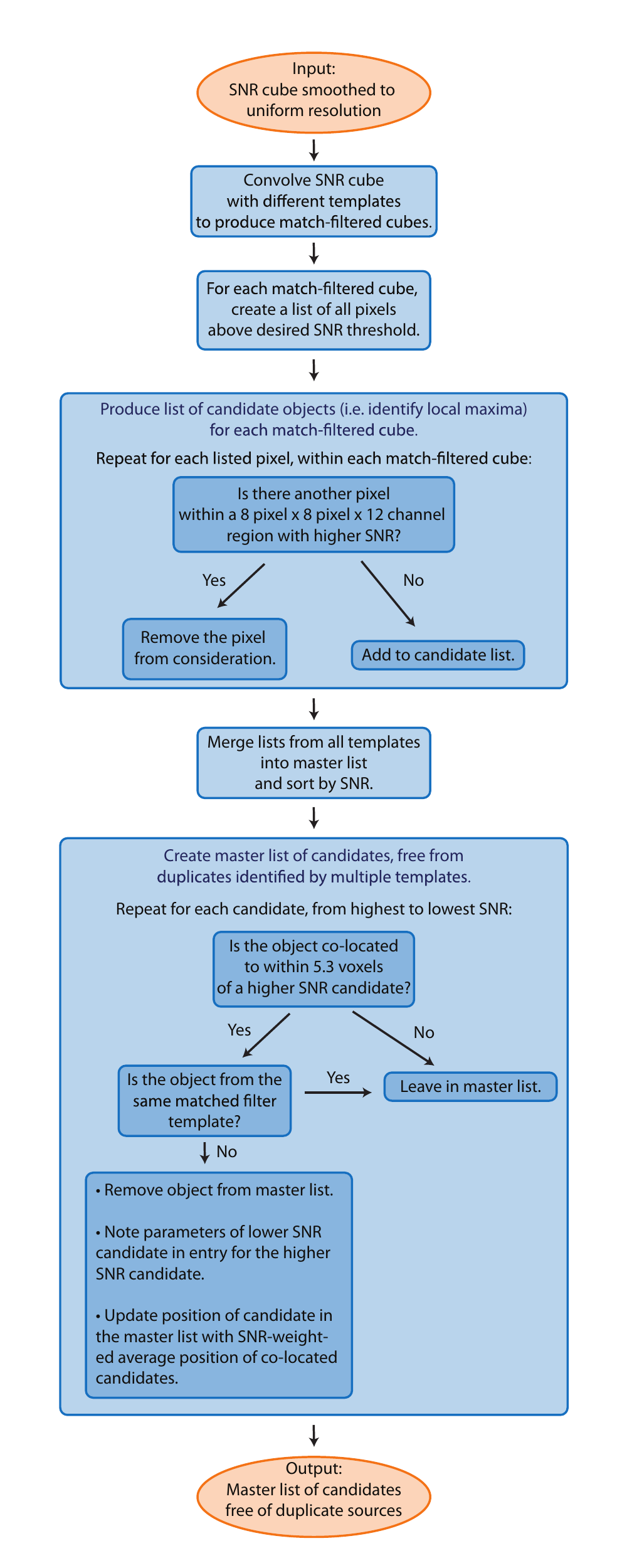}
}
\caption{Flowchart describing the detailed procedure of our line search algorithm, utilizing Matched Filtering in 3D. The input is a signal-to-noise ratio cube and the output is a list of line features, characterized by a feature signal-to-noise ratio, which accounts for the total signal-to-noise evaluated through the template that gives the highest value.}
\label{fig:flow_MF3D}
\end{figure}

The detailed steps of the blind search are summarized by the flowchart in Fig.~\ref{fig:flow_MF3D}, and in the following. As described in Section 2, in order to correctly mosaic different fields together we smooth every pointing to a common, larger, beam. This procedure reduces the SNR for point sources (see Fig.~\ref{fig:Noise}).  We therefore also run a single-pixel Matched Filtering search on the Natural-mosaic, which was obtained without any smoothing. While the lack of a common beam in the Natural-mosaic would, strictly speaking, imply that the spatial structure may not be accurately calculated, this effect is negligible in the COSMOS data, where the different pointings have roughly equal resolution.  Furthermore, the Natural-mosaic is sufficient for a search for un-resolved sources, where the flux at the peak pixel represents the total flux, and is therefore correctly recorded in the Natural-mosaic. We treat the result of this Natural-mosaic Matched Filtering step as an additional ``spatial template", one for which less smoothing was done than even the single-pixel search in the Smoothed-mosaic. We therefore refer to it as the ``-1 pixel" template. In the end, we combine the results from this search with those of the other templates, as detailed in the following.

The signal-to-noise ratio of a detection corresponds to the ratio of the height of the peak in the Matched-Filtered cube to the standard deviation of empty regions. We initially normalize our templates such that the sum of the squares of the template values equals one. For independent pixel noise (which applies to the frequency channels, but not to the spatial pixels), this normalization choice would imply that the noise after convolving would be the same as the noise before. This implies that the peak height corresponds to the total SNR of the candidate. For the 3D case, in particular for spatially extended templates, we need to account for the fact that the synthesized beam size results in small scale spatial noise correlations. While the calculation for the noise in the smoothed cube is close to the measured values (see Appendix B), we decide to measure the noise in the convolved cubes directly from the standard deviation of pixel values.
Since our dataset is mostly free of signal, we estimate the noise in each Matched-Filtered cube in the COSMOS field simply by taking the standard deviation of the whole cube, and normalize by dividing each pixel by this value. In the case of the GOODS-N mosaic though, the beam size is not uniform across the mosaic, even after smoothing. In particular, the beam size in the pointings that had higher native resolution, ends up being larger after smoothing than in the other pointings due to the particular nature of the $uv$ coverage  (see Appendix C for details). The main consequence of this slightly spatially varying beam, affecting the Smoothed-mosaic for the GOODS-N field, is that during Matched Filtering, the noise in the Matched-Filtered cubes is not uniform. This is expected, as the noise change during convolution is a function of the ratio of the pre- and post-convolution beam size.
We therefore measure the noise in the Matched-Filtered cubes, separately for two sets of pointings (GN29--GN42; and all the others), and use these sets to construct an approximate noise map for the Matched-Filtered mosaics in GOODS-N, for each template (see Appendix C for details).

%We use templates of 4, 8, 12, 16 and 20 channels in FWHM (x4MHz) and 3, 6, 7, 8.5, 10, 12 pixels in spatial FWHM (.5 arcsec per pixel), where the beam size is between 6 and 7 pixels (in COSMOS).

%give widths in km s$^{-1}$ too
\hspace{-30pt}
\begin{table}
\caption{Template Sizes for MF3D Line Search Technique}

\centering
\begin{tabular}{ l l l }
\hline
 & Spatial FWHM  & Frequency FWHM \\
 & (arcsec) &   (4 MHz-channels) \\
   \hline
    %COSMOS & 1.5, 3, 3.5, 4.25, 5, 6 & 4, 8, 12, 16, 20 \\
 %   COSMOS & 1.5, 3, 3.5, 4.25, 5, 6 & 137,  274,  411,  549, 686 \\
    COSMOS & 0, 1, 2, 3, 4 & 4,  8,  12,  16, 20 \\

   \hline
 % GOODS-N &  2, 3.5, 4, 4.75, 5.5, 6.5 & 4, 8, 12, 16, 20 \\
% GOODS-N &  2, 3.5, 4, 4.75, 5.5, 6.5 & 140,  282,  424,  565, 706 \\
 GOODS-N &  0, 1, 2, 3, 4 &4,  8,  12,  16, 20 \\
  \hline
\end{tabular}

\label{template_sizes}
 \begin{tablenotes}
 \item
 \textbf{Note} Gaussian template sizes utilized in our Matched Filtering in 3D. A spatial size of 0 stands for a single-pixel spatial extent, and implements the single-pixel search that is optimal for un-resolved sources. These sizes represent a uniform sampling of the parameter space that we conservatively expect to represent  CO sources. 4 MHz correspond to $\sim$35 km s$^{-1}$, mid-band.
 \end{tablenotes}
\end{table}

%The normalization of the templates is arbitrary, therefore the final cubes need to be normalized.

%With Matched Filtering in 1D we search for the highest peak in frequency at each spatial pixel, if the peak in any template-case crosses the SNR threshold we keep it and store the channel where the highest peak is located. Hence we do not allow having more than one peak per spatial pixel.
%Then we have to spatially clump high pixels, we do that by building lists of clumps with running average positions, where a new object gets added to 3D clumps if within $\sim$ 9 voxels radius, or creates a new clump if not added anywhere.

In order to blindly identify line features in our data and evaluate their significance using the Matched-Filtered cubes (one for each template), we need to locate the peaks and determine the template which provides the highest SNR to each candidate, thereby identifying the template that best matches the feature shape.
The first stage is to identify, in each Matched-Filtered cube, the peaks, i.e. the local maxima above some significance threshold. 
In order to find the significance and position of a peak, we select all voxels (i.e., volume pixels) in the data cube above a fixed threshold, and then retain those voxels that are local maxima by comparing to the values in a small surrounding box of 12 channels by 8 pixels by 8 pixels.

%Next we identify objects across different Matched-Filtered cubes obtained from different templates. We form a master list of all the objects selected from all templates, sorted by SNR. Then we go through each entry from the highest down and form clusters characterized by SNR-weighted average positions and with each cluster being tagged by the template where it peaks. We add an object to a cluster if within $\sim$5.3 voxel radius from the cluster center and only if the peak template differs from the peak of the current cluster, in order not to cluster together features identified  in the same template (since these should be treated as separate clumps). The strategy of going down a SNR-sorted list guarantees building up clusters from their highest significance members to the lower, in such a way that a broad object will have lower-SNR in the smaller templates and will be included as cluster members, but two neighboring point-like sources that have their highest SNR in the small templates will see the features identified by wider templates be added to both objects, representing that they both contribute to that SNR peak in the wider template while not capturing the property of interest. The objective is ensuring that we do not clump too much, i.e. putting separate objects into the same cluster, and that we do not split single objects into separate components.
Next, we cross-match objects identified in the different Match-Filtered cubes (obtained from the
different templates) in order to remove repeat identifications of the same object. We form a master list of all objects selected from all templates, sorted by SNR.  We then  parse through each entry, from highest to lowest SNR, and form clusters characterized by their SNR-weighted
average positions and the template with the highest SNR. We add a candidate object to a cluster if it resides within a 5.3 voxel radius of the cluster center (this threshold was found to be appropriate for our pixel size and channel width; more generally the frequency and spatial separation thresholds may need to be different) and only if the template under consideration differs from the other templates in the cluster (to avoid clustering features identified in the same template since they are most likely independent objects). By moving down the SNR-sorted list, we guarantee that clusters are built from their highest significance members to their lowest. This method ensures that spatially extended/broad objects that are also identified at lower significance in smaller templates are included in the appropriate cluster as members. For neighboring point-like sources (with high significance in the smallest templates), this method maintains both objects as separate clusters, and allows their corresponding low-significance/extended template candidate to be associated with both clusters. In this way, we avoid grouping separate objects into the same cluster, and we avoid splitting single objects into multiples. Each cluster then corresponds to a single galaxy candidate in our final catalog.

In order to choose the clustering thresholds, and to asses the independence of the result from their precise values,  we test how well the algorithms performs in not clumping too much or too little by computing distances to closest neighbors. We test this both for the first stage of clump-finding in the Matched-Filtered cubes to check the method to identify clump peaks works, and for the second stage of matching features across different templates.
We inspect the distribution of the neighbor distances, and check that they behave as expected, without splitting clumps into different components (which would show up as many objects having a very close neighbor that would look like part of the same clump to visual inspection), and without including different clumps (by changing the clustering thresholds and looking for any significant changes). We do not find significant issues in either phase of clustering.  This technique was therefore used to refine our choice of clustering thresholds. A few objects are objectively difficult to distinguish as one or more parts and so the algorithm performance is at a comparable level to what could be achieved through manual inspection. Overall, the method does very well in finding local peaks, appropriately splitting separate objects even when they are close together.  The second stage of associating entries across different templates, while more challenging to evaluate, appears to be largely insensitive to the precise value of the thresholds within a few voxels range.

\subsection{Comparison to Matched Filtering in 1D}

A simpler version of our line search algorithm, which we call Matched Filtering in 1D (MF1D),  corresponds to extracting a spectrum at each spatial pixel and running a spectral line search on each spectrum with 1D Gaussian templates. 
As emission lines at high-$z$ are typically approximated as Gaussians, we note that assumptions of square profile templates are less optimal matches in the frequency dimension and therefore do not maximize the SNR for candidate emission lines, although we find the difference to be small.

%One line searching technique that was recently utilized in \cite{ASPECS1}, is equivalent to Matched Filtering in 1D with square line templates. This choice of effective square templates is sub-optimal in the frequency dimension because it is expected to provide a worse match to the spectral line profile than a Gaussian shape, therefore not recovering the full SNR.

We have investigated the MF1D approach, which is frequently used in the case of single-dish data, in order to provide a check on the results of our line search and to evaluate its performance. 
\cite{ASPECS1} utilized a version of this method, which is effectively matched filtering in 1D with square line templates. The main difference between the method utilized there and our implementation consists in our estimating the noise through the standard deviation of the full SNR cube, rather than individual binned channel maps. This is not expected to cause a significant difference.
This method, like MF3D, also requires some prescription for recognizing clusters of significant voxels as belonging to the same candidate when they are close together. We achieved this by building lists of clumps with running average positions, and clustering up to a radius of 9 voxels.
In the case of unresolved sources, where the peak pixel contains the maximum signal-to-noise, this method performs just as well as our more general method, since the set of templates used in this technique is a subset of those in MF3D (``0-size" templates). However, it will miss a large fraction of the extended sources by underestimating their true SNR. Although we do not expect a large fraction of resolved sources, a blind search should be as agnostic as possible with regards to the properties of the galaxies that may be selected. Indeed, since CO(1--0)  traces the total cold, dense gas mass, it is precisely the tracer that may reveal extended gas reservoirs. One of our top candidates, COLDz.COS.3, harbors a very extended gas reservoir, with SNR peaking in the 2$^{\prime\prime}$ template.  A single-pixel search assigns this line a SNR=8.2 rather than 9.2, which would imply a discrepancy of -10\%.  While this error would not significantly affect the significance of this candidate, such an error would be enough to move a moderate significance candidate with SNR=5.5 , to 4.9, and therefore  would  effectively be missed by our search.
Another advantage of the MF3D method over the 1D is that it allows to capture a larger fraction of the signal, for broader lines, because in that case the peak signal may be substantially spread over several spatial pixels, in different frequency channels, due to noise.
While this spreading of the signal over different pixels, in different channels, causes an ambiguity between spectral-SNR and moment-map-SNR for single-pixel methods, this ambiguity is resolved when the full 3D information is taken into account through MF3D.
Therefore, we conclude that this method can be absorbed into our more general, improved MF3D framework and that it can be considered a subset of that technique.

\subsection{Comparison to the \texttt{Source Extractor} method}
We also considered modifications of existing source finding software, such as \texttt{Source Extractor}, which can effectively capture the spatial information of a line candidate while avoiding merging adjacent independent peaks \citep{SExtractor}. We used the spatial source detection part of SExtractor on individual channel maps with varying frequency binnings. We combined the detections across different binnings and at different frequencies and then established prescriptions to identify lines and their aperture-integrated SNR. These prescriptions made the results very dependent on the precise criteria used to evaluate the significance of a line.
The principle is somewhat similar to Matched Filtering. It requires binning  data cubes to multiple different velocity widths, and these binnings correspond to templates of different frequency width. Then the method  relies on SExtractor for the spatial source extraction (recognizing clusters of high pixels as one unique object). It also requires finding the correct binning that maximizes the signal-to-noise ratio.
A challenge for this method is the choice of an aperture size for the flux extraction in the channel maps, to be used in separately evaluating signal and noise.  Combining different aperture sizes, which  imitates the range of spatial templates in MF3D,  introduces additional difficulties with precisely evaluating aperture flux noise.
This hybrid technique is sub-optimal in the frequency dimension, because binning is equivalent to filtering with a rectangular function, which is a worse match to the expected spectral line profile than a Gaussian shape, although the difference is small.
Our tests show that this line search method, can have similar outcomes in selecting lines to Matched Filtering in 3D for our data. In particular, $>85\%$ of the top $\sim100$ candidates are matched between both methods. Comparing lists of candidates, we find that objects that were assigned a high SNR by the SExtractor method but were less significant in MF3D appeared to be less plausible candidates to visual inspection because of improbable line shapes. %We find however that visual inspection tends to prefer objects with a higher signal-to-noise ratio assigned by MF3D compared to those from the SeXtractor method, as expected based on the extra profile match requirement the former evaluates.

The SExtractor method provides a valuable check for our use of extended templates in MF3D. In particular, the extended templates in MF3D allow finding those sources that would be missed by MF1D. The SExtractor method, which is sensitive to extended structure, confirms our extended candidate selection.
Therefore, we conclude that our MF3D method coherently combines the results from single-pixel methods and other methods which are biased towards extended sources, like the use of SExtractor with fixed aperture sizes.

\subsection{Comparison to the {\sc spread} technique}
We have also explored {\sc spread}, an algorithm developed by \cite{Decarli14} for the PdBI blind field line search, to find emission lines in our VLA observations. This method corresponds to binning the data set in frequency, and identifying channel maps with an excess of signal compared to the Gaussian noise pixel intensity distribution. This method does not take advantage of the spatial information (neither spatial extent nor position), but only of the total flux. The excess signal in a channel map does not need to come from a single source, because the {\sc spread} statistic is a global value that characterizes the whole channel map.
This method did not perform reliably on our data set, since it relies on the small number pixel statistics on the tails of the noise distribution, which are necessarily subject to large fluctuations. The {\sc spread} statistic was able to isolate the same top candidate sources as our other methods, but it loses discriminating power below a SNR of $\sim8$, since the {\sc spread} statistic does not track SNR and loses the ability to locate moderate significance features.
We conclude that MF3D captures any useful information obtained from {\sc spread}.

\subsection{Comparison to  Duchamp}
 We also compare our method to the sophisticated line-searching tool {\em Duchamp}, which was developed for SKA-precursor data cubes \citep{Duchamp1}. {\em Duchamp} was extensively tested by \citep{Duchamp_test1,Duchamp_test2}, and found to provide a good blind search algorithm for both unresolved and extended emission. Because our survey is only expected to detect unresolved or slightly resolved CO emission, much of the power of {\em Duchamp} (e.g., ``a trous" wavelet reconstruction) is not optimized for our targets of interest. The smoothing (convolution) pre-processing offered by {\em Duchamp} is equivalent to Matched Filtering with Gaussian templates in the spatial dimension and Hanning templates in the frequency dimension, although {\em Duchamp} only allows specifying one template size at a time, and not combining results from different templates. We find that smoothing along the frequency axis is necessary in order to recover even the most significant line emitters in our cubes, as expected due to the wide line-widths relative to channel widths. On the other hand, {\em Duchamp} does not allow to smooth in the frequency and spatial dimensions simultaneously, thereby preventing optimal recovery of the full signal-to-noise for slightly extended sources. While the ``a trous" wavelet reconstruction is designed to perform well on extended structure of a general shape, it is not optimal for recovering only slightly extended spatial structure, and hence does not yield the same signal-to-noise recovery as Matched Filtering in this specific case of interest. {\em Duchamp} offers two choices for peak identification algorithms leading to candidate identification, with pixel clustering being predominantly carried out spatially rather than spectrally. While both of these algorithms perform equally well in recovering all of our top line candidates, the simple 3D peak identification algorithm implemented in MF3D simultaneously utilizes the full 3D information. 

Based on all these considerations, we find that the best use of {\em Duchamp} in our data is achieved by manually adopting different frequency-width smoothing templates and combining the resulting signal-to-noise ratios, to select unresolved line candidates of different velocity widths. This procedure directly mimics our MF3D method, and therefore we do not adopt {\em Duchamp} for the COLDz survey data.

\section{Matched-Filtering interferometric images}
We here discuss the analytical results of our investigation of Matched Filtering in 3D, in the specific case when it is adapted to interferometric images. We study idealized noise and source conditions, in order to derive an approximate relationship between the template spatial size and the ``matched" size of a feature that would display the highest SNR for that template.
The purpose is to demonstrate the effect of correlated interferometric noise on the sizes that are selected through this technique.

If for simplicity we assume the synthesized beam to be a circular Gaussian, with standard deviation $\sigma_b$, then the noise correlation function can be shown to be $\langle n({\bf x}) n({\bf x'})\rangle=\sigma_0^2 e^{-|{\bf x}-{\bf x'}|^2/4\sigma_b^2}$, where $\sigma_0$ is the noise of the image.  Let us define our idealized  data as containing a Gaussian source of peak intensity {\it s} and convolved size $\sigma_A$, in addition to additive, zero-mean Gaussian noise in the image. We also assume the spatial template to be a circular Gaussian of size $\sigma_h$, i.e. $h({\bf x})=\frac{1}{2\pi\sigma_h^2} e^{-x^2/2\sigma_h^2}$.
The expectation value of the template-convolved image at the (assumed known) position of the source is then given by: $\frac{s\sigma_A^2}{\sigma_A^2+\sigma_h^2}$. Furthermore, the standard deviation of the convolved image is given by $\frac{\sigma_0 \sigma_b}{\sqrt{\sigma_b^2+\sigma_h^2}}$. Therefore, the signal-to-noise ratio measured in the Matched-Filtered image is given by $SNR=\frac{s}{\sigma_0} \frac{\sigma_A^2 \sqrt{\sigma_b^2+\sigma_h^2}}{\sigma_b (\sigma_A^2+\sigma_h^2)}$. For a fixed source size $\sigma_A$, this SNR has a maximum at $\sigma_h^2=\sigma_A^2-2\sigma_b^2$ or $\sigma_h=0$ (i.e. the delta function limit of a Gaussian, corresponding to a single-pixel template) in case $\sigma_A^2<2\sigma_b^2$, i.e, if the intrinsic deconvolved source size is smaller than the beam size.

We have run simulations to compare these  analytical results to the discretized case of pixels, and did not find significant differences. We also explored the effect of the realistic implementation of Matched Filtering, i.e. where the source position is not known a-priori but the peak of the convolved image is taken instead. The main result appears to be that for $\sigma_A^2$ up to $2\sigma_b^2$ the signal-to-noise is almost flat as a function of template size.  Therefore, a single-pixel template and slightly extended templates maximize the signal-to-noise with a smooth and slow transition, as the source size becomes more important relative to the beam size.
For the purpose of our measurement, a precise formula for the match between template size and source size is not needed, and a probabilistic assignment based on artificial source recovery suffices. In particular, the results from our artificial sources show that in the full 3D case the matching may be complex and it depends on signal-to-noise as well as on line velocity width. The extra dependence on the line width can be understood as due to the use of the peak value, rather than the value at the known source position, because a wider-velocity line allows for a larger area over which the peak may be found (due to the combination to positive noise and real signal).
To conclude, the matching of sources and templates at the basis of Matched Filtering can be approximately estimated from the previous calculation. For unresolved or  slightly resolved sources, the SNR is a  weak function of templates size.  As the $\sigma_A^2\sim2\sigma_b^2$ threshold is approached and crossed, the SNR becomes a rapidly increasing function of template size, with a clear peak for extended templates. Therefore, in order to avoid missing extended sources in blind line searches in interferometric data, we recommend the inclusion of extended (hence 3D) templates, as described in this work.

\section{Accounting for the beam inhomogeneity in our GOODS-N mosaic}
In order to mosaic pointings together, it is preferable to smooth all pointings to a common beam. This is not straightforward for the wide-area part of this survey (in the GOODS-N field), due to the large number of pointings observed over the course of several months, which caused a range of array configurations to be utilized.

For the pointings in the COSMOS field, each pointing was observed in every track.  Therefore, the mosaic has uniform beam size properties. In the GOODS-N field, on the other hand, the beam differences potentially cause non-uniformity in the mosaic. This is most significant for pointings GN29 to GN42, which were mostly observed in the DnC configuration.
In preparation for mosaicking the individual pointings are all smoothed in the image plane to a common beam size with {\sc casa} \texttt{imsmooth}, but pointings for which a larger amount of smoothing was required end up with slightly larger beams than the target beam size.

\begin{figure}[htb]
\centering{
  \includegraphics[width=.5\textwidth]{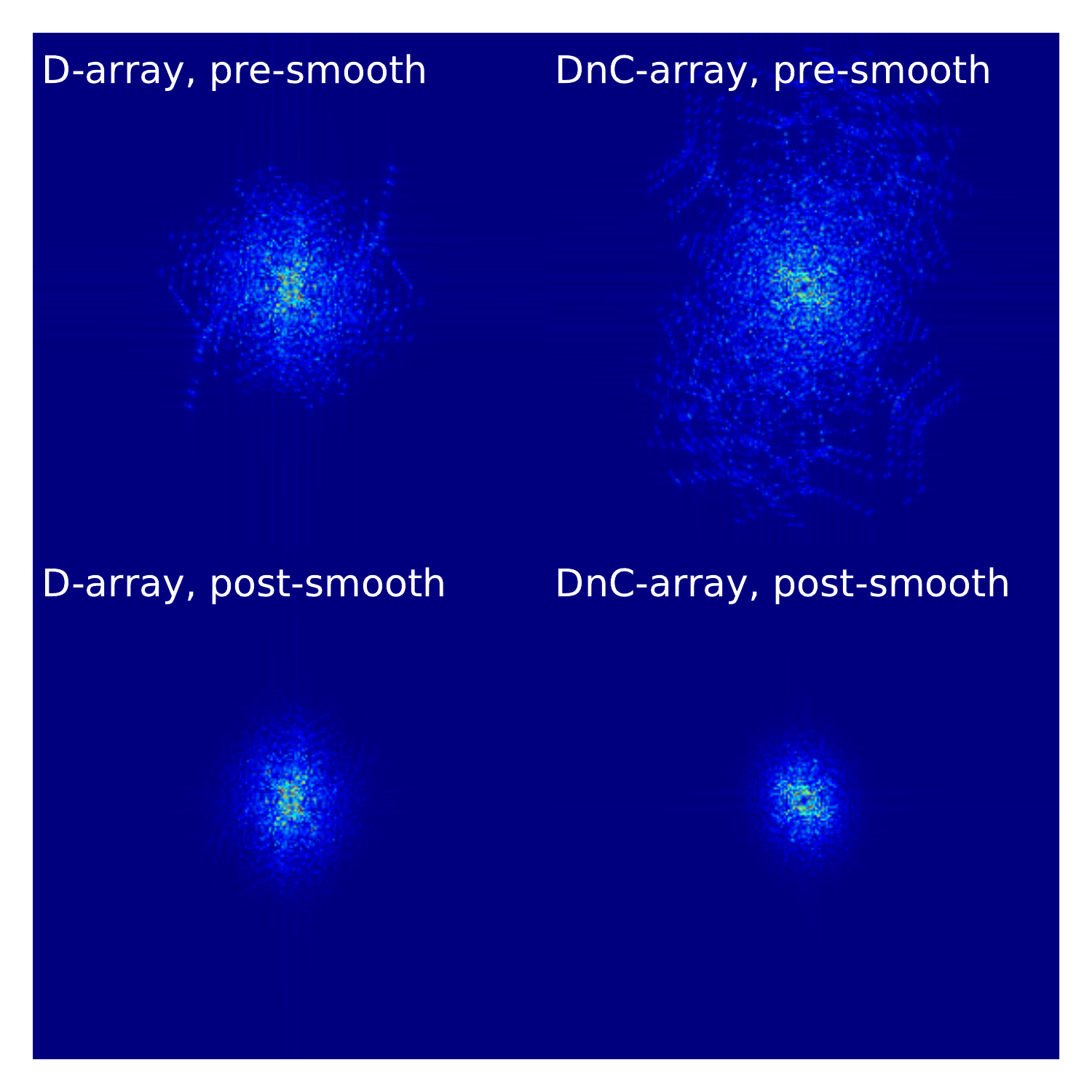}
}
\caption{Absolute value of the Fourier transform of channel maps (in this example, channel 100), of individual pointings (in this example, pointing GN3, which has D-array data only, and GN34, which has mostly DnC array data), before and after smoothing. The post-smoothing image shows that, although the smoothing is supposed to bring the two pointings to a common beam, instead it makes the resolution of the DnC array pointings coarser, i.e. it produces a larger beam than in the pointings with D-array data only.}
\label{fig: Smoothing_GN}
\end{figure}

The reason why smoothing DnC data seems to necessarily produce slightly different beams than D-configuration data can be appreciated from a look at the Fourier Transform of a typical image from these pointings, effectively their $uv$ coverage, in Fig.~\ref{fig: Smoothing_GN}. Smoothing multiplies the $uv$-plane by a tapering Gaussian of the appropriate size, calculated from the size of the starting beam size and the target beam size. However, the D-configuration $uv$ coverage, does not look the same as a Gaussian-tapered DnC-configuration $uv$ coverage, hence the final beam is always going to be an imperfect match (unless both datasets are smoothed to a very large beam, at which point the initial shape of either $uv$-coverage does not matter).

While the slight spatial inhomogeneity of the beam size is inconsequential in producing the Signal-to-Noise ratio cube (as the noise in the mosaic can be calculated analytically and accounted for by Eqn.~\ref{mosaic_noise}), the beam size difference causes spatially varying noise in the Matched-Filtered cubes (see Appendix B).
The main difference is between the set of 14 pointings (GN29--GN42) and the rest, so we also mosaic and Match-Filter them separately, in addition to working with mosaics with and without this set. Exploiting the improved uniformity within these sub-mosaics, we can measure the noise post-Matched-Filtering. The objective is using the noise in the Matched-Filtered sub-mosaics to calculate the noise in the Matched-Filtered full GOODS-N mosaic. The Signal-to-Noise ratio in the full mosaic is related to the sub-mosaics by:

\begin{equation}
\begin{split}
%SNR_{TOT}=\frac{\sqrt{\sum_A r_i^2 A_i^2}}{\sqrt{\sum_{TOT} r_i^2 A_i^2}} SNR_A+\frac{\sqrt{\sum_B r_i^2 A_i^2}}{\sqrt{\sum_{TOT} r_i^2 A_i^2}} SNR_B \doteq f_A SNR_A+f_B SNR_B
SNR_{TOT}=\\
&\hspace*{-60pt} \frac{\sqrt{\sum_A  A_i^2/\sigma_i^2}}{\sqrt{\sum_{TOT} A_i^2/\sigma_i^2}} SNR_A+\frac{\sqrt{\sum_B A_i^2/\sigma_i^2}}{\sqrt{\sum_{TOT} A_i^2/\sigma_i^2}} SNR_B \doteq \\
& \hspace*{-60pt} f_A SNR_A+f_B SNR_B,
\end{split}
\label{appendix_eqn}
\end{equation}
where $TOT=A \bigcup B$ represent the set of GN29--GN42 pointings and the set of the remaining pointings. Match-Filtering then corresponds to convolving with template {\it h}. This  can be expressed as:

\begin{equation}
\begin{split}
(f \cdot SNR) \ast h=\int f(x) SNR(x) h(y-x) dx \simeq \\
& \hspace*{-60pt} f(y) \int SNR(x) h(y-x),
\end{split}
\end{equation}
 with $f(x)\simeq f(y)+ O(f'\cdot FWHM_h) $. To zeroth order, we can take {\it f} as constant over the scale of template {\it h}, which allows to pull it out of the integral. This approximation is appropriate, because the fraction functions {\it f} change slowly over the size of a template.    Therefore, the noise after convolving with the template is:
\begin{equation}
std[ (f_A(x) SNR_A) \ast h ]  \simeq f_A \cdot std[ SNR_A \ast h ],
\end{equation}
and we can calculate the noise in the Matched-Filtered mosaic by summing the standard deviations from the two terms in Eqn.~\ref{appendix_eqn} in quadrature. This method only requires measuring the noise in the Matched-Filtered sub-mosaics, in which  the noise is uniform, and the fraction functions: $f_A$ and $f_B$, which can be calculated.
Therefore, this process allows us to calculate noise maps of the Matched-Filtered cubes in GOODS-N, thereby accounting for the noise inhomogeneity due to spatially varying beam sizes.

\begin{figure}[htb]
\centering{
 \includegraphics[width=.4\textwidth]{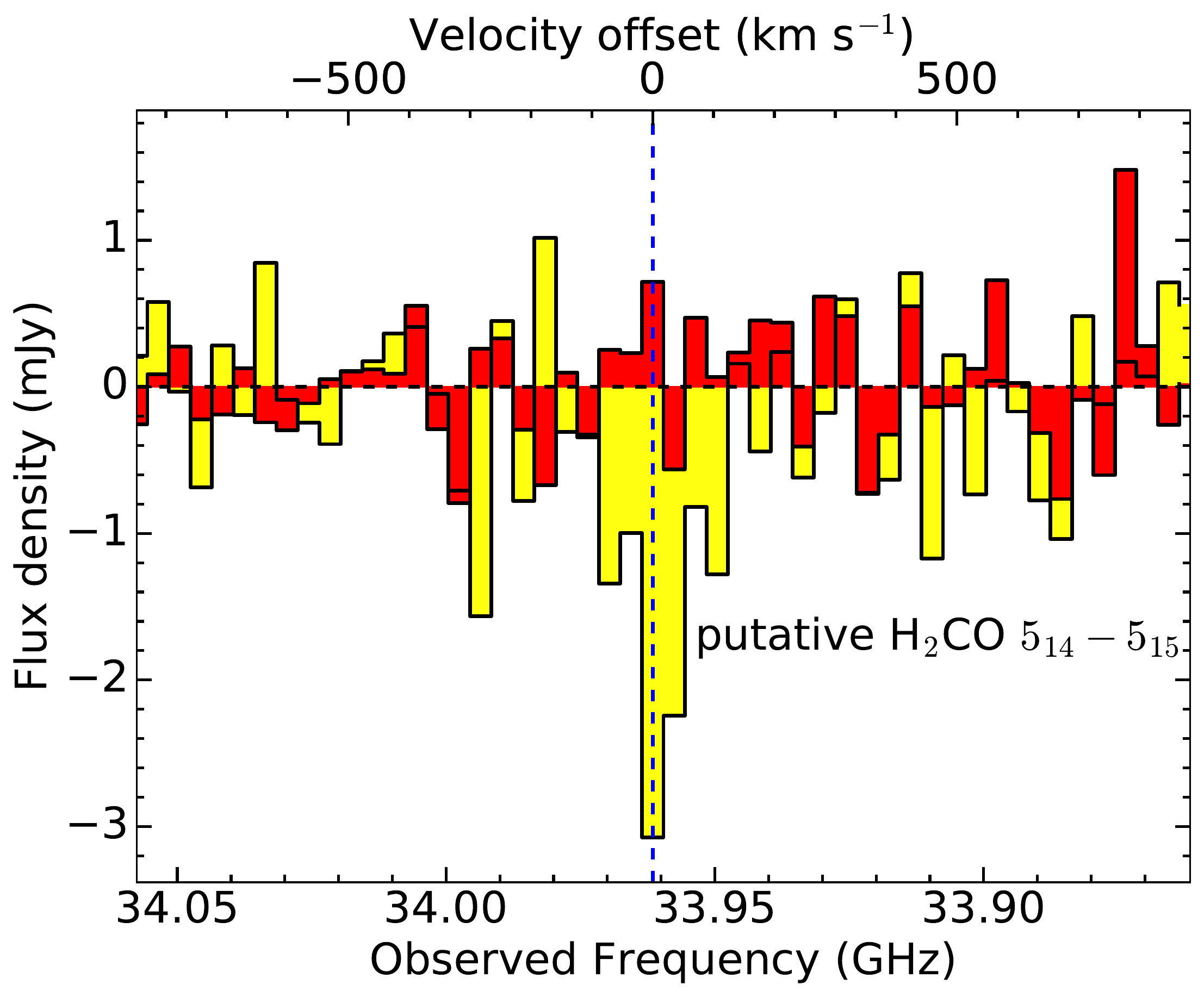} 
 }
 \includegraphics[width=.28\textwidth]{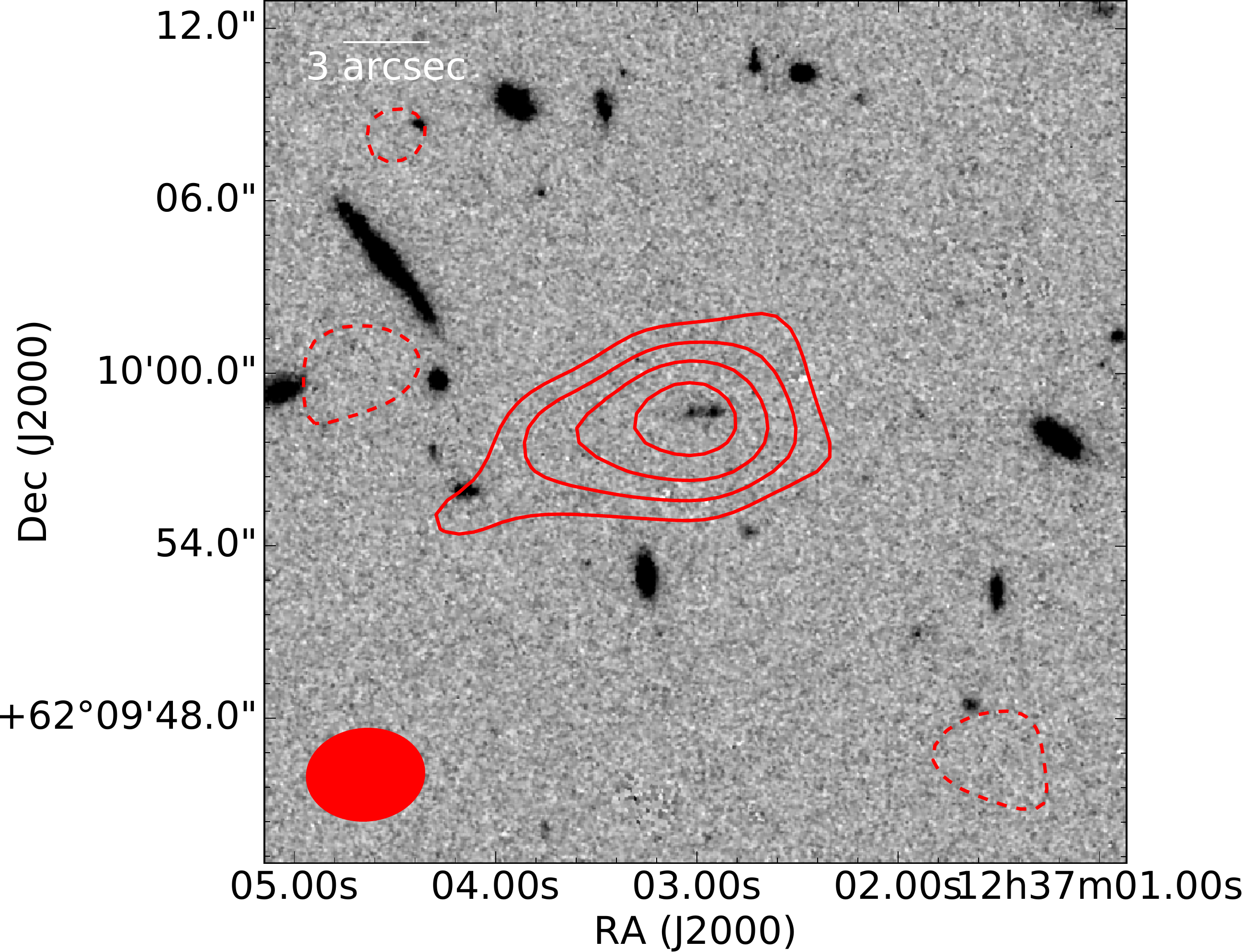}\includegraphics[width=.23\textwidth]{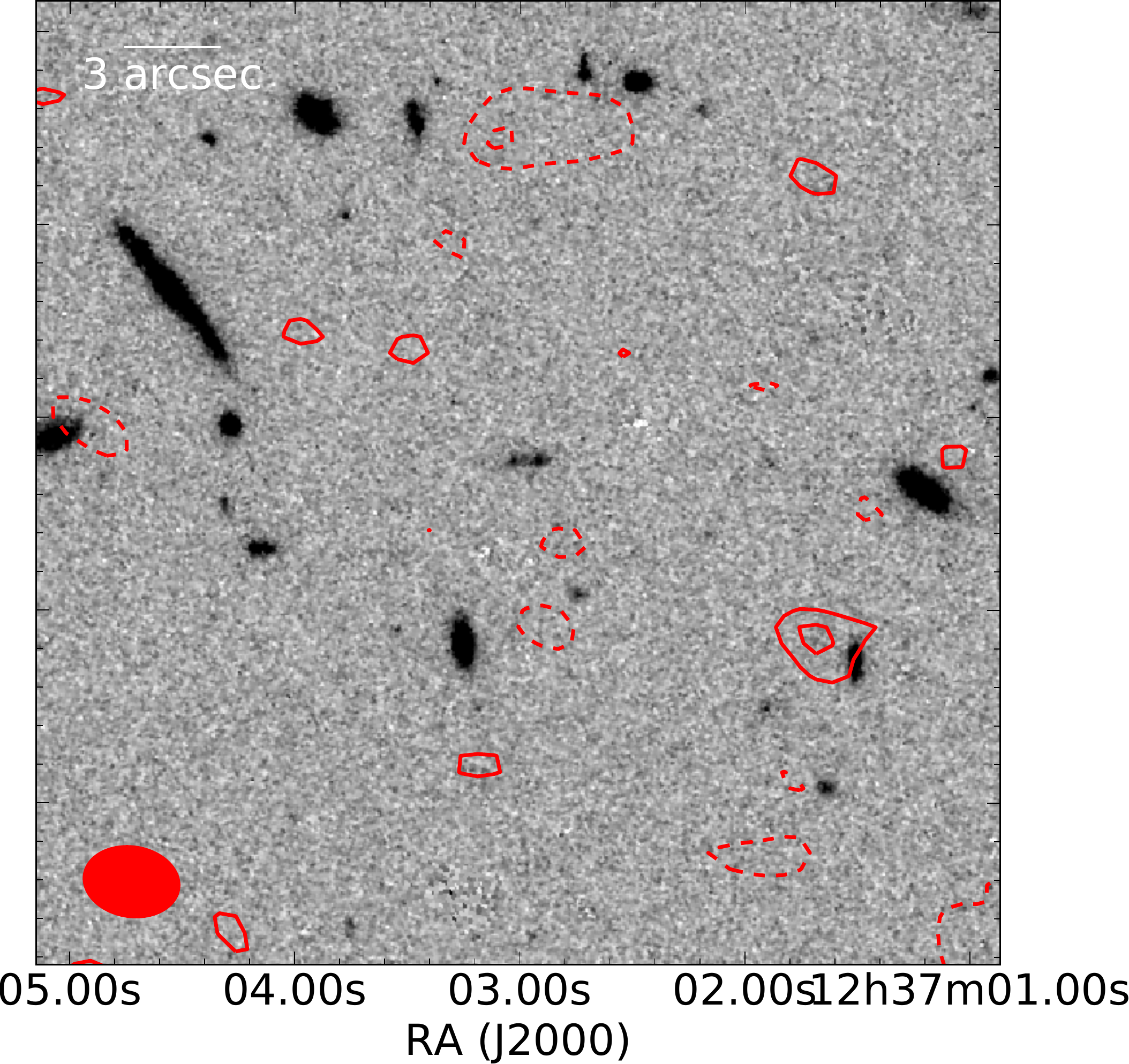}

\caption{Top: Aperture spectra of the most significant negative feature in the GOODS-N data in the original data (yellow histogram; where the feature was selected) and in the newer observations (red; pointing GN57). Bottom: The feature, a putative formaldehyde absorption line against the CMB (contours), appeared to be compatible with the 72.4 GHz ($5_{14}$--$5_{15}$) line of formaldehyde at the photo-$z\sim1.13$ of the galaxies shown in the {\em HST} H-band image.  The left image shows the line map in the original data, the right image shows the same frequency range in the newer data, where the line is not present, which suggests that it was simply due to noise. The contours are shown in steps of $1\sigma$ starting from $\pm2\sigma$ with negative signal as solid contours to show absorption.}
\label{fig:formald_spectra_before}
\end{figure}

\section{Search for Negative features as potential Formaldehyde absorption}
\subsection{Putative feature}
To better understand the characteristics of the noise in our survey data, we have also used our MF3D line searching algorithm to detect negative line features. In order to constrain spurious line features due to noise we take advantage of the symmetry around zero of interferometric noise, in the absence of strong sources in the field.
Although negative line features can usually be assumed to be due to noise, line absorption against the uniform CMB has been suggested to be a potential source of such negative lines. In particular, formaldehyde in dense molecular gas in  galaxies has been confirmed, at low-$z$, to have the potential to produce such absorption against the CMB \citep{Zeiger10,Darling12}.

The most significant negative feature in the initial GOODS-N data cube (pointings GN1--GN56) had a high significance of $\sim6.6\sigma$, and it appeared to be coincident with a pair of local interacting galaxies, GOODS J123702.92+620959.0 with a photo-$z$ of $z$=$1.13\pm0.05$ (Fig.~\ref{fig:formald_spectra_before}). Intriguingly, the strong absorption feature would be consistent with the 72.4 GHz ($5_{14}$--$5_{15}$) line of formaldehyde (${\rm H_2CO}$) at $z\sim$1.13. The energy-level structure of formaldehyde allows collisional population anti-inversion in dense molecular clouds, making the line excitation temperature lower than the CMB temperature and producing absorption against the CMB itself. Formaldehyde silhouettes of galaxies in absorption offer both a novel probe of cosmological size and distance, and a measurement of dense gas masses, density and excitation properties \citep{Darling12}.
Therefore, in order to investigate the possibility that our most significant negative feature may be a real absorption line against the CMB, we obtained additional VLA data, both at the same frequency (as an additional pointing, GN57 as part of our main survey) and at 22.6 GHz, in order to target the $4_{13}$--$4_{14}$ line of formaldehyde.
%Pointing GN57 was observed for 3 hours (127 min on source) in D-array configuration as part of the VLA COLDz program 14A-214 on 18 December 2015. The phase calibrator used was J1302+5748 and the flux was calibrated by observing 3C286. The spectral setup employed uses two dithered sets of spectral windows, with a relative shift of 16 MHz, in order to fully cover the 8 GHz bandwidth available.
We observed this additional tuning with the VLA K-band (project ID: 15B-370; PI: Pavesi) on 6 November 2015. The observations lasted approximately 3 hours (130 min on source) in D-array configuration, with a spectral setup consisting of a single tuning of the two 1-GHz 8-bit samplers (2 GHz total, dual polarization), with central frequencies of 21.58 GHz and 22.5815 GHz for the two intermediate frequencies (IFs), respectively. The same calibrators were observed as for the main survey observations. We calibrated the data using {\sc casa} v.4.5 using the VLA pipeline and minor manual flagging, and we imaged the visibilities using natural weighting. The data cube was produced with 1 MHz channels, corresponding to $\sim$13 km s$^{-1}$, which is small compared to the expected linewidth. The data cube has a beam size of $4.4^{\prime\prime}\times 3.4^{\prime\prime}$, and an {\it rms} noise of $\sim$0.2 mJy beam$^{-1}$ in 1 MHz-wide channels.

\begin{figure}[htb]
 \includegraphics[width=.5\textwidth]{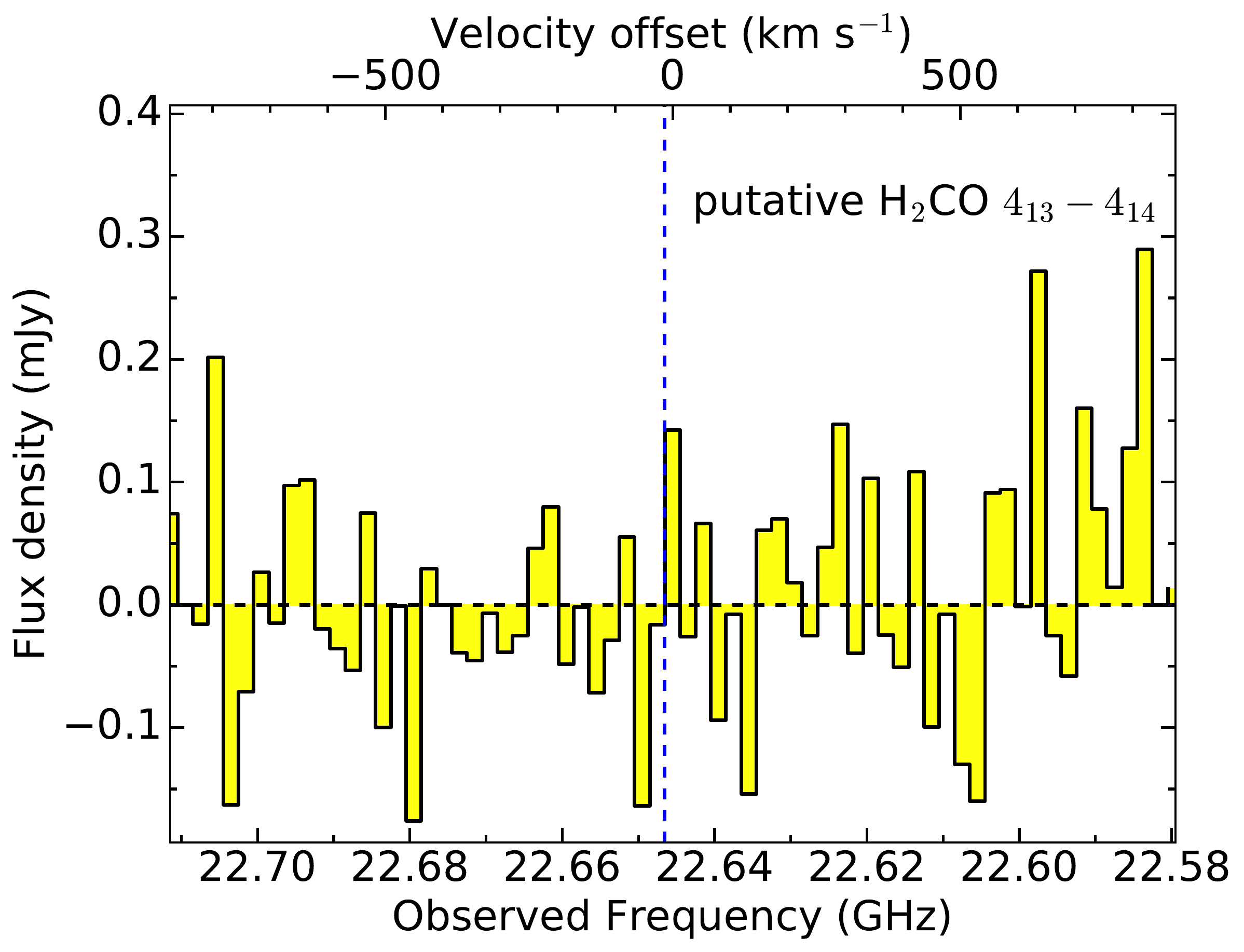}
\caption{VLA K-band spectrum at the expected redshifted frequency of the 48.3 GHz $4_{13}$--$4_{14}$ line of formaldehyde, at $z\sim1.13$.  If the absorption feature we detected in our original data had been a real formaldehyde absorption  line we would expect to detect strong absorption, which is not seen.
}
\label{fig:formald_spectra_after}
\end{figure}

We did not detect the lower frequency line (Fig.~\ref{fig:formald_spectra_after}), and the new observations in pointing GN57 at the same frequency do not show any evidence for absorption at the same position and frequency (Fig.~\ref{fig:formald_spectra_before}). We therefore rule out the presence of an absorption line, and we conclude that the original feature was simply due to noise.
By excluding the possibility that the most significant negative feature may correspond to real absorption we strengthen our confidence in the assumption that all (or at least most) negative line features are due to noise, which is crucial for the purity assessment of our positive line candidates.

\subsection{${\rm H_2CO}$ deep field limits}
Our lack of detections of significant formaldehyde absorption lines allows us to place some of the first constraints on the cosmic abundance of such absorption lines. By assuming a line FWHM of 200 km s$^{-1}$, we derive median $6\sigma$ limits (with no negative line candidates found above this threshold) of 0.18 mJy beam$^{-1}$ and 0.55 mJy beam$^{-1}$ for the COSMOS and GOODS-N fields, respectively, corresponding to $\Delta T_{\rm Obs}$ of $-0.03$ K and $-0.11$ K, at the average frequency and beam size of our survey. We note that the beam size of our observations ($\sim3^{\prime\prime}$) is likely to be larger than the absorbing molecular regions ($\sim0^{\prime\prime}.25-1^{\prime\prime}.25$ at $z\sim1$; \citealt{Darling12}),  implying a dilution of the expected signal strength due to the beam filling factor of $\sim0.025-0.1$.
We use the absence of significant negative detections to infer a probability distribution for their space abundance, by assuming a uniform, uncorrelated distribution of sources over the cosmic volume covered by our survey, and therefore, a Poisson number count. The probability distribution for the space abundance is then an exponential distribution with a mode at zero, and a mean equal to the inverse of the  volume sampled (the 68th percentile upper limit to the space density is listed in Table~\ref{formald_table}).
We use the model results from \cite{Darling12} to derive, from our $\Delta T_{\rm Obs}$ limit,  a constraint on the line optical depth. These models imply that, at $z\sim1$, the maximal expected temperature decrement with respect to the CMB is $\Delta T_{\rm Obs}/(1-\exp^{-\tau})\sim-1.2\,$K, for the $5_{14}$--$5_{15}$ and $4_{13}$--$4_{14}$ lines covered by our survey. This implies limits on the line optical depth of $\tau\lesssim$0.025 and 0.1 for the COSMOS and GOODS-N fields, respectively. Although these values are comparable to the optical depths previously measured for the lower frequency formaldehyde transitions, our results may be weaker by an order of magnitude or more, due to the beam filling factor \citep{Mangum08,Darling12,Mangum13}.

%68 percentile is just 1.14/Vol

\begin{table*}
\caption{Formaldehyde Lines, Redshift Ranges and Volumes Covered by the COLDz Survey}
%\begin{center}
\centering
\begin{tabular}{ c c c c c c c c }
\hline
  Transition & $\nu_0$ & $z_{min}$ & $z_{max}$ & $\langle z \rangle$ & Volume&$\Delta T_{\rm Obs}$ limit&Volume density  \\
   & [GHz] & & & & [$\rm{Mpc}^3$]&[K]&  [$\rm{Mpc}^{-3}$]\\
   \hline
   \multicolumn{6}{c}{COSMOS}\\
 $4_{13}$--$4_{14}$&48.285& 0.24&0.56&0.44&1,850&$-0.03$&  $<6.2\times10^{-4}$ \\
 $5_{14}$--$5_{15}$&72.409 &0.86&1.34 &1.12&9,253&$-0.03$& $<1.2\times10^{-4}$ \\
  \hline
\multicolumn{6}{c}{GOODS-N}\\
  $4_{13}$--$4_{14}$&48.285&0.27&0.61& 0.47&13,690&$-0.11$&  $<8.3\times10^{-5}$ \\
 $5_{14}$--$5_{15}$&72.409 &0.90&1.42&1.18&62,329&$-0.11$&$<1.8\times10^{-5}$  \\
  \hline
\end{tabular}
%  \end{center}
\tablecomments{The quoted $\Delta T_{\rm Obs}$ limits correspond to $6\sigma$ over a line FWHM of 200 km s$^{-1}$. The volume density limit represents the 68\%-quantile of the probability distribution for the space abundance.}
\label{formald_table}
\end{table*}

\section{Description of the individual line candidates}

\begin{table*}
\tiny
%\begin{deluxetable}
\caption{{\rm Catalog of the line candidates identified in our analysis which have not been independently confirmed to date. We advise caution in interpreting these lower significance candidates on a per-source basis until they are independently confirmed. Columns are: (1) Line ID. (2-3) Right ascension and declination (J2000).  (4) Central frequency and uncertainty, based on Gaussian fitting. (5) CO(1--0) redshift and uncertainity, unless otherwise noted. (6) Velocity integrated line flux and uncertainty. (7) Line Full Width at Half Maximum (FWHM), as derived from a Gaussian fit. (8) SNR measured by MF3D. (9) Presence of a spatially coincident optical/NIR counterpart (10) Comments.}} \label{tab_lines_appendix}
\begin{center}
\begin{tabular}{cccccccccl}
\hline
ID  & RA	  & Dec 	& Frequency	 & Redshift & Flux	     & FWHM     & $S/N$ & Opt/NIR & Comments \\  
    & (J2000.0)   & (J2000.0)	& [GHz] 	&  & [Jy\,km\,s$^{-1}$] & [km\,s$^{-1}$] &		   & c.part? &          \\  
 (1) & (2)	  & (3) 	& (4)		  & (5) 	     & (6)	& (7)		   &  (8)    & (9)  &(10)    \\  
\hline
\multicolumn{9}{l}{\bf COSMOS} \\
COLDz.COS.4  & 10:00:22.34& +02:34:14.0 & $ 34.887\pm0.007$ & $2.3041\pm0.0007$&$0.12\pm0.04$ & $ 600\pm150$  &     5.71  & possible &  \\
COLDz.COS.5  & 10:00:17.63 & +02:34:36.0 & $34.814 \pm0.005$ &$2.3110\pm0.0005$& $0.08\pm0.03$ & $ 360\pm100$  &     5.62  & N & \\
COLDz.COS.6  & 10:00:23.27& +02:34:22.0 & $ 31.989\pm0.003$ &$2.6034\pm0.0003$& $0.037\pm0.013$ & $250\pm70$  &     5.59  & possible & \\
COLDz.COS.7  & 10:00:21.60& +02:33:56.0 & $35.823\pm0.002$ & $2.2178\pm0.0002 $ &$0.12\pm0.04$ & $140\pm40$  &     5.56  & possible? &  extended\\
COLDz.COS.8  & 10:00:25.07& +02:35:56.0 & $ 35.291\pm0.004$&$2.2663\pm0.0004$ & $0.24\pm0.09$ & $ 250\pm70$  &     5.56  & N &  extended\\
COLDz.COS.9  & 10:00:22.44& +02:36:16.0& $36.593\pm0.003$ &$2.1501\pm0.0003$ & $0.13\pm0.04$ & $ 220\pm50$  &     5.57  & N & extended \\
COLDz.COS.10 & 10:00:23.44& +02:36:29.0 & $34.132\pm0.001$ &$2.3772\pm0.0001$ & $0.11\pm0.03$ & $92\pm18$  &     5.56  & possible & extended\\
COLDz.COS.11 & 10:00:20.43 & +02:34:56.00 & $ 37.81\pm0.002$ & $ 2.0487\pm0.0002$ & $ 0.05\pm0.02$ & $ 120\pm40$ & 5.49 & N & blended $z$=0.3, slightly extended\\
COLDz.COS.12 & 10:00:17.53 & +02:35:11.00 & $ 35.354\pm0.002$ & $ 2.2605\pm0.0002$ & $ 0.025\pm0.009$ & $ 160\pm40$ & 5.43 & N &  \\
COLDz.COS.13 & 10:00:14.26 & +02:35:02.50 & $ 32.423\pm0.004$ & $ 2.5552\pm0.0004$ & $ 0.09\pm0.03$ & $ 300\pm90$ & 5.43 & N &  \\
COLDz.COS.14 & 10:00:21.73 & +02:35:57.00 & $ 35.005\pm0.002$ & $ 2.293\pm0.0001$ & $ 0.018\pm0.008$ & $ 90\pm30$ & 5.42 &  N & blended $z$=0.9 \\
COLDz.COS.15 & 10:00:20.20 & +02:35:31.50 & $ 34.681\pm0.006$ & $ 2.3237\pm0.0005$ & $ 0.033\pm0.015$ & $ 340\pm110$ & 5.41 & possible  & matched photo-$z$ \\
COLDz.COS.16 & 10:00:25.50 & +02:35:35.00 & $ 36.934\pm0.004$ & $ 2.121\pm0.0004$ & $ 0.07\pm0.03$ & $ 260\pm80$ & 5.34 & possible & matched photo-$z$  \\
COLDz.COS.17 & 10:00:19.70 & +02:35:01.50 & $ 33.917\pm0.007$ & $ 2.3986\pm0.0007$ & $ 0.06\pm0.02$ & $ 570\pm150$ & 5.33 & N &  \\
COLDz.COS.18 & 10:00:24.60 & +02:34:38.00 & $ 31.296\pm0.002$ & $ 2.6832\pm0.0002$ & $ 0.037\pm0.012$ & $ 180\pm40$ & 5.32 & possible  & matches photo-$z$, slightly extended \\
COLDz.COS.19 & 10:00:21.97 & +02:34:54.50 & $ 37.083\pm0.002$ & $ 2.1085\pm0.0002$ & $ 0.046\pm0.017$ & $ 130\pm30$ & 5.29 & N & M star nearby \\
COLDz.COS.20 & 10:00:17.03 & +02:34:59.50 & $ 31.437\pm0.002$ & $ 2.6667\pm0.0003$ & $ 0.021\pm0.008$ & $ 170\pm50$ & 5.28 & N &  \\
COLDz.COS.21 & 10:00:23.47 & +02:34:58.50 & $ 36.349\pm0.004$ & $ 2.1712\pm0.0004$ & $ 0.02\pm0.02$ & $ 110\pm80$ & 5.28 & N & slightly extended \\
COLDz.COS.22 & 10:00:22.90 & +02:34:10.00 & $ 35.301\pm0.005$ & $ 2.2653\pm0.0004$ & $ 0.12\pm0.04$ & $ 350\pm90$ & 5.27 & possible & very uncertain photo-$z$, extended \\
COLDz.COS.23 & 10:00:20.57 & +02:34:01.00 & $ 38.504\pm0.001$ & $ 1.9937\pm0.0001$ & $ 0.14\pm0.04$ & $ 60\pm10$ & 5.26 & possible & matches photo-$z$, slightly extended \\
COLDz.COS.24 & 10:00:14.99 & +02:35:41.00 & $ 35.362\pm0.002$ & $ 2.2597\pm0.0002$ & $ 0.07\pm0.02$ & $ 130\pm30$ & 5.25 & possible & close separation and photo-$z$\\
COLDz.COS.25 & 10:00:21.07 & +02:34:30.50 & $ 33.579\pm0.005$ & $ 2.4328\pm0.0006$ & $ 0.049\pm0.018$ & $ 410\pm110$ & 5.25 & possible & close separation and photo-$z$, extended\\
 \hline
 \hline
 \multicolumn{9}{l}{\bf GOODS-N} \\
COLDz.GN.1 & 12:36:59.79 & +62:11:09.50 & $ 37.485\pm0.003$ & $ 2.0751\pm0.0002$ & $ 0.405\pm0.137$ & $ 200\pm50$ & 6.38 & possible & slightly extended \\
COLDz.GN.2 & 12:36:27.94 & +62:14:09.78 & $ 32.518\pm0.002$ & $ 2.5448\pm0.0002$ & $ 0.109\pm0.033$ & $ 210\pm50$ & 6.14 & N & \\
COLDz.GN.4 & 12:36:54.77 & +62:17:28.00 & $ 35.937\pm0.002$ & $ 2.2076\pm0.0002$ & $ 0.382\pm0.12$ & $ 180\pm40$ & 6.08 & N & extended\\ %809-1116 8 need to check \\
COLDz.GN.5 & 12:37:00.00 & +62:15:21.00 & $ 37.229\pm0.005$ & $ 2.0962\pm0.0005$ & $ 0.283\pm0.073$ & $ 520\pm100$ & 6.06 & N & photo-$z$ at lower $z$  \\
COLDz.GN.6 & 12:37:01.50 & +62:12:35.50 & $ 32.878\pm0.003$ & $ 2.5059\pm0.0003$ & $ 0.116\pm0.038$ & $ 260\pm60$ & 6.0 & possible &  \\
COLDz.GN.7 & 12:36:28.89 & +62:13:00.80 & $ 30.615\pm0.002$ & $ 2.7652\pm0.0003$ & $ 0.189\pm0.047$ & $ 290\pm50$ & 5.97 & N &  \\
COLDz.GN.8 & 12:36:37.31 & +62:15:03.39 & $ 37.737\pm0.004$ & $ 2.0546\pm0.0003$ & $ 0.823\pm0.272$ & $ 270\pm70$ & 5.89 & possible & close photo-$z$, very extended \\
COLDz.GN.9 & 12:36:56.35 & +62:18:19.50 & $ 36.348\pm0.001$ & $ 2.1713\pm0.0001$ & $ 0.077\pm0.022$ & $ 100\pm20$ & 5.88 & possible &  \\
COLDz.GN.10 & 12:36:33.87 & +62:15:29.36 & $ 32.841\pm0.002$ & $ 2.51\pm0.0002$ & $ 0.143\pm0.05$ & $ 120\pm30$ & 5.83 & possible & slightly extended \\ %1101-879 2 need to check \\
COLDz.GN.11 & 12:36:53.20 & +62:14:34.49 & $ 35.073\pm0.001$ & $ 2.2866\pm0.0001$ & $ 0.323\pm0.101$ & $ 90\pm20$ & 5.81 & N & extended \\
COLDz.GN.12 & 12:37:03.50 & +62:12:52.00 & $ 37.93\pm0.007$ & $ 2.039\pm0.0006$ & $ 0.317\pm0.114$ & $ 490\pm130$ & 5.81 & possible & photo-$z$=4 \\
COLDz.GN.13 & 12:36:43.18 & +62:14:22.44 & $ 36.743\pm0.005$ & $ 2.1372\pm0.0005$ & $ 0.199\pm0.071$ & $ 390\pm100$ & 5.8 & N & \\
COLDz.GN.14 & 12:36:59.07 & +62:14:48.00 & $ 34.133\pm0.006$ & $ 2.3771\pm0.0006$ & $ 0.605\pm0.192$ & $ 490\pm120$ & 5.65 & possible & very close photo-$z$, extended \\
COLDz.GN.15 & 12:36:41.67 & +62:15:47.93 & $ 33.34\pm0.009$ & $ 2.4574\pm0.0009$ & $ 0.178\pm0.069$ & $ 640\pm190$ & 5.64 & possible & blended photo-$z\sim$3 \\
COLDz.GN.16 & 12:36:49.42 & +62:12:17.98 & $ 32.807\pm0.002$ & $ 2.5136\pm0.0002$ & $ 0.062\pm0.022$ & $ 140\pm40$ & 5.63 & possible & local foreground? \\%884-496 -1 need to check\\
COLDz.GN.17 & 12:37:01.22 & +62:13:04.50 & $ 33.2\pm0.002$ & $ 2.472\pm0.0002$ & $ 0.084\pm0.036$ & $ 140\pm50$ & 5.63 & N & \\
COLDz.GN.18 & 12:36:51.12 & +62:15:54.99 & $ 36.938\pm0.003$ & $ 2.1206\pm0.0003$ & $ 0.152\pm0.043$ & $ 270\pm60$ & 5.62 & possible& matched  photo-$z$ \\
COLDz.GN.19 & 12:37:08.45 & +62:14:23.48 & $ 36.017\pm0.002$ & $ 2.2005\pm0.0002$ & $ 0.207\pm0.09$ & $ 120\pm40$ & 5.61 & N & local foreground, extended \\ %618-747 4 need to check \\
COLDz.GN.20 & 12:37:04.88 & +62:17:44.49 & $ 30.727\pm0.003$ & $ 2.7514\pm0.0003$ & $ 0.205\pm0.062$ & $ 280\pm60$ & 5.6 & possible & close photo-$z$   \\
COLDz.GN.21 & 12:36:54.93 & +62:11:29.50 & $ 35.265\pm0.004$ & $ 2.2687\pm0.0004$ & $ 0.194\pm0.057$ & $ 390\pm90$ & 5.6 & N &  \\
COLDz.GN.22 & 12:37:00.14 & +62:11:58.50 & $ 33.441\pm0.003$ & $ 2.447\pm0.0003$ & $ 0.075\pm0.037$ & $ 140\pm50$ & 5.59 & N & \\
COLDz.GN.23 & 12:36:56.71 & +62:13:19.50 & $ 37.925\pm0.002$ & $ 2.0394\pm0.0002$ & $ 0.156\pm0.051$ & $ 160\pm40$ & 5.59 & N & \\ %782-619 -1 need to check \\
COLDz.GN.24 & 12:36:46.85 & +62:12:18.97 & $ 31.165\pm0.008$ & $ 2.6988\pm0.0009$ & $ 0.222\pm0.095$ & $ 560\pm180$ & 5.56 & N & slightly extended \\ %920-498 0 need to check\\
COLDz.GN.25 & 12:37:03.35 & +62:08:59.00 & $ 35.598\pm0.002$ & $ 2.2381\pm0.0002$ & $ 0.119\pm0.043$ & $ 130\pm30$ & 5.55 & possible & close photo-$z$ \\%689-98 -1 need to check\\
COLDz.GN.26 & 12:37:00.29 & +62:16:31.50 & $ 36.733\pm0.006$ & $ 2.1381\pm0.0005$ & $ 0.169\pm0.056$ & $ 490\pm120$ & 5.54 & possible  & very faint counterpart\\
COLDz.GN.27 & 12:36:45.09 & +62:18:00.46 & $ 37.373\pm0.001$ & $ 2.0843\pm0.0001$ & $ 0.096\pm0.025$ & $ 90\pm20$ & 5.54 & possible & close photo-$z$ \\ %944-1181 -1 need to check\\
COLDz.GN.28 & 12:36:47.77 & +62:12:57.47 & $ 32.289\pm0.003$ & $ 2.57\pm0.0003$ & $ 0.113\pm0.04$ & $ 210\pm60$ & 5.51 &  N & different spec-z, $z$=2.932, slightly extended  \\
COLDz.GN.29 & 12:37:06.38 & +62:16:34.49 & $ 32.861\pm0.002$ & $ 2.5078\pm0.0002$ & $ 0.089\pm0.027$ & $ 190\pm40$ & 5.51 & possible &  \\
COLDz.GN.30 & 12:36:50.43 & +62:10:29.48 & $ 31.297\pm0.002$ & $ 2.6832\pm0.0002$ & $ 0.236\pm0.088$ & $ 150\pm40$ & 5.51 & N & slightly extended \\ %870-279 0 need to check\\
\hline
\end{tabular}
%\tablecomments{$^a$ CO(2--1)  redshift.}
\end{center}
%\end{deluxetable}
\end{table*}

In this section, we briefly describe the remaining CO line candidates and potential counterpart associations. These candidates are currently not independently confirmed, and thus, are only used in our statistical analysis. Because we only expect a small fraction of these candidates to correspond to real CO line emission, we advise caution in interpreting these lower significance candidates on a per-source basis until they are independently confirmed. Quoted photometric redshift ranges are the $1\sigma$ uncertainties reported in the COSMOS2015 \citep{Laigle16} and the CANDELS catalogs \citep{Brammer,Skelton14,Momcheva16}.  The positional search radius considered is $3^{\prime\prime}$, which is dictated by the positional uncertainties (which are larger for extended sources) and the possibility of real physical offsets in the stellar emission, e.g., due to differential dust obscuration.
The visual counterpart inspection was carried out utilizing {\em HST}  (H-band in GOODS-N and I-band in COSMOS, where H-band was not available) and IRAC 3.6 $\mu$m images (band 1), which are shown in Figs. \ref{fig:spectra_COS2} and \ref{fig:spectra_GN1}. We have also inspected images from the other IRAC bands and find no evidence for additional counterpart matches relative to IRAC band 1.

\subsection{COSMOS}

\textbf{COLDz.COS.4}: This is the highest signal-to-noise candidate in COSMOS without a secure counterpart. There is a potential match at $0.5^{\prime\prime}\pm0.6^{\prime\prime}$ to the NW in the COSMOS2015 catalog, with an uncertain photo-$z$=1.5--2.3, matching the CO(1--0) redshift of $z$=2.30. Two I-band and 3.6$\mu$m sources are aligned with the elongated CO candidate emission (Fig.~\ref{fig:spectra_COS2}).

\textbf{COLDz.COS.5}: No counterpart is found in the COSMOS2015 catalog or the images at the position of this candidate.

\textbf{COLDz.COS.6}: The images show an IRAC 3.6$\mu$m source $3^{\prime\prime}\pm0.2^{\prime\prime}$ to the SE of the candidate which has a photo-$z$=2.9--3 and might therefore be associated with our candidate although the offset appears significant (the CO(1--0) redshift is $z$=2.60). At the CO position there is an I-band source with photo-$z$=0.44--0.48, although its 3.6$\mu$m image is contaminated by the brighter, higher-z galaxy. It is unclear if the CO candidate may be related.

\textbf{COLDz.COS.7}: The CO(1--0) redshift of this candidate is $z$=2.22.  It is at the position of an {\em HST} I-band source, which is not in the COSMOS2015 catalog, and is not visible in the 3.6$\mu$m image. 

\textbf{COLDz.COS.8}: Candidate is at the position of a faint {\em HST} I-band and IRAC 3.6$\mu$m source, which is listed in the COSMOS2015 catalog $2.3^{\prime\prime}\pm0.4^{\prime\prime}$ to the N ($z_{phot}$=0.2--0.7).

\textbf{COLDz.COS.9}: Candidate shows spatially extended CO emission, centered on an {\em HST} I-band source with a photo-$z$=0.1--0.8, which is therefore unlikely to be associated with the candidate.

\textbf{COLDz.COS.10}: Candidate is spatially extended and is co-spatial with multiple faint {\em HST} I-band galaxies. The COSMOS2015 catalog only reports a faint galaxy $1.3^{\prime\prime}\pm0.7^{\prime\prime}$ to the SW, with a very uncertain photo-$z$ of 0.8--4.3, which may be associated with our candidate.

\textbf{COLDz.COS.11}: Candidate is near the position of a low-$z$ galaxy ($z_{phot}$=0.32--0.35). There is a brighter IRAC 3.6$\mu$m source $1.6^{\prime\prime}\pm0.4^{\prime\prime}$ to the NW,  ($z_{phot}$=1.5--1.6), which may be related to the CO candidate with a CO redshift of 2.0.

%\textbf{COLDz.COS.12}: This is a very spatially extended feature, unlikely to be real. It appears within $3^{\prime\prime}$ of a photo-$z$=5.4 galaxy which is to the W, which could be associated. Assuming CO(2--1)  would place this candidate at z=5.34.

\textbf{COLDz.COS.12}: No counterpart is found in the COSMOS2015 catalog or the images at the position of this candidate.

\textbf{COLDz.COS.13}: No counterpart is found in the COSMOS2015 catalog or the images at the position of this candidate.

\textbf{COLDz.COS.14}: Candidate is affected by foreground contamination which prevents any counterpart assessment. In particular, there is a bright photo-$z$=0.9 galaxy at $1.6^{\prime\prime}\pm0.3^{\prime\prime}$ to the NW.

\textbf{COLDz.COS.15}: Candidate has a potential counterpart match. The COSMOS2015 catalog lists a galaxy $1.8^{\prime\prime}\pm0.4^{\prime\prime}$ to the NW with a  photo-$z$=1.8--2.8, which is very faint in the I band and IRAC 3.6$\mu$m images. Assuming CO(1--0) would place this candidate at $z$=2.32.

\textbf{COLDz.COS.16}: Candidate has a potential counterpart match, but it appears confused with a bright galaxy $1.6^{\prime\prime}$ to the SE, with a photo-$z$=1.0--1.2. The potential counterpart has photo-$z$=1.3--2.6 and is located about $2^{\prime\prime}\pm0.5^{\prime\prime}$ to the NE.

\textbf{COLDz.COS.17}: No counterpart is found in the COSMOS2015 catalog or the images.

\textbf{COLDz.COS.18}: Candidate has a potential counterpart match, $1.6^{\prime\prime}\pm0.4^{\prime\prime}$ to the N, with a photo-$z$=2.4--2.5; assuming CO(1--0) would place it at $z$=2.68. This candidate is contaminated by a local bright galaxy to the NE.

\textbf{COLDz.COS.19}:  Candidate does not appear to have a counterpart. An M star is located $0.8^{\prime\prime}\pm0.7^{\prime\prime}$  to the NE, and partly prevents counterpart identification.

\textbf{COLDz.COS.20}: Candidate does not have a counterpart. The COSMOS2015 catalog lists two galaxies at separations of $2.2^{\prime\prime}\pm0.3^{\prime\prime}$ and $2.3^{\prime\prime}\pm0.3^{\prime\prime}$, respectively. The first galaxy has a photo-$z$=0.6--0.9 and the second one is at photo-$z=$1.9--2.5. This latter galaxy may be associated with our candidate, which has a CO(1--0) redshift of $z_{10}$=2.67.

\textbf{COLDz.COS.21}: No counterpart is found in the COSMOS2015 catalog or the images, at the position of this  spatially extended candidate, but the IRAC 3.6$\mu$m images are contaminated by bright nearby stars and galaxies.

\textbf{COLDz.COS.22}: A faint galaxy is visible in the {\em HST} I-band image, $1.5^{\prime\prime}\pm0.7^{\prime\prime}$ to the NW, which may be associated with our line candidate. The catalog lists a very uncertain photo-$z$=1.5--5.5, which is compatible with the CO(1--0) redshift of $z_{10}$=2.27.

\textbf{COLDz.COS.23}: Candidate has a potential counterpart association. This is a galaxy $1.5^{\prime\prime}\pm0.5^{\prime\prime}$ to the SW, which is compatible with the position of at least part of the slightly spatially extended line emission. The photometric redshift for this galaxy is photo-$z$=1.7--2.8, which is compatible with the CO(1--0) redshift of $z_{10}$=1.99.

\textbf{COLDz.COS.24}: Candidate has a potential counterpart association. The potential counterpart is only $0.8^{\prime\prime}\pm0.9^{\prime\prime}$ to the N and has a photometric redshift of $z_{\rm phot}$=1.9--2.0 which is close to the CO(1--0) redshift of $z_{10}$=2.26. A second galaxy is seen, $1.6^{\prime\prime}\pm0.9$ to the E, which has a photometric redshift of photo-$z$=0.89--0.92 and which contaminates the emission in the IRAC 3.6$\mu$m images.

\textbf{COLDz.COS.25}: This spatially extended candidate has a potential counterpart. This potential counterpart is $1.4^{\prime\prime}\pm0.7^{\prime\prime}$ to the SE and has a photometric redshift of photo-$z$=2.6--3.0 which is close to the CO(1--0) redshift of $z_{10}$=2.43. The IRAC 3.6$\mu$m images are contaminated by a nearby star, which makes it difficult to identify faint sources reliably.

\subsection{GOODS-N}
\textbf{COLDz.GN.1}: This spatially extended line candidate has potential counterpart matches. There are multiple galaxies which are compatible with the line emission position, blended in the IRAC 3.6$\mu$m image but visible in {\em HST} H-band, with photometric redshifts in the CANDELS catalog \citep{Skelton14}. The closest catalog match has a separation of only $0.4^{\prime\prime}\pm0.6^{\prime\prime}$ to the NE and has an uncertain photo-$z$=0.8--2.2. The catalog lists three more galaxies within $3^{\prime\prime}$ (separations of $1.3^{\prime\prime}\pm0.6^{\prime\prime}$, $2^{\prime\prime}\pm0.6^{\prime\prime}$ and $2.6^{\prime\prime}\pm0.6^{\prime\prime}$), with photo-$z$s of 1.5--1.7, 0.9--2.5 and 1.1--1.8 respectively.  The CO(1--0) redshift  of our candidate is $z_{10}$=2.08, which makes it compatible with at least two of these potential counterparts.

\textbf{COLDz.GN.2}: This is the highest SNR candidate in GOODS-N without a clear counterpart. The CANDELS catalog lists a faint source $2.6^{\prime\prime}\pm0.3^{\prime\prime}$ to the SE, with uncertain photo-$z$=1.0--2.0  \cite{Skelton14}. The CO(1--0) redshift of our candidate ($z$=2.54) makes it a possible, although unlikely counterpart.

\textbf{COLDz.GN.4}: Candidate is unlikely to have a counterpart. There are no galaxies in the CANDELS catalog within $3^{\prime\prime}$, and no galaxies are visible in the {\em HST} H-band or IRAC 3.6$\mu$m images. 

\textbf{COLDz.GN.5}: There are two galaxies in the images within $2^{\prime\prime}$ of the line candidate, with separations of 1.5$^{\prime\prime}\pm0.3^{\prime\prime}$ and 1.7$^{\prime\prime}\pm0.3^{\prime\prime}$, respectively. They are unlikely to be counterparts because they have photometric redshifts of  $z_{phot}=$1.1--1.3 and 0.4--0.5 respectively, while the CO(1--0) redshift of our candidate is $z_{10}$=2.1.

\textbf{COLDz.GN.6}: Candidate has a potential match $2.8^{\prime\prime}\pm0.4^{\prime\prime}$ to the SE, in the direction where the CO emission is slightly spatially extended. The catalog lists a photo-$z$ of 2.4--2.5 which is compatible with the  CO(1--0) redshift of $z_{10}$=2.51, suggesting a possible counterpart match.

\textbf{COLDz.GN.7}: Candidate has an unlikely, but possible match $2.9^{\prime\prime}\pm0.3^{\prime\prime}$ to the SE, which appears to be at a significant offset. The galaxy has a photo-$z$=1.8--2.0, which is not compatible with the CO(1--0) redshift of $z_{10}$=2.76,  therefore we do not consider this to be a match.

\textbf{COLDz.GN.8}: This spatially extended CO candidate has a possible match $3.0^{\prime\prime}\pm0.8^{\prime\prime}$ to the SE, with an uncertain photo-$z$ of 1.4--2.4 which is compatible with the  CO(1--0) redshift of $z_{10}$=2.05. This is a potential match, because the line emission appears to be very spatially extended, and may be compatible with coming from a dust-obscured part of the optical galaxy.

\textbf{COLDz.GN.9}: Candidate is unlikely to have a counterpart. It appears near a spec-$z$=0.516 galaxy, which is $2.8^{\prime\prime}\pm0.3^{\prime\prime}$ to the NW. The catalog also lists a faint, photo-$z$=1.9--2.1 galaxy $2.4^{\prime\prime}\pm0.3^{\prime\prime}$ to the SW (which appears to be significantly offset from the CO line emission), which could be consistent with the CO(1--0) redshift of $z_{10}$=2.17.

\textbf{COLDz.GN.10}: This slightly spatially extended candidate is unlikely to have a counterpart. The closest catalog association is $1.5^{\prime\prime}\pm0.5^{\prime\prime}$ to the NE and has a photo-$z$ of 4.4--5. The CO(2--1)  redshift for our candidate would be $z_{21}$=6.0, and is therefore an unlikely match. The CANDELS catalog lists two more galaxies, just below $3^{\prime\prime}$ to the NE with photo-$z$s of 1.7--2.0 and 1.9--2.4, respectively, which may be compatible with the CO(1--0) redshift of $z_{10}$=2.51.

\textbf{COLDz.GN.11}: Candidate appears spatially extended and elongated. No objects are seen in the {\em HST} H-band and IRAC 3.6 $\mu$m images. The CANDELS catalog lists a galaxy 1.8$^{\prime\prime}\pm0.7^{\prime\prime}$ to the NW which has a photo-$z$ of 0.6--1.6. This is inconsistent with the CO(1--0) redshift of $z_{10}$=2.29, so a match is unlikely.

\textbf{COLDz.GN.12}: Candidate has no likely match. The catalog lists a faint galaxy, $1.7^{\prime\prime}\pm0.4^{\prime\prime}$ to the NE, with photo-$z$=4.1--4.4, which may potentially be associated.  The counterpart status is difficult to evaluate due to blending with  the bright local (spec-$z$=0.784) galaxy at a separation of just $2.4^{\prime\prime}\pm0.4^{\prime\prime}$.

\textbf{COLDz.GN.13}: Candidate is spatially extended and elongated, and is unlikely to have a counterpart association. The CANDELS catalog lists two potential matches within $3^{\prime\prime}$, with separation 1.7$^{\prime\prime}\pm0.5^{\prime\prime}$ and 3$^{\prime\prime}\pm0.5^{\prime\prime}$, respectively. The photometric redshifts listed by the catalog are $z_{phot}=$0.2--0.4 and 0.6--1.4, respectively, which make them unlikely counterparts given the CO(1--0) redshift of our candidate ($z_{10}$=2.14).

\textbf{COLDz.GN.14}: This is a spatially extended CO candidate, and it has a possible counterpart, which is faint but visible in the IRAC 3.6 $\mu$m image. We identify this counterpart with the catalog listing of a photo-$z$=2.4--2.7 galaxy which is displaced by $2.8^{\prime\prime}\pm0.7^{\prime\prime}$ to the SW. This counterpart is compatible with the CO(1--0) redshift of $z_{10}$=2.38. The offset may not be significant, because the IR-detected galaxy appears to be compatible with the position of this spatially extended candidate.

\textbf{COLDz.GN.15}: Candidate may have a counterpart. The 3.6 $\mu$m image is partly blended with a  spec-z=0.453 galaxy $2.3^{\prime\prime}\pm0.6^{\prime\prime}$ to the W which makes the identification difficult. The catalog lists two possible counterparts, with photo-$z$=1.6--1.8 and 3.1--3.9, offset respectively $0.9^{\prime\prime}\pm0.6^{\prime\prime}$ and $1.7^{\prime\prime}\pm0.6^{\prime\prime}$ to the NW and NE, which are not compatible with the CO(1--0) redshift of $z_{10}$=2.46.

\textbf{COLDz.GN.16}: Candidate may have a counterpart. The image is partly blended with a  spec-$z$=0.961 galaxy $2.3^{\prime\prime}\pm0.6^{\prime\prime}$ to the NE, which makes identification difficult. The CANDELS catalog lists three more galaxies within $2^{\prime\prime}$ and $3^{\prime\prime}$ from our candidate, with photo-$z$=0.9--2.0, 2.2--2.4 and 1.4--2.3, all of which  may be compatible with the CO(1--0) redshift of $z_{10}$=2.51.

\textbf{COLDz.GN.17}: Candidate appears spatially extended and elongated. The catalog lists a galaxy $1.5^{\prime\prime}\pm0.5^{\prime\prime}$ to the SE, which is visible in the IRAC 3.6 $\mu$m images. This galaxy has a photo-$z$ of 1.2--1.9, which is only somewhat inconsistent with the CO(1--0) redshift of $z_{10}$=2.47. Therefore a counterpart association cannot be ruled out.

\textbf{COLDz.GN.18}:  Candidate appears to be closely associated with other, lower significance, candidates which are visible in the line maps.
It has a potential match, a faint galaxy with photo-$z$=1.3--2.2 only $0.8^{\prime\prime}\pm0.5^{\prime\prime}$ to the SE, which is compatible with the CO(1--0) redshift of $z_{10}$=2.12. The catalog also lists three galaxies $0.4^{\prime\prime}\pm0.5^{\prime\prime}$, $1.8^{\prime\prime}\pm0.5^{\prime\prime}$ and $2.7^{\prime\prime}\pm0.5^{\prime\prime}$ to the SE, with  photo-$z$s of 0.3--0.9, 2.5--2.8 and 4.3--5.2, respectively, which may be associated in case of incorrect photometric redshifts.

\textbf{COLDz.GN.19}: This spatially extended candidate is blended with a local foreground galaxy at $z$=0.564. No continuum emission is detected in our data at this position and therefore we exclude the possibility that the line candidate may be spurious  and due to noise superposed to continuum emission. The presence of the bright foreground contaminates the {\em HST} H-band and the IRAC 3.6 $\mu$m images making it difficult to evaluate the counterpart status. Lensing of a faint $z$=2.20 galaxy by the foreground galaxy is a possibility.

\textbf{COLDz.GN.20}: Candidate may have a counterpart. It appears close to a foreground galaxy, which partly contaminates the IRAC 3.6 $\mu$m image. The {\em HST} H-band image shows a potential match which the catalog identifies as a galaxy $1.16^{\prime\prime}\pm0.3$ to the NW with a  photo-$z$ of 1.6--2.5. The association is not ruled out by our CO(1--0) redshift  of $z_{10}$=2.75.

\textbf{COLDz.GN.21}:  Candidate is unlikely to have a counterpart. The images show a galaxy $1.0^{\prime\prime}\pm0.6^{\prime\prime}$ to the NE, which has a grism-$z$ of 0.86--0.94 from \cite{Momcheva16}. This makes it incompatible with the CO(1--0) redshift of $z_{10}$=2.27.

\textbf{COLDz.GN.22}: No counterpart is found in the images or the CANDELS catalog at the position of this candidate.

\textbf{COLDz.GN.23}: No counterpart is found in the images or the CANDELS catalog at the position of this candidate.

\textbf{COLDz.GN.24}: This spatially extended candidate may have a counterpart, which is visible in the IRAC 3.6 $\mu$m image but not in the {\em HST} H-band image. The separation is $2.9^{\prime\prime}\pm0.5^{\prime\prime}$ to the S, but this may not be significant due to the extent of the emission. The photo-$z$ is uncertain and ranges from 0.9 to 2.3, therefore an association is not strongly ruled out by our CO(1--0) redshift of $z_{10}$=2.7.

\textbf{COLDz.GN.25}: Candidate has a potential counterpart, $1.7^{\prime\prime}\pm0.7^{\prime\prime}$ to the NE. The galaxy is well visible in H-band and IRAC 3.6 $\mu$m images and has a photo-$z$ of 1.9--2.2, which is compatible with the CO(1--0) redshift of $z_{10}$=2.24.

\textbf{COLDz.GN.26}: This spatially extended CO candidate has potential matches, which appear very faint in the IRAC 3.6 $\mu$m image. The catalog lists two galaxies at $1.8^{\prime\prime}\pm0.5^{\prime\prime}$ and $1.9^{\prime\prime}\pm0.5^{\prime\prime}$ to the NE, with uncertain photo-$z$s of  0.6--4.2 and 2.1--3.7 respectively, which makes them compatible with the CO(1--0) redshift of $z_{10}$=2.14. The spatial offset may not be significant because of the spatial extent of the emission, and these are therefore potential counterpart matches.

\textbf{COLDz.GN.27}:  Candidate has three potential counterparts in the catalog with close photo-$z$s. The first is  $2.3^{\prime\prime}\pm0.4^{\prime\prime}$ to the SE, with a photo-$z$ of 1.5--2.3. The second is $2.7^{\prime\prime}\pm0.4^{\prime\prime}$ to the SE, with a photo-$z$ of 1.5--2. Both of these are  compatible with the CO(1--0) redshift of $z_{10}$=2.08. The third potential counterpart is located $2.9^{\prime\prime}\pm0.4^{\prime\prime}$ to the NE and has a photo-$z$ in the range 5.2--5.7 which is consistent with the CO(2--1)  redshift of $z_{21}$=5.16.

\textbf{COLDz.GN.28}: The {\em HST} H-band and IRAC 3.6 $\mu$m images show a potential match $1.6^{\prime\prime}\pm0.3^{\prime\prime}$ to the S, but this galaxy was reported to have a spec-$z$=2.932  \citep{Skelton14}. Assuming CO(1--0) would imply $z_{10}$=2.57, which implies either a lack of counterpart or an incorrect spectroscopic redshift.

\textbf{COLDz.GN.29}: Candidate has a possible counterpart. Both the {\em HST} H-band and the IRAC 3.6 $\mu$m images show multiple sources within 2$^{\prime\prime}$. The catalog lists two faint galaxies, with photo-$z$=0.8--3.5 and 1.1--1.7 just $1.4^{\prime\prime}\pm0.4^{\prime\prime}$ and $1.5^{\prime\prime}\pm0.4^{\prime\prime}$ to the SE and NE respectively. The first of these is compatible with the CO(1--0) redshift of $z_{10}$=2.51.

\textbf{COLDz.GN.30}: This spatially extended candidate does not appear to have counterparts. The IRAC 3.6 $\mu$m image is blended with a bright foreground galaxy, and the only catalog association (offset by $2.7^{\prime\prime}\pm0.9^{\prime\prime}$ to the SW) has a photo-$z$=0.7--0.9, which is incompatible with the CO(1--0) redshift of $z_{10}$=2.68.

\begin{figure*}[tbh]
\includegraphics[scale=1.7]{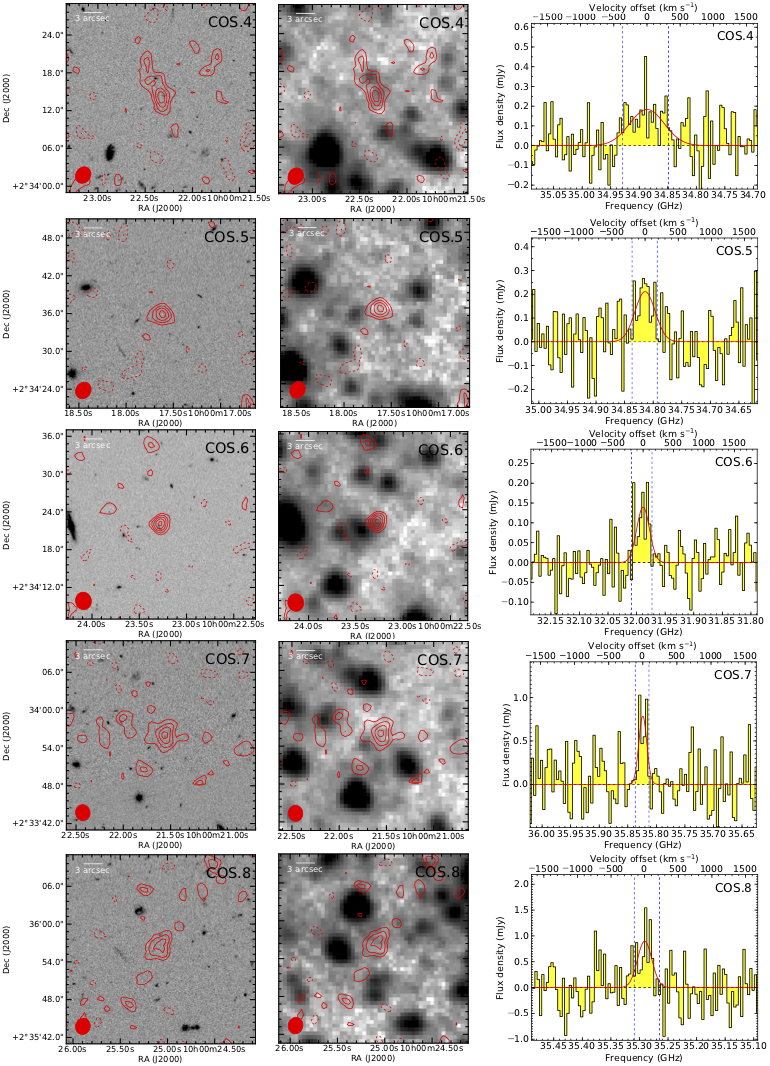} 
\caption{Additional candidate integrated line map overlays (contours) over {\em HST} I-band (left) and IRAC channel 1 images (middle) from SPLASH (grayscale; \citealt{Steinhardt14}).  {\em HST} images were obtained from the online IRSA/IPAC database. Contours are shown in steps of $1\sigma$ starting at $\pm2\sigma$. 
Right: Line candidate aperture spectra (``histograms") and Gaussian fits (red curves) to the line features. The observed frequency resolution is the same as in Fig.~\ref{fig:spectra_COS1}.
The velocity range used for the overlays is shown by the dashed blue lines.}
\label{fig:spectra_COS2}
\end{figure*}

\begin{figure*}[tbh]
\ContinuedFloat
\includegraphics[scale=1.8]{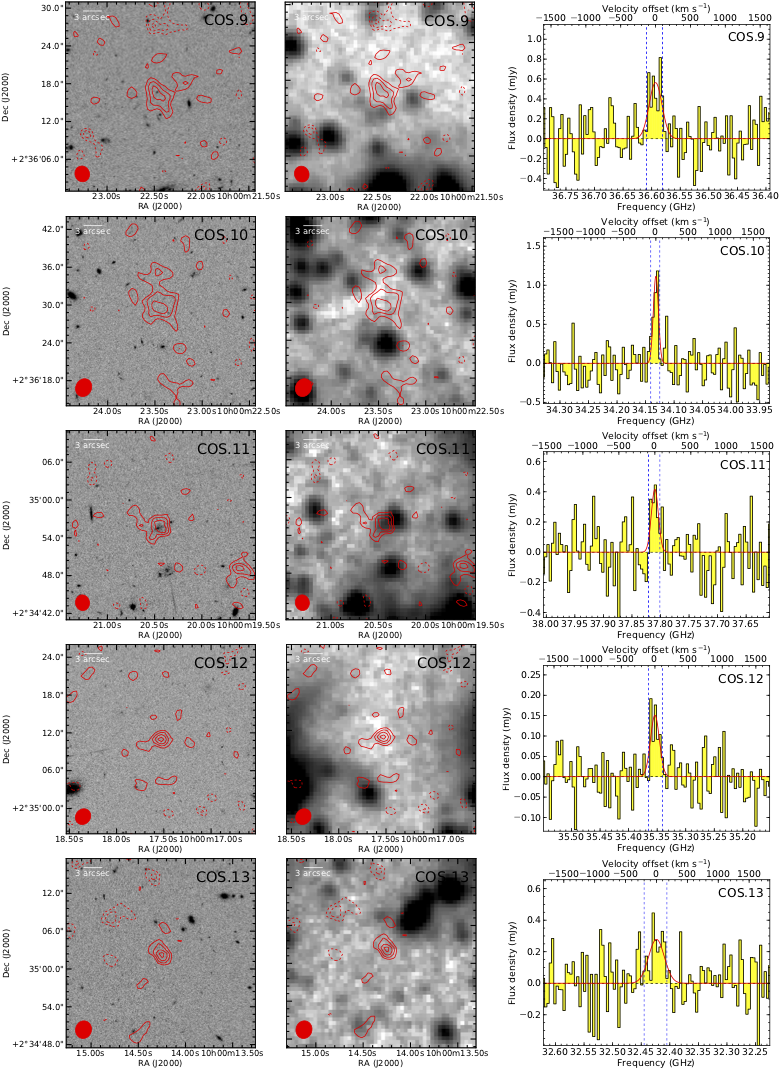} 
\caption{(continued)}
\label{fig:spectra_COS3}
\end{figure*}

\begin{figure*}[tbh]
\ContinuedFloat
\includegraphics[scale=1.8]{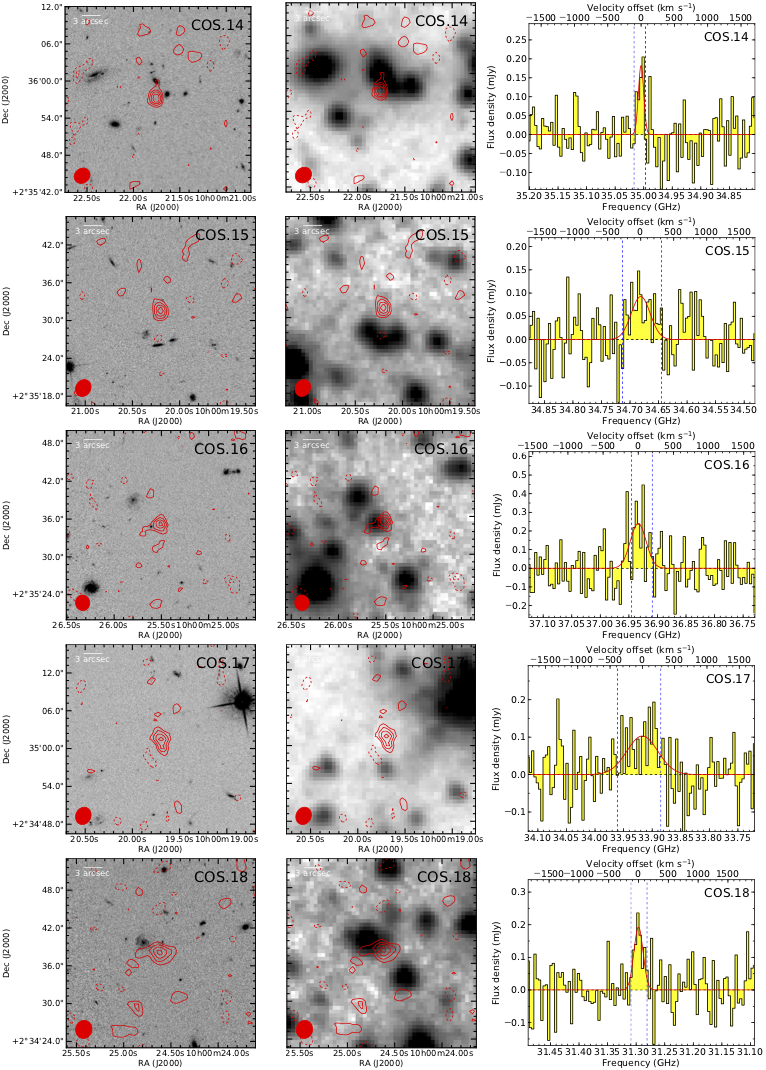} 
\caption{(continued)}
\label{fig:spectra_COS4}
\end{figure*}

\begin{figure*}[tbh]
\ContinuedFloat
\includegraphics[scale=1.8]{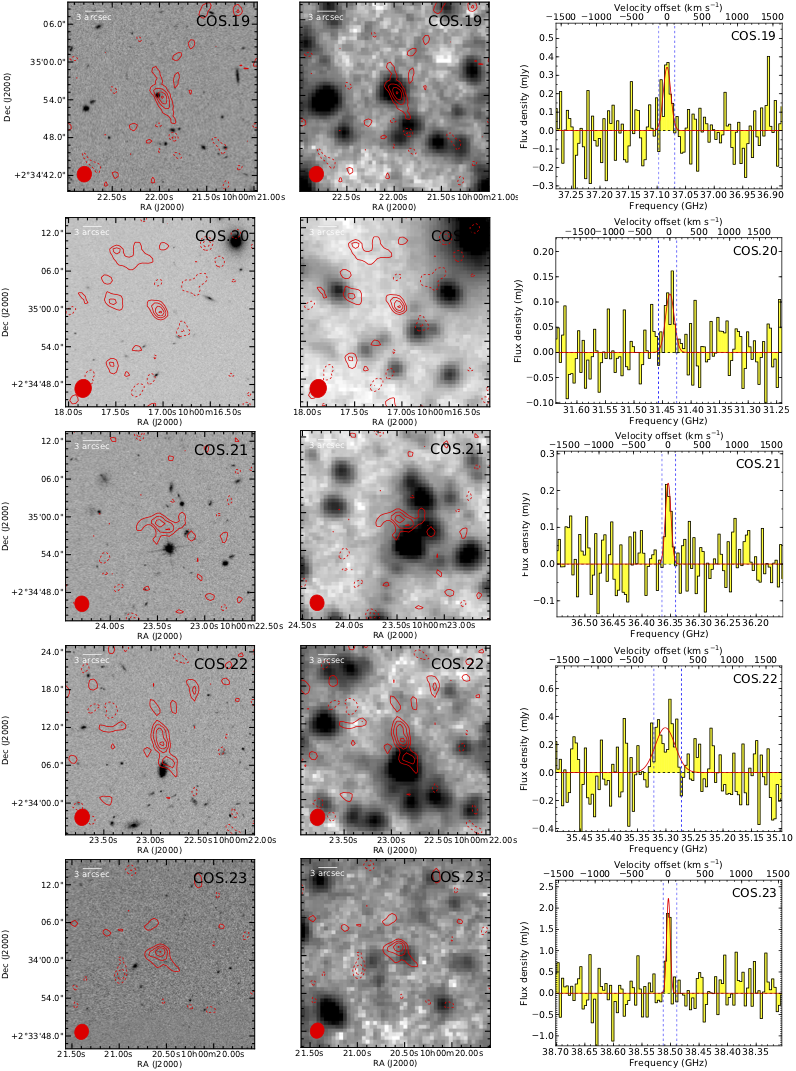} 
\caption{(continued)}
\label{fig:spectra_COS5}
\end{figure*}

\begin{figure*}[tbh]
\ContinuedFloat
\includegraphics[scale=1.8]{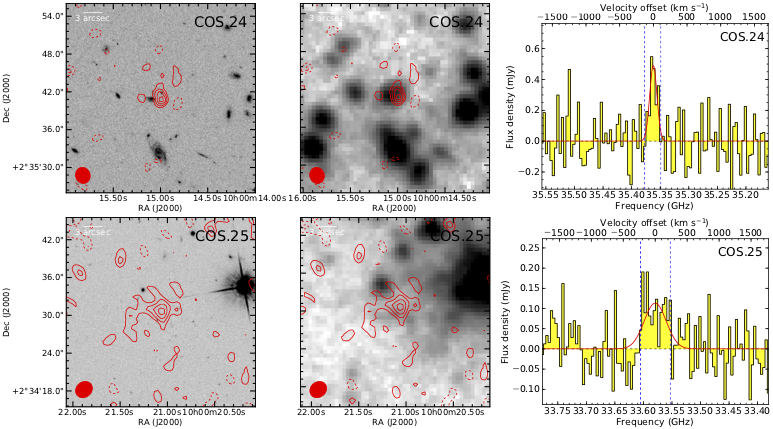} 
\caption{(continued)}
\label{fig:spectra_COS6}
\end{figure*}

\begin{figure*}[tbh]
\includegraphics[scale=1.7]{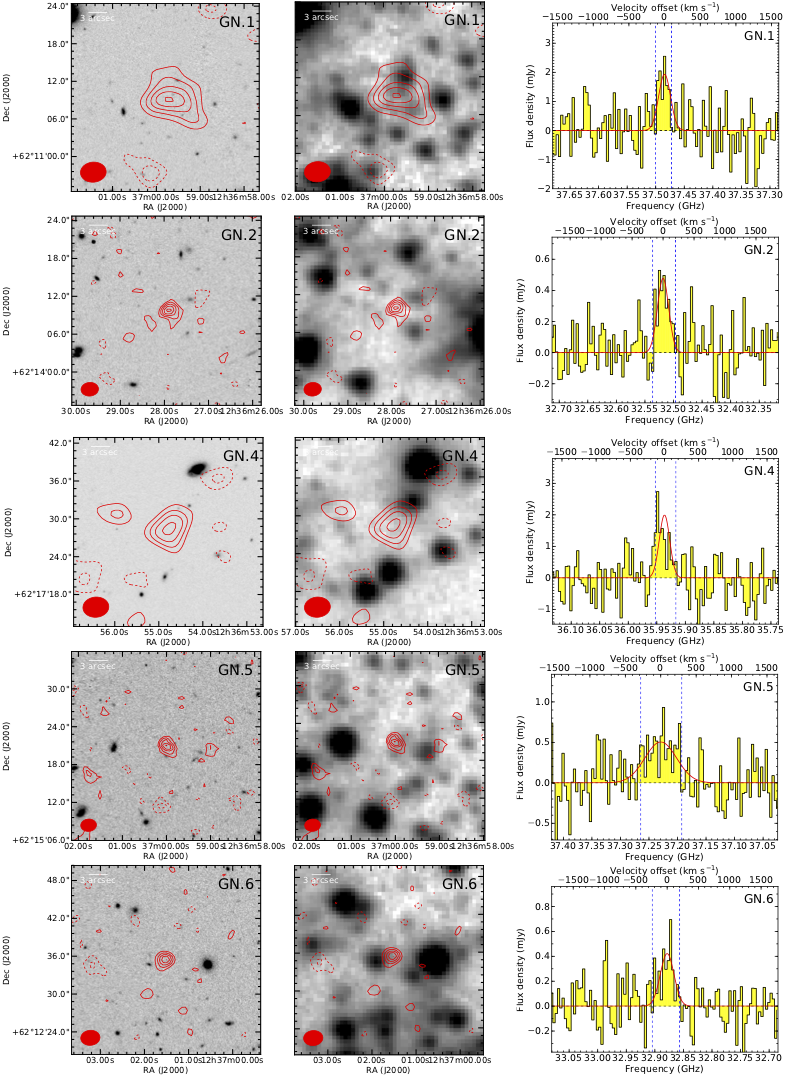} 
\caption{Additional candidate integrated line map overlays (contours) over {\em HST} H-band (left) and IRAC channel 1 images (middle; grayscale).  {\em HST} and {\em Spitzer} images were obtained from the  CANDELS database. The CO line data were taken from the Natural-mosaic or Smoothed-mosaic when the line emission is unresolved/resolved, respectively. Contours are shown in steps of $1\sigma$ starting at $\pm2\sigma$. 
Right:  Line candidate single-pixel/aperture spectra (``histograms"; for unresolved/resolved emission) and Gaussian fits to the line features (red curves). The observed frequency resolution is the same as in Fig.~\ref{fig:spectra_COS1}.  The velocity range  used for the overlays is shown by the dashed blue lines.}
\label{fig:spectra_GN1}
\end{figure*}

\begin{figure*}[tbh]
\ContinuedFloat
\includegraphics[scale=1.8]{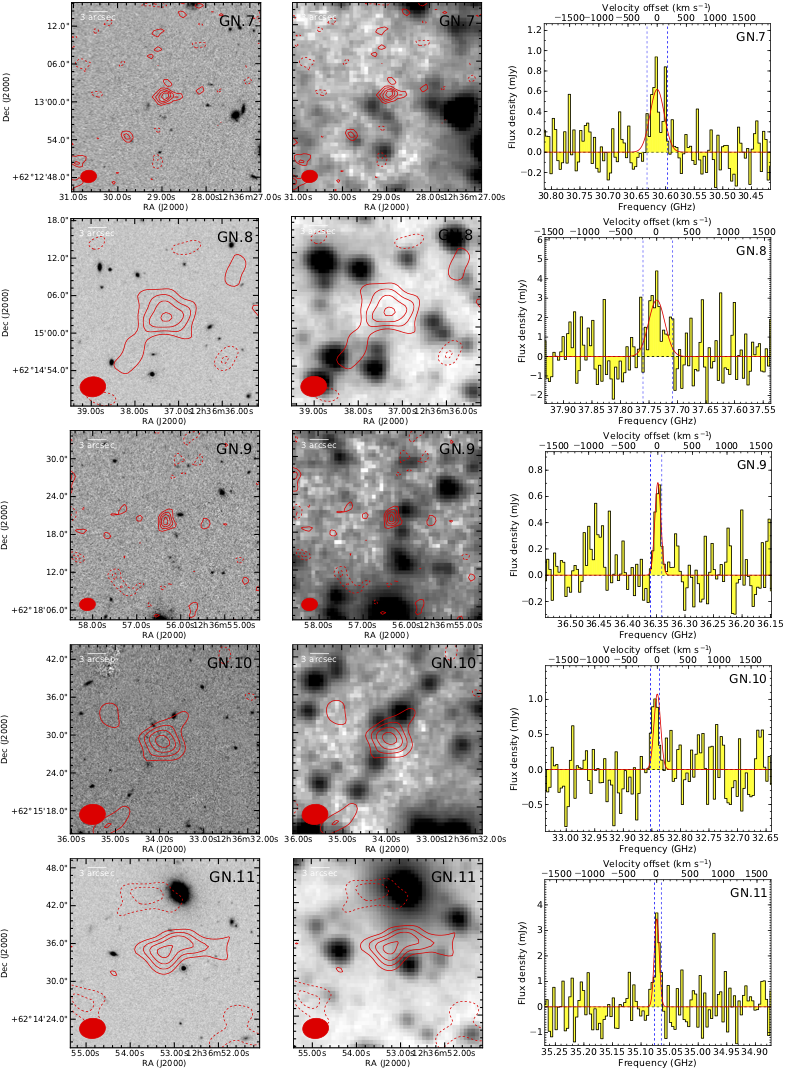} 
\caption{(continued)}
\label{fig:spectra_GN2}
\end{figure*}

\begin{figure*}[tbh]
\ContinuedFloat
\includegraphics[scale=1.8]{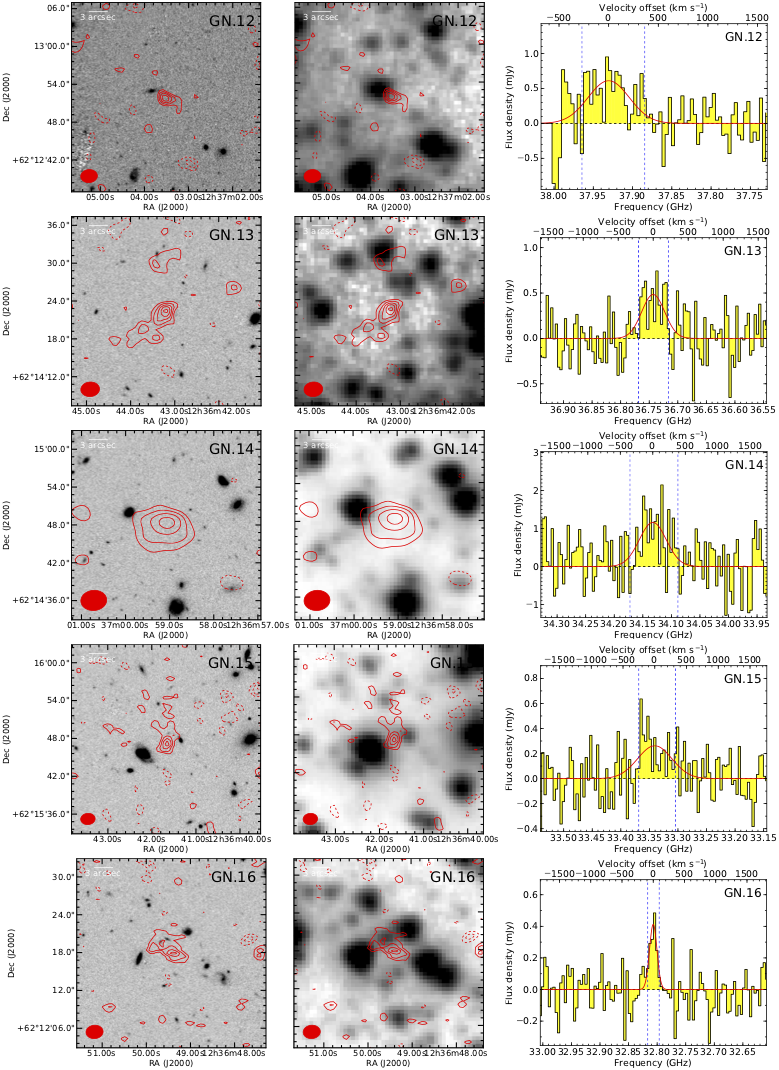} 
\caption{(continued) }
\label{fig:spectra_GN3}
\end{figure*}

\begin{figure*}[tbh]
\ContinuedFloat
\includegraphics[scale=1.8]{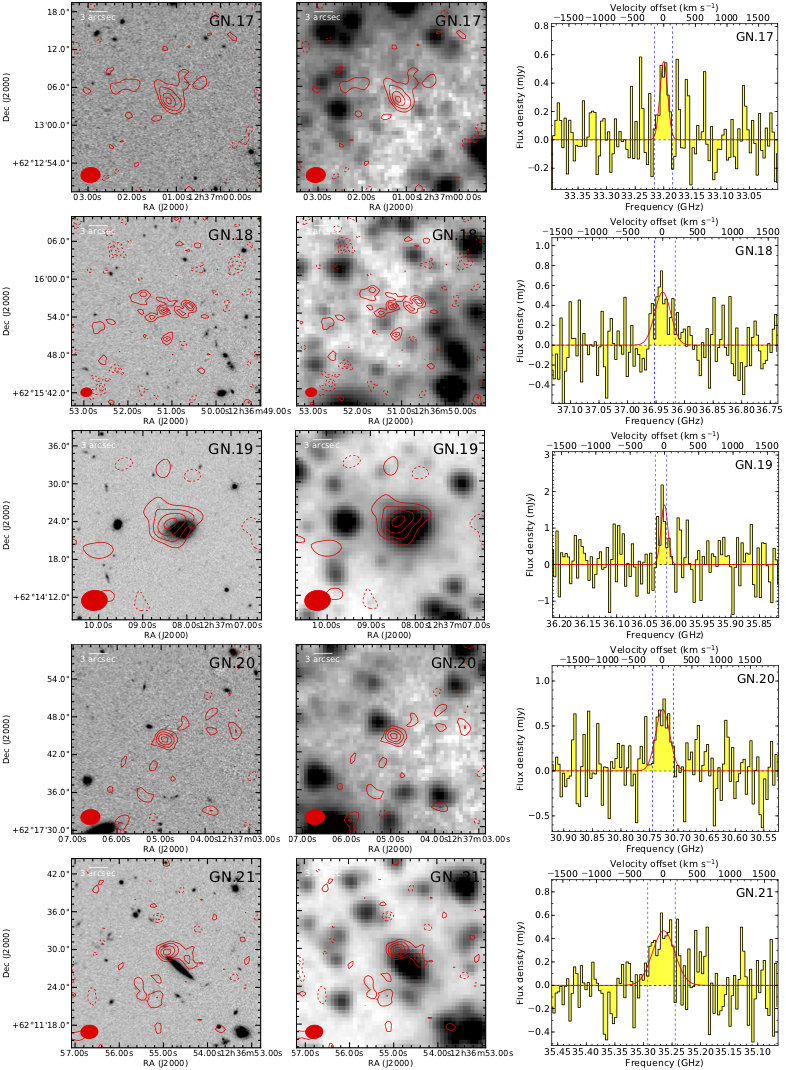} 
\caption{(continued)}
\label{fig:spectra_GN4}
\end{figure*}

\begin{figure*}[tbh]
\ContinuedFloat
\includegraphics[scale=1.8]{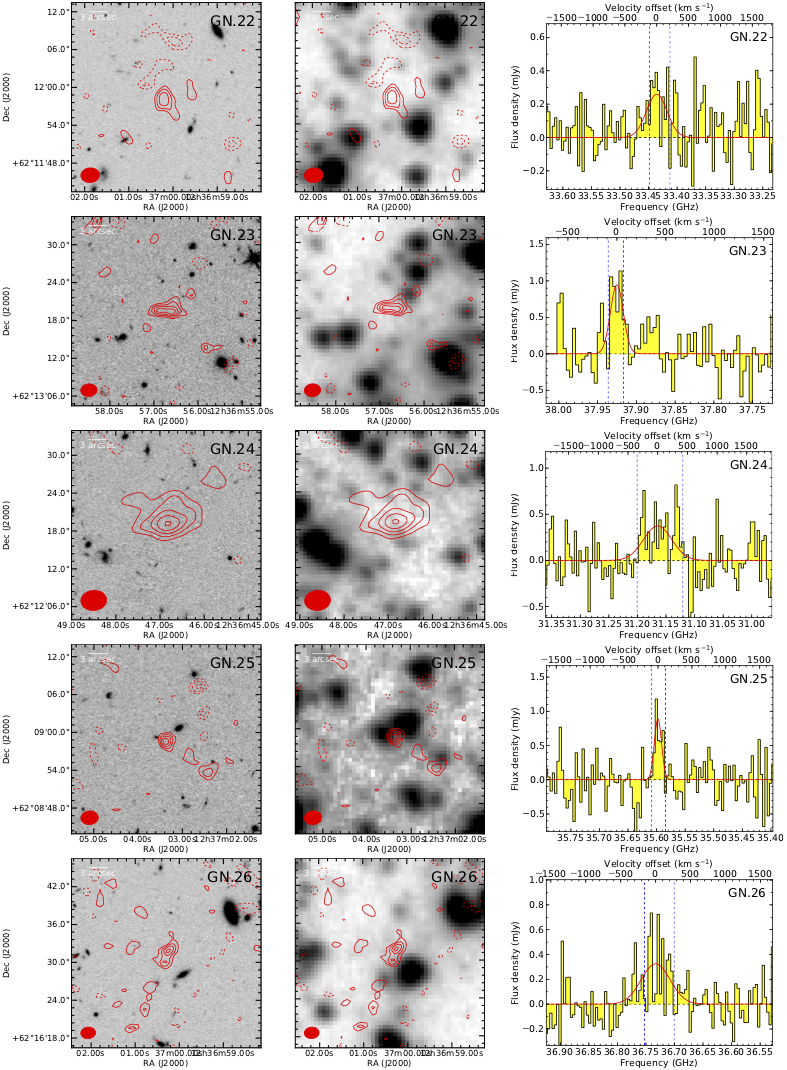} 
\caption{(continued) }
\label{fig:spectra_GN5}
\end{figure*}

\begin{figure*}[tbh]
\ContinuedFloat
\includegraphics[scale=1.8]{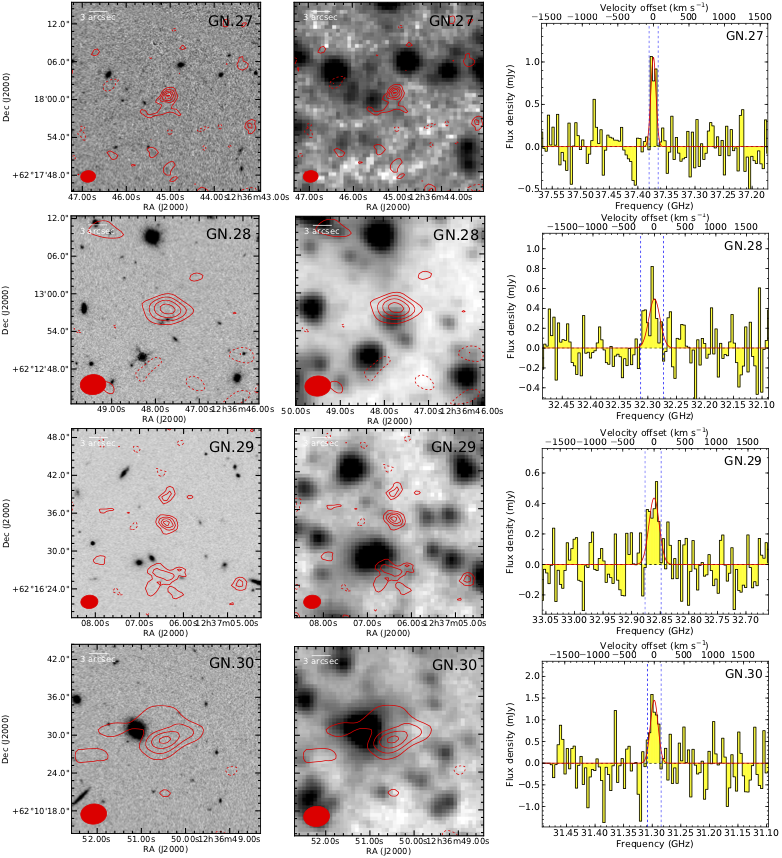} 
\caption{(continued)}
\label{fig:spectra_GN6}
\end{figure*}

\clearpage

\section{Statistical properties of the candidate CO emitter sample}
In order to extract as much statistical information as possible from our CO candidate list, we have to evaluate: 1) the probability of each line candidate to be real, 2) the line luminosity probability function and, 3) for each luminosity bin,  the completeness of our line search, i.e. the probability that a galaxy  would in fact be detected by our line search, as a function of the line emission luminosity, spatial size and velocity width.

In the following sub-sections, we will describe the methods we have developed to evaluate each of these separate components, which enter the luminosity function calculation (Paper~\rom{2}).

\subsection{Reliability analysis}

\begin{figure}[htb]
\centering{
 \includegraphics[width=.5\textwidth]{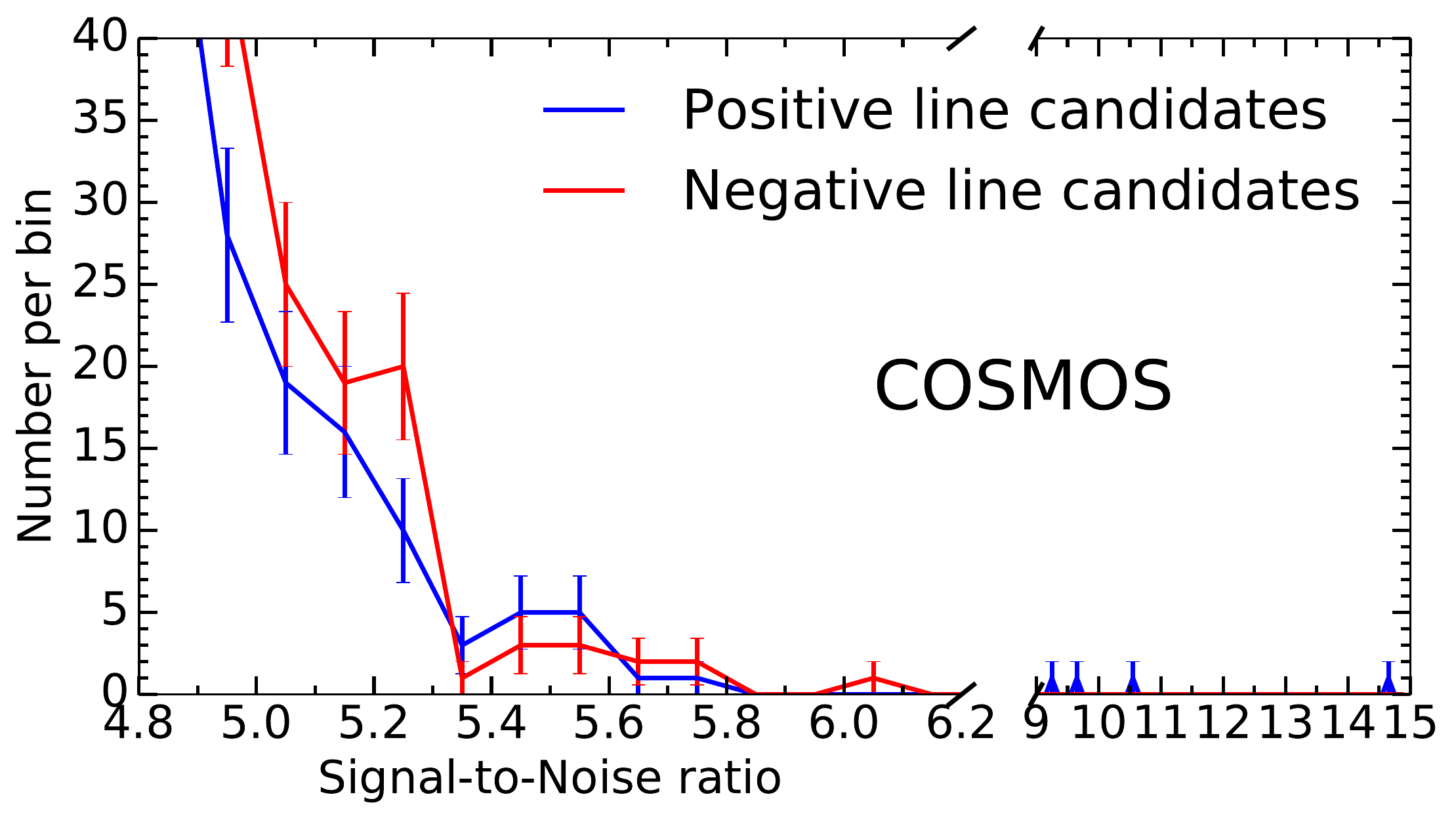}
 \includegraphics[width=.5\textwidth]{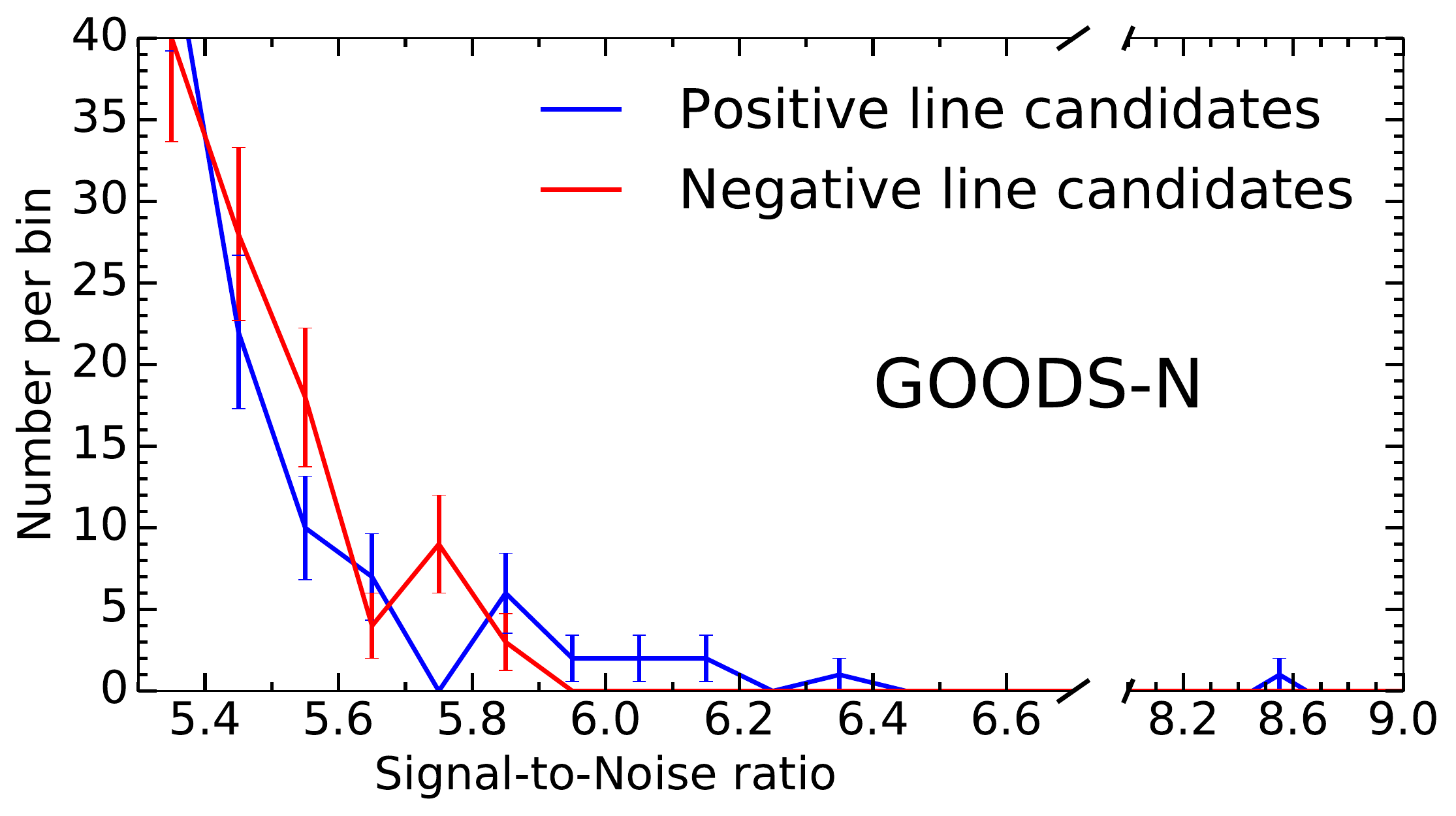}
}
\caption{Signal-to-noise ratio distributions of our line search candidates. The blue line shows the histogram of positive line features and the red shows negative line features. Poisson errors per bin are shown. The statistical evaluation of the ``excess" of positive over negative line features at a given SNR provides a measure of how many candidates may be expected to be real sources.}
\label{fig:SNR_histo}
\end{figure}

The purpose of a reliability (also called purity or fidelity) analysis is to consistently assign probability estimates to each line candidate to represent a real line source.  In this section, we attempt to provide a general solution to the problem of evaluating purities in the case of blind interferometric line searches, which builds the foundation for our analysis.

The most accurate way to tackle this problem is to utilize the symmetry around zero of the noise distribution provided by interferometric data. This is subject to the caveat of imperfect calibration and of sidelobes of bright sources which, however, should be negligible in our case, because the continuum sources in our field are not very bright ($<0.3\sigma$ and $<2\sigma$ per 4 MHz channel for the brightest source in COSMOS and GOODS-N, respectively). An alternative approach would be to try and reproduce many instances of the noise distribution by a well defined, simplified noise model, and to evaluate the rate of false positive detections as a function of SNR. However, this procedure may be strongly dependent on precisely capturing the statistical correlation properties of the noise \citep[e.g.,][]{frontiers}.
We therefore run an equivalent blind line search for negative line features in our data, in order to estimate the contamination due to noise. We show the comparison of the distributions of SNR for positive and negative lines in Fig.~\ref{fig:SNR_histo}, which were used in the following to estimate the reliability for each positive line candidate.
In the following, we apply Bayesian techniques to obtain estimates of the purities that are subject to well controlled assumptions.

The basic idea is to estimate the significance of the excess of positive over negative features at a given SNR. Any excess can be considered indication that a fraction of the positive features may correspond to real line signal. 
Some previous studies have taken a ``cumulative" approach to this problem, and used the ratio of the number of positive and negative features with SNR greater than the SNR of the line under consideration,  utilizing this ratio to estimate  purities (e.g., \citealt{ASPECS1}). This may cause a substantial bias for purities that refer to individual candidates. In particular, the presence of high SNR real candidates would raise the purity of moderate SNR positive features.
We therefore choose a ``differential" approach, but we also choose not to use bins in SNR. This choice is motivated by the small number of candidates in the bins of interest, which would make the results highly dependent on the precise binning of the SNR axis.
We therefore model the occurrence rate of lines as an inhomogeneous Poisson process along the SNR axis, with a parametrized mean occurrence rate per unit SNR interval (see Section 14.5 of \citealt{Gregory_book}, for an introduction). We can then use the machinery of Bayesian inference to study the posterior probability distribution for the rate of real sources and noise spikes, and therefore infer purities for each line candidate.

In order to derive our final likelihood function, we first consider a case where we group  line candidates in bins of SNR. While the result of this calculation already has wide applicability and offers certain benefits (e.g., by avoiding any parametric assumptions for the source and noise distributions), binning introduces an unnecessary dependence on bin choice, and does not allow to capture the intrinsic continuity of the source and noise rates as a function of SNR. Therefore,  we will follow the standard procedure and take the limit in which the bins are small, such that each bin contains at most one detection, thereby eliminating the bias introduced by binning (e.g., \citealt{GregoryLoredo}). In each SNR bin, the task at hand is to determine the probability distribution for the fraction of line detections that are real sources rather than noise. 

As a starting point, we infer a model for the noise distribution by fitting a Poisson process to the distribution of negative line features. Complex modeling for the noise feature occurrence rate is not necessary for estimating purities because in the moderate SNR regime of interest, the uncertainty will be dominated by  shot noise due to the small number of candidate features. We therefore assume the Poisson rate (i.e., the expected number of negative lines per bin) to be well described by the tail of a Gaussian as a function of SNR, centered at zero.
 We fit for the normalization and width, of this Gaussian and thereby obtain a probabilistic description of the noise.  The adopted two-parameter Gaussian tail model  provides an excellent fit to the distribution of negative features.  We stress that this method does not rely on the assumption of a Gaussian noise distribution, but it rather represents a convenient fitting function which takes advantage of the smoothness of the underlying noise distribution as a function of SNR. This method avoids using discontinuous bins or discontinuous cumulative functions, and allows us to exploit the symmetry between positive and negative noise features to generalize the noise realization provided by the negative features, and to estimate the probability of any positive line candidate to also be due to noise.

In the following, we derive   purities using SNR bins.  We then  consider the continuum limit, as explained above. The quantity of interest is  the probability of having $N_{s,i}$ real sources in the i-th SNR bin, given that we observed $N_{o,i}$ lines, $p(N_{s,i}| N_{o,i}, \mu_{b,i})$. Here, $\mu_{b,i}$ is the mean number of noise lines expected in the i-th bin. By explicitly introducing the dependence on the real source rate (for the Poisson process), $\mu_{s,i}$, we can calculate this probability as follows:
\begin{equation}
\begin{split}
p(N_{s,i}| N_{o,i}, \mu_{b,i})= \\
& \hspace*{-60pt} \int \textrm{d}\mu_{s,i} \, p(N_{s,i} | N_{o,i}, \mu_{b,i}, \mu_{s,i})\, p(\mu_{s,i} | N_{o,i}, \mu_{b,i}).
\end{split}
\label{eqn_purity1}
\end{equation}

The first term, i.e. the probability of $N_{s,i}$ real sources once we assume a source rate, is the same as the product probability for $N_{s,i}$ sources given a source rate $\mu_{s,i}$ times $N_{o,i}-N_{s,i}$ noise features, given a noise rate of $\mu_{b,i}$:
\begin{equation}
\begin{split}
p(N_{s,i} | N_{o,i}, \mu_{b,i}, \mu_{s,i})= \\
& \hspace*{-60pt} \frac{Pois(N_{s,i},\mu_{s,i})\cdot Pois(N_{o,i} - N_{s,i},\mu_{b,i})}{\sum_{k=0}^{N_{o,i}} [Pois(k,\mu_{s,i})\,Pois(N_{o,i}-k,\mu_{b,i})]}.
\end{split}
\label{eqn_pur3}
\end{equation}
Here, $Pois(N, \mu)$ stands for the Poisson probability for N events, given a mean $\mu$, and the denominator in the previous expression is a normalization factor. The second term in Eqn.~\ref{eqn_purity1} is the probability for the source rate, given the observed number and noise rate, and it is therefore given by  
\begin{equation}
\begin{split}
p(\mu_{s,i} | N_{o,i}, \mu_{b,i})\propto \\
& \hspace*{-60pt} p(N_{o,i} | \mu_{s,i}, \mu_{b,i})\,p(\mu_{s,i})=Pois(N_{o,i}, \mu_{s,i}+\mu_{b,i}) \,p(\mu_{s,i})
\end{split}
\label{eqn_pur4}
\end{equation}
by a straightforward application of Bayes theorem.

We then follow the standard prescription for inhomogeneous Poisson processes,  considering it as the case where the equally-distributed bins are so small that each bin either contains a single line or not.  In this section, we use the term rate of the Poisson process to indicate the number of line feature occurrences per unit SNR interval. In the limit of small bins, containing at most one line detection, the probability for a Poisson rate $\mu$, (which can be assumed to take the form of a parametric function of SNR) given the list of detection SNR, is calculated by the standard formula for the likelihood of an inhomogeneous Poisson process:
\begin{equation}
\begin{split}
\log p(\{SNR_i\} | \mu)=\\
& \hspace*{-60pt} \sum_i \log \mu(SNR_i)-\int_a^b \textrm{d}(SNR') \, \mu(SNR').
\end{split}
\label{eqn_purity2}
\end{equation}
Here, $\{SNR_i\}$ refers to the list of line detection signal-to-noise ratios, the {\it a} and {\it b} integration limits reflect the range of SNR that is considered for fitting, and $\mu$ is our parametric model function for the rate of lines as a function of SNR.

In the next steps, we use the occurrence rate of background, noise lines, measured from the negatives by maximizing the likelihood for the noise model. A more complex approach would include the full probability distributions for the noise model parameters in the purity evaluation.
We have tested this approach and confirmed that it does not affect our purity results. In particular, using MCMC samples from the probability distribution for the noise model parameters, we have evaluated the purity of one of our moderate SNR candidates.  We found that the median purity coincides with the purity evaluated with our simpler method, and that the relative scatter in the purity introduced by this uncertainty on the noise model is $\lesssim10\%$.  This is much smaller than our conservative estimate of the systematic uncertainty, which we adopt in the following.
Therefore, we maximize the probability for the complete set of negative lines (range of the integral $SNR \in [4,\infty)$), while assuming a Gaussian tail model for the rate function $\mu_b=N \, \exp(-\frac{SNR^2}{2\sigma_b^2})$, in order to determine the parameters $N$ and $\sigma_b$.
To determine the purity/reliability of each object, we calculate the probability that its ``small bin" contains one real source and zero noise lines.  Eqn.~\ref{eqn_pur3} therefore gives  
\begin{equation}
p(N_{s}=1 | N_{o}=1, \mu_b, \mu_s)=\frac{\mu_s}{\mu_s+\mu_b},
\end{equation}
 and hence Eqn.~\ref{eqn_purity1} becomes
\begin{equation}
\begin{split}
\mathrm{purity_k}=p(SNR_k\, \mathrm{is\, real} | \{SNR_i \}, \mu_b)=\\
& \hspace*{-180pt} \int \textrm{d}\mu_s(SNR_k) \, \frac{\mu_s(SNR_k)}{\mu_s(SNR_k)+\mu_b(SNR_k)}\, p(\mu_s | \{SNR_i\}, \mu_b).\\
\end{split}
\label{eqn_pur10}
\end{equation} 
The last term is  important, and represents the probability distribution for the source rate parameters (replacing Eqn.~\ref{eqn_pur4}). It can be written as the product of the  probability in Eqn.~\ref{eqn_purity2} (for a rate equal to $\mu_b+\mu_s$) multiplied by priors on the source rate parameters (i.e., the last term above). 

In order to compute these purities, we therefore implement a posterior probability function for the source rate $\mu_s$, computed by Eqn.~\ref{eqn_purity2}, as a function of the model parameters.  We sample it using an MCMC technique, making use of the python package \texttt{emcee} \citep{emcee}.
The  integral  in Eqn.~\ref{eqn_pur10}, which corresponds to the purity of the {\it k}-th detection,   is equivalent to averaging the ratio 
\begin{equation}
\frac{\mu_s(SNR_k)}{\mu_s(SNR_k)+\mu_b(SNR_k)}
\end{equation}
over these MCMC samples of source rate parameters. It may be seen as a weighted average of this ratio, weighted by the posterior probability for $\mu_s$.

The simple parametrization adopted for $\mu_s (SNR)$ is $\mu_{s0}(\frac{SNR}{6})^{-\alpha}$. Thus, we normalize the occurrence rate of real sources at a SNR=6 and allow for a shallow power-law increase of the rate toward lower SNR values, as we expect that there may be more real faint sources than bright sources. We impose uniform, unconstraining priors on $\mu_{s0}$ and $\alpha$. This parametrization is intended to only accurately describe the source rate over a small range of SNR, because the line candidates of dominant interest for the purity estimation are those with $5<$SNR$<6.5$.

\begin{figure}[htb]
\centering{
 \includegraphics[width=.5\textwidth]{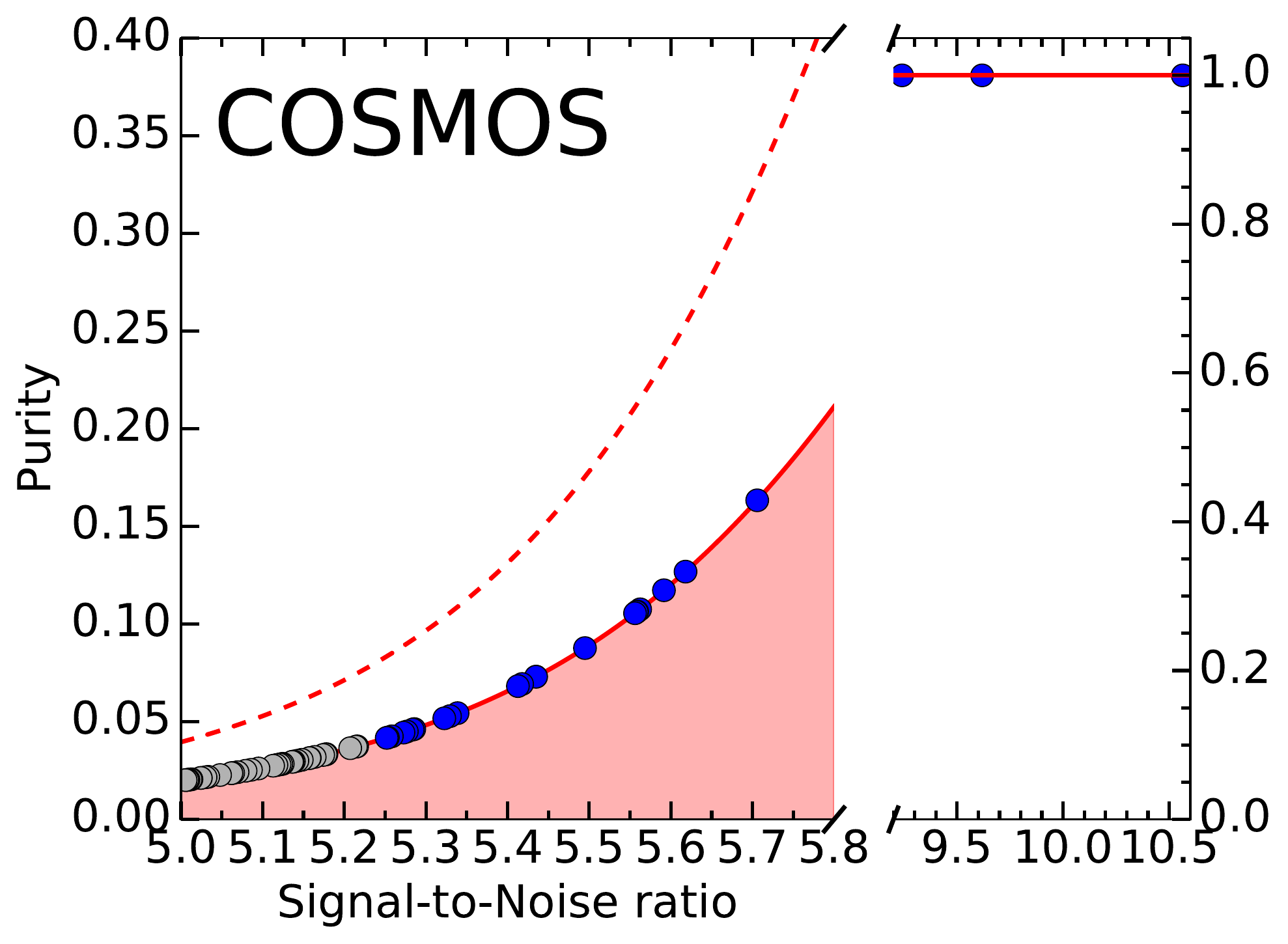}
 \includegraphics[width=.5\textwidth]{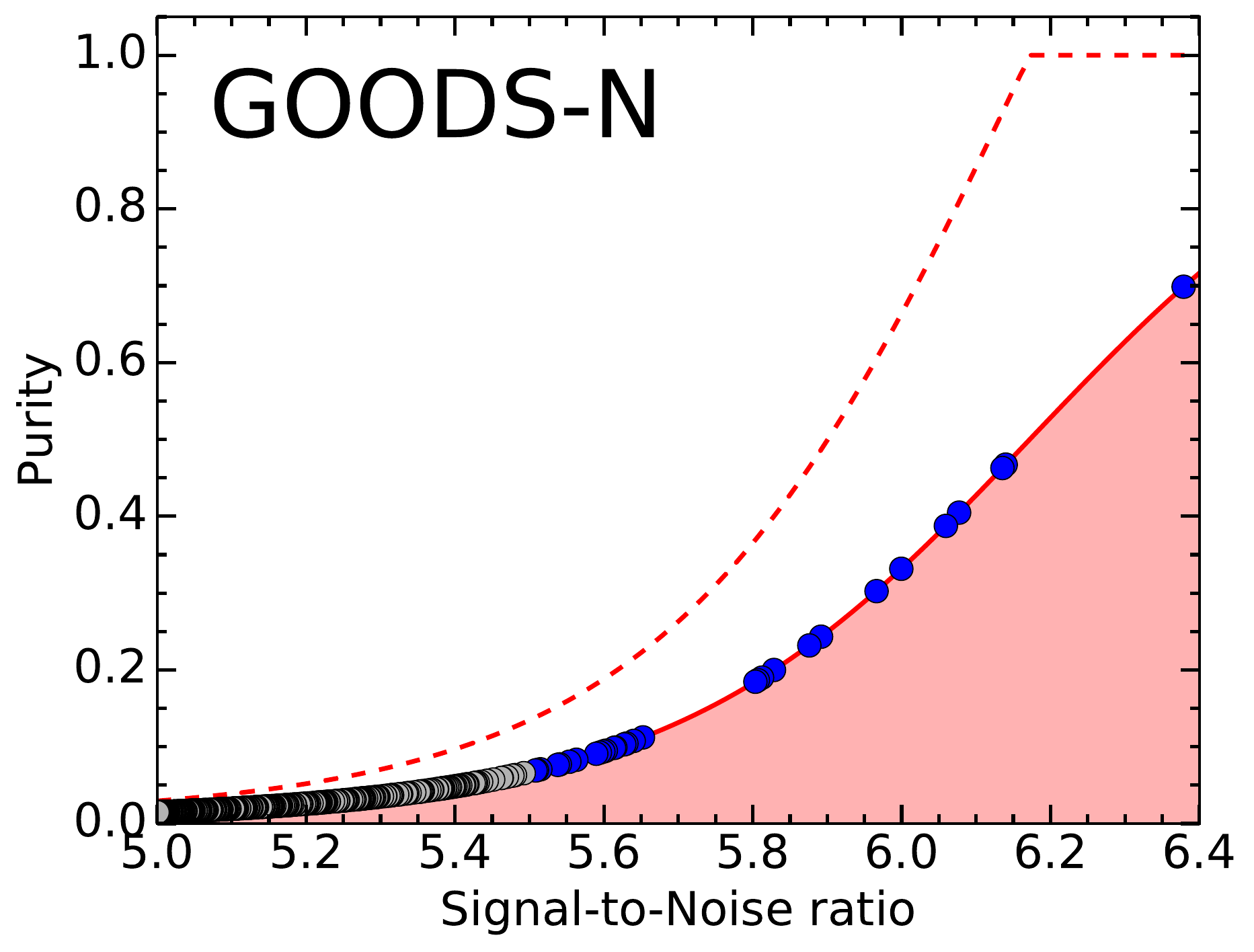}
}
\caption{Estimated purity as a function of candidate SNR. The blue points indicate the higher significance candidates reported in this work. The purity estimation utilizes an extended  candidate list down to a SNR of 5. The red shaded area indicates the SNR range that we consider for calculating purities. We also show the more conservative prescription adopted in applying purity corrections, i.e. assuming a uniform distribution for the purity, treating the calculated value as an upper limit. Our alternative prescription instead uses the calculated purities as having a Gaussian uncertainty of 100\% (the upper $1\sigma$ limit is indicated by the dashed line). We do not show the highest SNR line detections corresponding to AzTEC-3 and GN10, because they correspond to CO(2--1)  line emission, and they were not included in the purity estimation because of their high SNR (see text for details).}
\label{fig:purities}
\end{figure}

By applying this procedure, we face a choice of the SNR range to be fitted. In the COSMOS field, we start by including all the line candidates with $SNR>5$. This results in purities of 100\% for the top candidates (with secure counterparts) and $<7\%$ for the next objects down the list. This is caused by the large gap between SNR=5.7 and 9, where no candidates were found, and which favors a low source rate, for our assumed model for the source distribution. 
Our simplified Poisson model, with slowly varying source rates as a function of SNR, may only be assumed to be an accurate description of the data over a limited range in SNR. We therefore also attempt to exclude the brightest sources, and the large SNR gap without detections in the model fitting.
Therefore, to obtain an upper limit on the purities, we exclude the brightest candidates and only fit the range $5<SNR<5.8$. This yields an upper limit on the purities of up to $\sim10\%-20\%$ of the top few remaining objects to be real (Fig.~\ref{fig:purities}).
In the GOODS-N field, there is no gap in the SNR distribution of the line candidates (the highest SNR source is GN10 at z$>$5). Therefore we include all candidates in the range $5<SNR<6.4$.
%Although the prior on $\alpha$, i.e. the SNR dependence exponent of the source rate, is inconsequential for COSMOS, it is constraining for the GOODS-N purities. This is due to the  slight excess of positive lines over negative in the full range $5<SNR<6$ range, which may suggest a steep increase in the real source rate occurrence. Nonetheless, we consider it unlikely that the real source rate may be more than $\sim15\times$ higher at SNR=5 than at SNR=6 based on the apparent excess in Fig.~\ref{fig:SNR_histo}, which therefore translates to an upper bound for the power law index $\alpha$ of $\sim15$.

The procedure we have described would attribute a purity of 70\% for the candidate COLDz.GN.1, of 50\% to GN19 (which we manually correct to be 100\% because we know it to be a real line) and in the 30\%--50\% range for the other SNR$\sim$6 candidates, subsequently decreasing to about 7\% at SNR=5.5 (Fig.~\ref{fig:purities}).

When utilizing these purities to assemble the CO luminosity function, we consider two possible alternative strategies which allow to estimate the effects of the systematic uncertainties introduced by our purity computation. 
In the first approach, we   treat these purities as having 100\% uncertainty, i.e. we will draw purities (for the Monte Carlo sampling used to estimate the allowed range of the luminosity function) as independent random numbers, normally distributed around the estimated values with standard deviation equal to the purity estimate themselves and truncating at zero. The alternative approach is to implement these purities as upper limits, and to draw purities from a uniform distribution between zero and the calculated values.
The latter provides a more conservative purity estimation. Therefore, the luminosity function constraints are  somewhat lower in this method, although compatible between the two methods. This conservative approach attempts to implement the additional information coming from the lack of clear multi-wavelength counterparts to our moderate SNR candidates.
We will present the detailed results of both approaches in Paper~\rom{2}.

The SNR thresholds adopted in Section~4 and Table~\ref{tab_lines} correspond to approximate purities of $\sim$4\% and $\sim$7\% for COSMOS and GOODS-N, respectively. We emphasize that previously employed definitions of purity have differed significantly. In particular, we attempt to assess the fidelity defined by \cite{ASPECS1} for our candidate selection. The comparison is not straightforward, because the definition of fidelity used in that work relies on the details of their line search algorithm, but an approximate implementation of their method indicates an equivalent fidelity of approximately 80\%--90\% for COSMOS and 50\%--60\% for GOODS-N in their method.
%The resulting luminosity function is not strongly dependent on the precise SNR or purity threshold that we choose, therefore we choose a cutoff of SNR=5 for COSMOS, which corresponds to a purity of approximately 2\%. Then a choice of a cutoff at SNR=5.3 in GOODS-N, which allows a comparable number of line candidates to be utilized from the two fields, is sufficient in order to achieve a good match between the overlapping parts of the two independent estimates of the luminosity function provided by the two fields.

\subsection{Estimating noise tail extent from data cube sizes}
 Due to the short-scale noise correlation intrinsic to interferometric noise (over the synthesized beam length-scale), the calculation of the highest expected SNR due to noise (both positive and negative) is not straightforward as the counting of ``independent elements" is non-trivial. \cite{Vio16} and \cite{Vio17} have independently discussed a similar analysis of this case. We have reached the same conclusions, although we take a slightly different approach as we describe below.
A detailed analysis of extreme value statistics in the case of smooth Gaussian random fields (which is a good approximation for interferometric noise) was developed by \cite{Bardeen86} and \cite{Bond87}, among others, and was expanded upon by \cite{Colombi11}. Here we only summarize the main results as relevant to our data, and discuss the implications.
The objective is a description of the probability distribution function for the highest SNR in a data cube which is uniquely due to noise, and how this varies as a function of cube ``size".
If we consider the original data cube, then noise is not correlated across different channels and a noise realization is equivalent to a 2D case, with spatial correlation only, and an area equivalent to the total area across the full cube (i.e., the sum of the areas over the independent channels). In this case, the approximate cumulative distribution function for the highest SNR ($\nu$) to be expected from such a noise realization is given by:
\begin{equation}
P(\nu_{max}<\nu)\simeq\exp(-\frac{1}{4\sqrt{2\pi}}N_{n} \nu e^{-\nu^2/2}),
\end{equation}
where $N_n$ is the ``naive" counting given by the total area divided by the ``beam area" (defined by a radius equal to the beam standard deviation).
The second case regards the case where correlation of the noise across channels has been introduced (for example by convolution with a spectral template in order to Matched-Filter) and is also relevant to line searches in the form of the noise properties of Matched Filtered cubes. In this case the correlation takes place in 3D and a slightly different approximate formula describes the cumulative distribution function for the highest SNR ($\nu$) to be expected:
\begin{equation}
P(\nu_{max}<\nu)\simeq\exp(-\frac{1}{6\pi\sqrt{2}}N_{n} \nu^2 e^{-\nu^2/2}),
\end{equation}
where $N_n$ is the ``naive" counting given by the total cube volume divided by the ``effective beam volume" (an ellipsoid with a radius equal to the beam standard deviation in the spatial dimension and the standard deviation of the template used in the spectral dimension).
We have verified that the highest significance noise peaks measured as negative features in our data are compatible with these probabilistic predictions. We note that these distribution functions are quite broad, and only predict the highest SNR expected due to noise to be approximately in the SNR=5.5--6 range for our deeper mosaic and SNR=5.7--6.4 for our wider mosaic. This is a manifestation of the strong intrinsic stochasticity of the noise tails.
We also note that the ``effective number of independent elements" implied by these estimates is $\sim10$ and $\sim20$ times higher than the naive counting in the 2D and 3D cases, respectively, and that these ratios are themselves increasing functions of data cube size. The naive counting of independent elements would therefore lead to a significant underestimation of the extent of the noise tails.  This conclusion is compatible with the results of  \cite{Vio16} and \cite{Vio17}. However, we here report equations that explicitly describe the distribution of the maximum SNR to be expected from noise rather than implicitly, through the probability distribution function of local maxima. The analysis above is only approximately equivalent to the analysis presented in \cite{Vio17}, because they express the distribution function of interest as a function of $N_p$, i.e., the number of local maxima in the noise realization, which is itself a random variable with its own probability distribution.

\subsection{Artificial source analysis}

\begin{table}
  \caption{Artificial sources injected sizes}
  
\centering
\begin{tabular}{ l l l }
\hline
% & Spatial FWHM  & Frequency FWHM \\
% & (arcsec) &   (4 MHz-channels, $\sim$35 km s$^{-1}$) \\
 %  \hline  
  %  COSMOS & (convolved) 3.1, 4.3, 5.6 & 5.8 ($\sim$200), 11.7 ($\sim$400), 17.5 ($\sim$600) \\
%    COSMOS & (intrinsic) $\sim$0.5, $\sim$3.0, $\sim$4.7 & 5.8 ($\sim$200), 11.7 ($\sim$400), 17.5 ($\sim$600) \\
  % \hline
 %GOODS-N & (intrinsic)  1.0, 2.5, 4.5 & 5.8 ($\sim$200), 11.7 ($\sim$400), 17.5 ($\sim$600) \\
 & COSMOS & GOODS-N\\
 \hline
 Intrinsic spatial &$\sim$0.5, $\sim$3.0, $\sim$4.7 &1.0, 2.5, 4.5 \\
size (arcsec)\\
 Convolved spatial & 2.6, 4.0, 5.3 &2--3, 3--4,4.8--5.4\\
size (arcsec)\\
 \hline
 Frequency width & 23.2, 46.8, 70& 23.2, 46.8, 70\\
 (MHz)\\
 Velocity width &$\sim$200, $\sim$400, $\sim$600 &$\sim$200, $\sim$400, $\sim$600\\
 (km s$^{-1}$) \\
  \hline
\end{tabular}

\label{injected_sizes}
\begin{tablenotes}
\item
{\bf Note:} Gaussian sizes utilized for the injected artificial sources. All sizes refer to the Gaussian FWHM.  The convolved sizes are the injected sizes in the Natural-mosaic. These are fixed in COSMOS while in GOODS-N, because of the larger beam differences across the mosaic, we injected sources of fixed intrinsic sizes and convolved them to the local beam size, appropriate for each mosaic position.
\end{tablenotes}
\end{table}

In order to estimate the completeness and biases introduced by our line search and flux extraction methods, we perform an extensive probabilistic analysis of artificially injected sources into our maps. The main goals of this analysis is to establish a probabilistic connection between recovered candidate properties and intrinsic properties such as spatial size, velocity width and line flux. This will provide some control over the uncertainties that affect the analysis of the CO luminosity function (Paper~\rom{2}). We also develop a method to correct the luminosity function by the completeness of our line search, which avoids a purely ``per-source" completeness estimation as far as possible (due to the bias of ``per-source" corrections), while avoiding assumptions that would significantly affect the result.

Since the large majority of the data cube contains very little signal, we use the data themselves as our model for the noise, and inject artificial sources of varying size, velocity width and fluxes at random positions in the data cube (Table~\ref{injected_sizes}). We inject sources in each cube (500 in COSMOS and 2500 in GOODS-N), estimating that this will not cause crowding of the field, therefore not causing overlaps between different sources, during the line search and effectively simulating the recovery of each injected source individually.
We then analyze each injected cube following the same steps of Matched Filtering in 3D that we applied to the real data, and in the end, we search for the injected sources to determine the recovered SNR, and the line parameters that would have been measured.
We define the ``flux-factor" as the ratio of the measured line flux to the injected flux. Therefore,  the distribution of flux-factors captures both flux corrections, and uncertainties on our flux estimations. The purpose of the flux-factor analysis is not just to correct for potential biases in our flux extraction procedure, but also to estimate the uncertainty of the flux recovery. We subsequently utilize these flux probability distributions to inform our luminosity function estimates (Paper~\rom{2}).
We ignore dependencies on frequency or position of the injected sources flux-factors (determined as function of local SNR) and completeness, thereby obtaining average values that correctly sample the data for an approximately uniform distribution of real sources in our cube.

 %This can introduce small biases because there will be slightly fewer artificial sources sampling the higher frequency end of the mosaic cube (where the field of view is slightly smaller), but this should be a negligible effect because the field of view only changes by ..\%.

Both the completeness and the flux-factors are dependent on the source size and line FWHM.  Since we only inject sources of three spatial sizes and three frequency widths, we develop a probabilistic framework to relate each  line candidate to the different injected sizes (Table~\ref{injected_sizes}). For each detected candidate, based on the template size and velocity width where their signal-to-noise peaks, we determine a probability distribution of belonging to each category of ``injected" spatial size and line width, therefore matching in a continuous and probabilistic way the measured sizes to a discrete grid of intrinsic properties, as explained in detail in the following.

\subsubsection{Flux-factors based on artificial sources}

\begin{figure*}[htb]
\centering{
 \includegraphics[width=.48\textwidth]{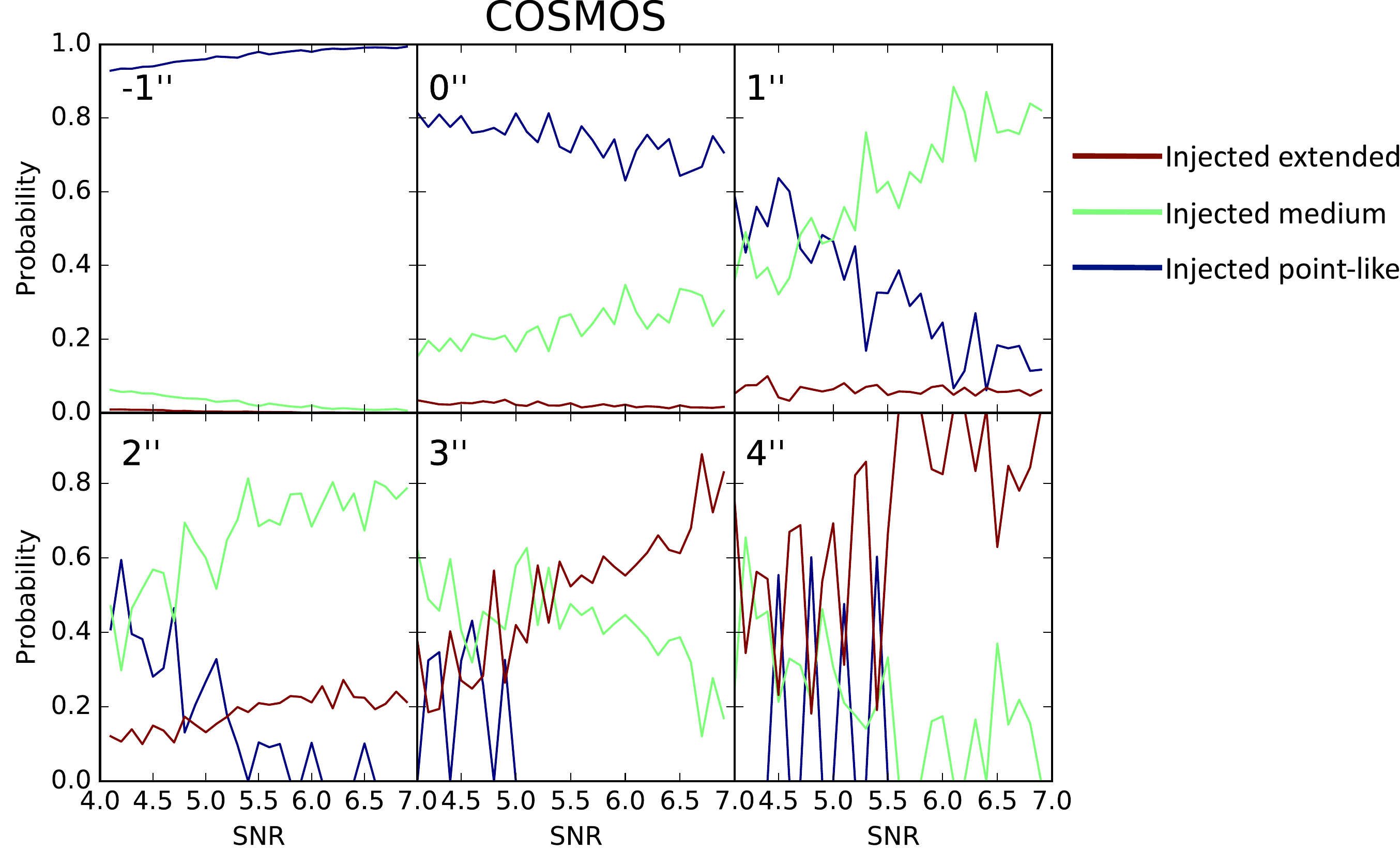}
  \includegraphics[width=.48\textwidth]{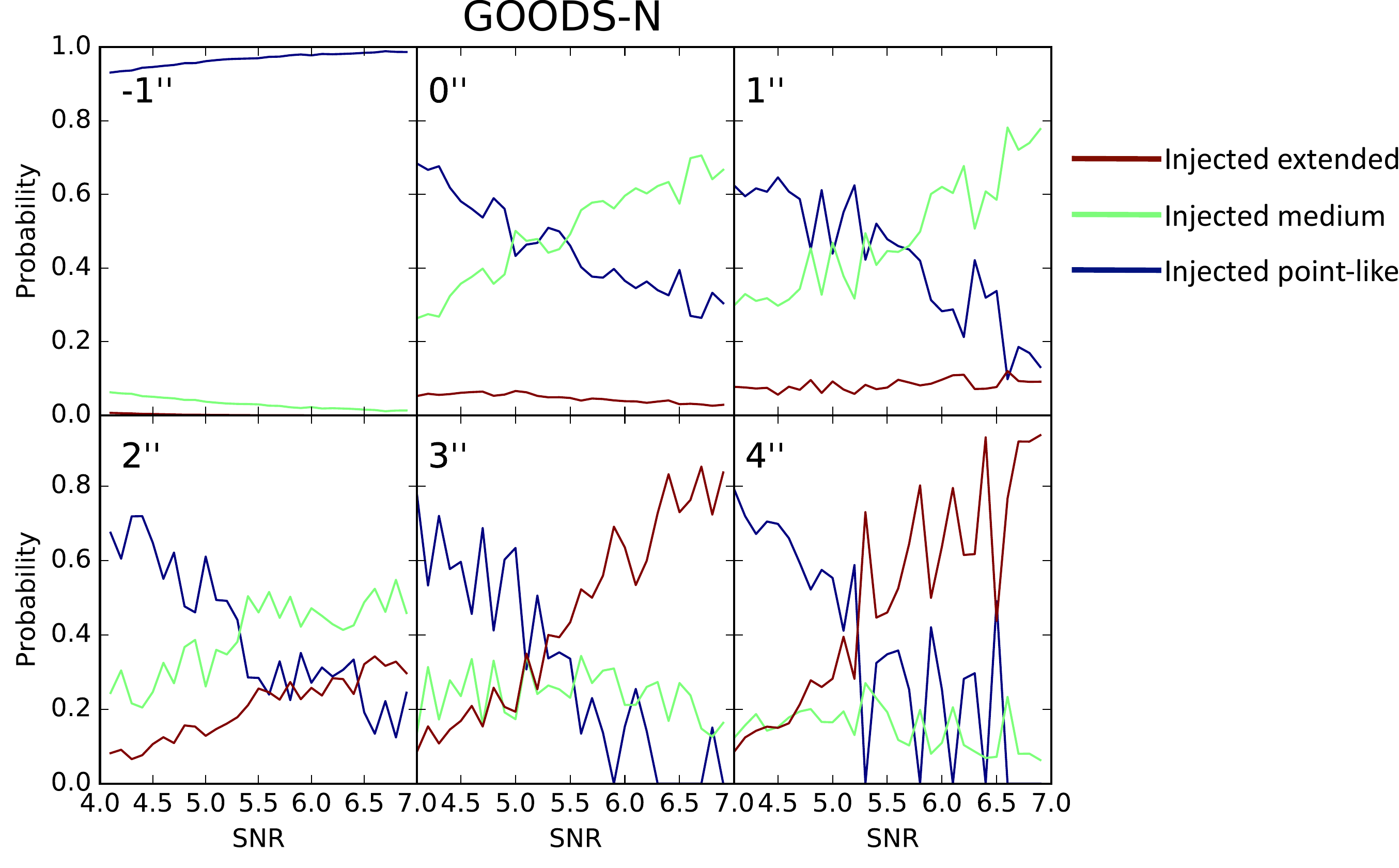}
}
\caption{Probability of injected spatial size (proxy for ``real" size) as a function of SNR, for different measured sizes (injected sizes are color-coded: smallest in blue; intermediate size in green; significantly extended in red; sizes of the artificial sources listed in Table~\ref{injected_sizes}). Measured sizes are indicated by the spatial size of the peak template (see Table~\ref{template_sizes}). Left: Results for the COSMOS field. Right: Results for the GOODS-N field. A measured (peak-template) size of -1$^{\prime\prime}$ corresponds to a source which achieves its peak SNR in the Natural-mosaic, i.e. before any smoothing, at the native resolution, while the 0$^{\prime\prime}$ template is to a point-source template applied to the Smoothed-mosaic.}
\label{fig:spatial_prob}
\end{figure*}

\begin{figure*}[htb]
\centering{
 \includegraphics[width=.48\textwidth]{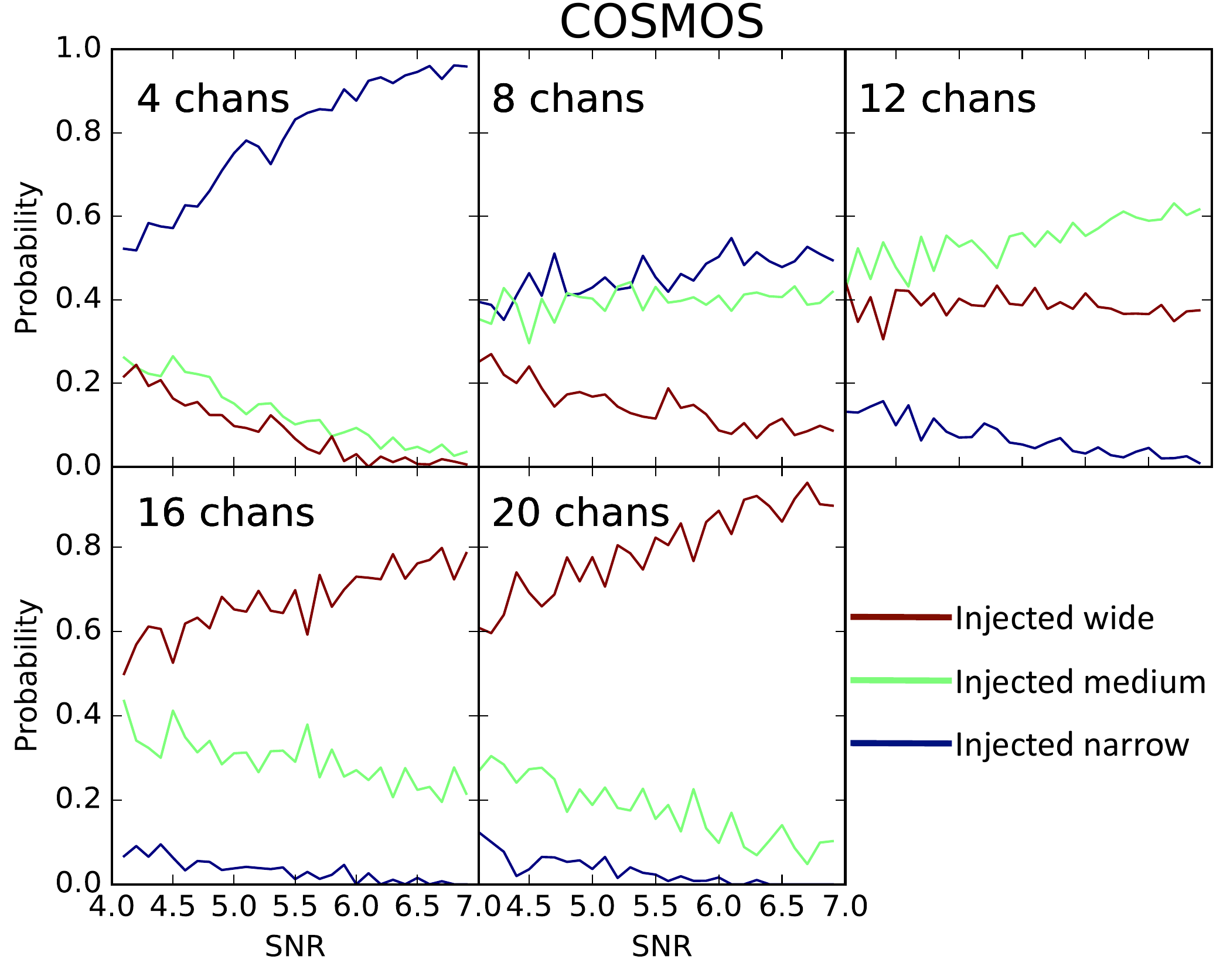}
  \includegraphics[width=.48\textwidth]{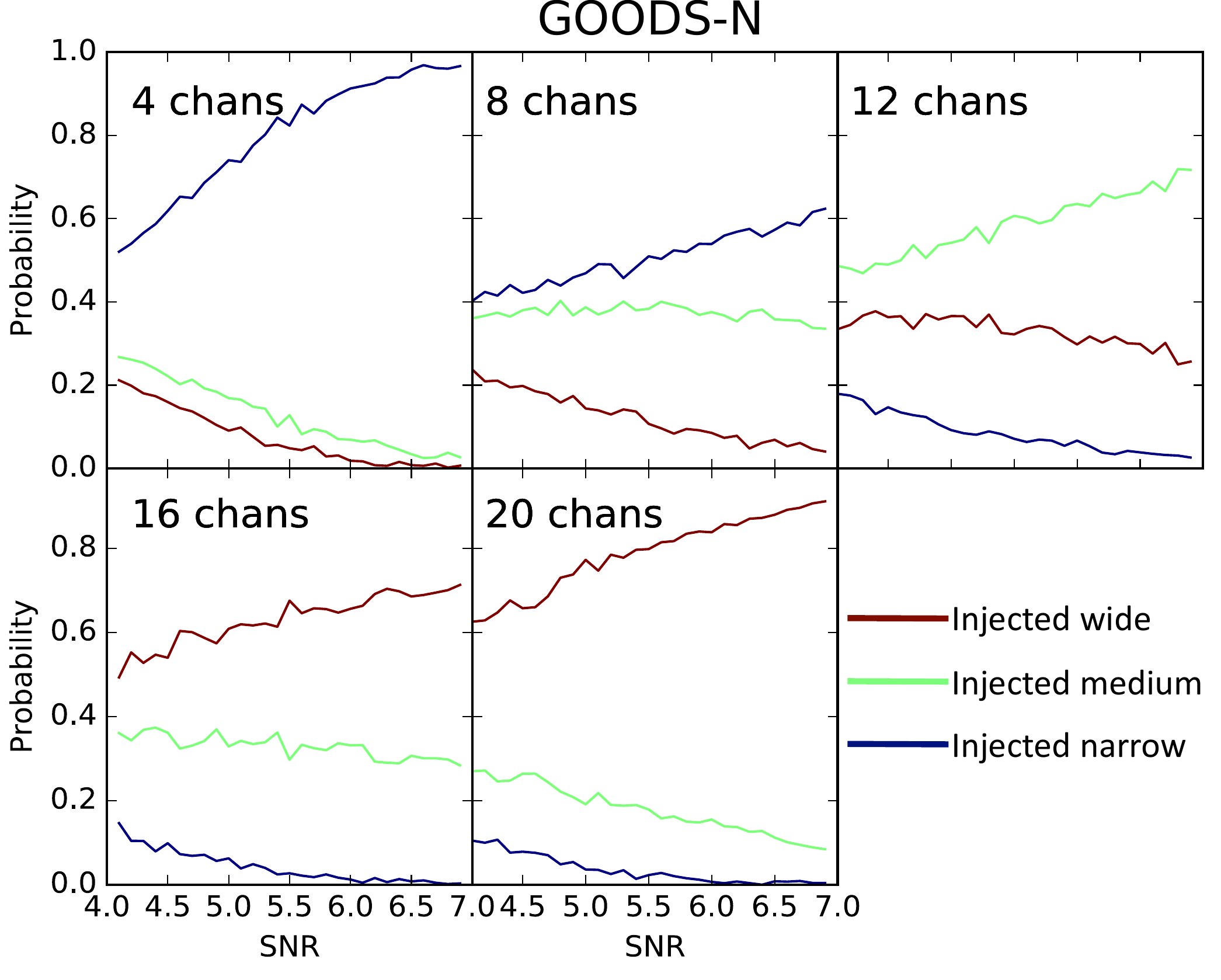}
}
\caption{Probability of injected velocity width (proxy for ``real" width) as a function of SNR, for different measured widths (injected sizes are color-coded: narrowest in blue; intermediate size in green; wide in red; FWHM of the artificial sources are listed in Table~\ref{injected_sizes}) in the COSMOS  (left) and GOODS-N (right) fields. The measured FWHM are indicated by the frequency size of the peak template (see Table~\ref{template_sizes}).}
\label{fig:FWHM_prob}
\end{figure*}

\begin{figure*}[htb]
\centering{
  \includegraphics[width=\textwidth]{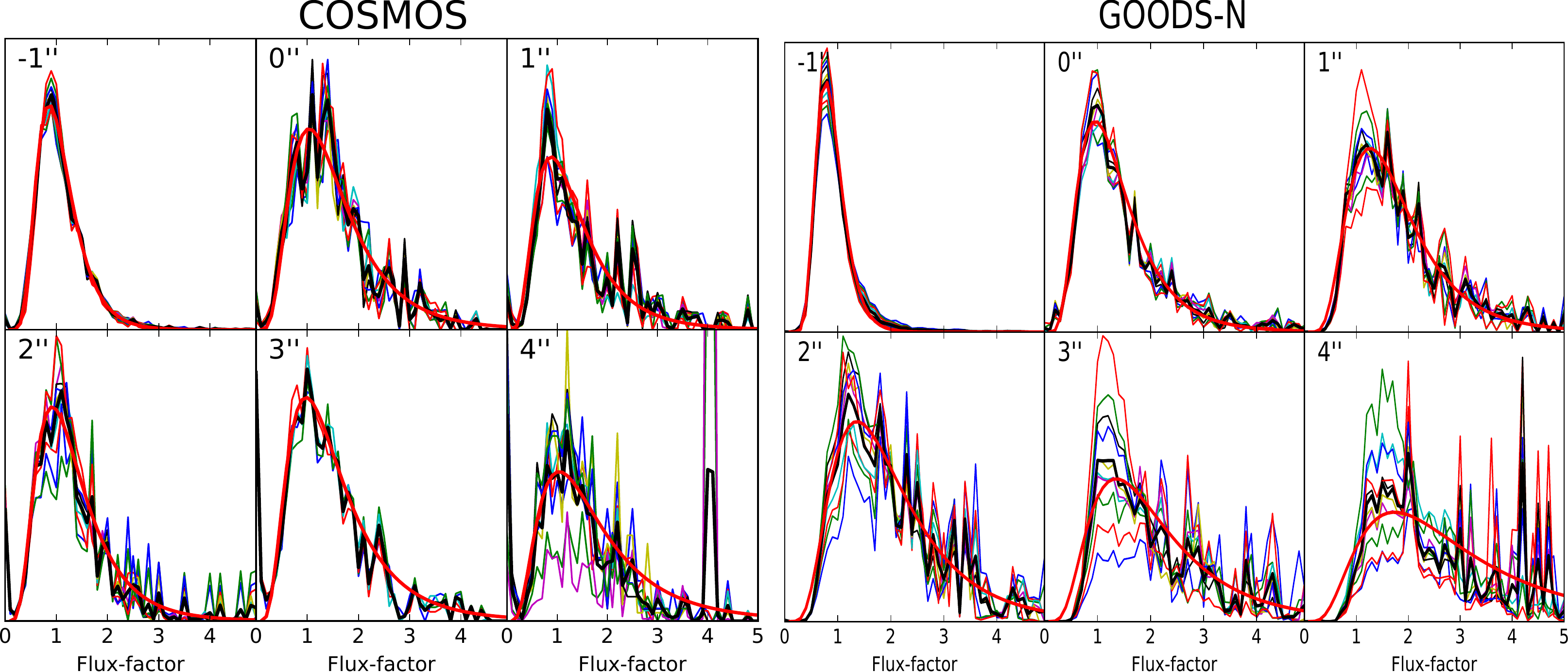}
}
\caption{Probability distribution functions of the ``flux-factor" (i.e., the ratio of measured-to-injected line flux) for the range 5$<$SNR$<$6 (in steps of 0.1$\sigma$), conditional to the measured spatial size, as indicated by the spatial size of the peak template. 
We also show in thick, black the mean distribution taken over the full SNR range, to obtain a more representative estimate. We fit this probability distribution by a log-normal  distribution, and show the best fit model in red. On the left, we show the results for the COSMOS field and on the right for the GOODS-N field. A measured (peak-template) size of -1$^{\prime\prime}$ corresponds to a source which achieves its peak SNR in the Natural-mosaic, i.e. before any smoothing, at the native resolution, while the 0$^{\prime\prime}$ template is to a point-source template applied to the Smoothed-mosaic.}
\label{fig:flux_prob}
\end{figure*}

The artificial source analysis allows us to estimate how well our measured fluxes correspond to the injected flux, for candidates of different SNR, spatial size and velocity width. The objective of the flux-factor analysis is to characterize the uncertainty and bias of our flux estimates, in order to correctly estimate the uncertainty of our luminosity function measurement.

We confirm that aperture fluxes have a slight bias toward higher fluxes, because positive noise adjacent to a candidate source tend to enlarge the fitted sizes, and therefore to contribute spurious flux to the candidate \citep{Condon97}. We thus need to estimate the magnitude of this bias, at the SNR range of interest ($\sim5$--6),  to correct the measured fluxes accordingly.
A correction factor relies on an estimate of how likely a measured extended source is to be due to noise rather than real extended structure. In order to determine this bias, we need to assume an expected approximate size distribution for our sources, to be combined with information from the artificial sources, regarding how the flux is affected by the interplay of real and measured sizes.

We  use the artificial sources to determine the probabilities of spatial extension (probability of being like spatial bin 0, 1 or 2 of the injected sources, Table~\ref{injected_sizes}) given the measured size, as traced by the size of the spatial template which gives the highest SNR. This probability is  used in the following to relate the measured properties of each line candidate to the flux-factor and completeness, which are computed for the bins of injected properties.
We use Bayes theorem to relate the probability of a given real size, conditional to a measured size: {\it p}(real-injected size $|$ measured size) to the probability distribution that we can measure from the artificial sources, which is the probability of measuring a given size, conditional to a certain injected size {\it p}(measured size $|$ injected size), by employing a prior on the expected real size distribution\footnote{Note that the artificial sources can only be used to estimate distributions conditional to a given injected size, because the relative frequency of injected sizes that was utilized (uniform) is not representative of the expected distribution of real sizes.}. 
We must employ a prior for the probabilities of the real sizes that captures our expectation that most sources would be unresolved, while allowing for a fraction of resolved sources coming from extended gas reservoirs, merging and/or blended objects. We adopt 88\%, 10\% and 2\% for the size bins in Table~\ref{injected_sizes}, respectively. We stress that although these relative fractions are uncertain, their precise choice does not significantly affect any of the results, because their effect is simply to modulate our assignment between line candidates and injected sources. The main effect of this assignment is in the estimation of completeness corrections, where the uncertainty introduced by these priors is small compared to the systematic uncertainty introduced by choosing a weighting based on the detected size distribution.

We estimate {\it p}(measured size $|$ injected size), by measuring the fraction of injected sources of a given size, recovered at different match-filter spatial template sizes. In this way, we compute the final posterior probability for a given measured size, to originate from an unknown ``injected" size, by combining this with the prior (Fig.~\ref{fig:spatial_prob}):
\begin{equation}
\begin{split}
p(\textrm{real/injected size} | \textrm{measured size, SNR}) \propto \\
&\hspace*{-200pt} p(\textrm{real size}) \times p(\textrm{measured size} | \textrm{injected size, SNR}).\\
\end{split}
\end{equation}

We follow the same procedure for relating the measured velocity width (from the peak template) to the injected line widths, by assuming a flat prior for the line width over the three injected bins (Table~\ref{injected_sizes}). These are required to compute the completeness of line candidates by relating their measured properties to the injected sources.
The derived probability distribution functions are shown in Fig.~\ref{fig:FWHM_prob}. 

In order to calculate flux-factors from the artificial sources, we employ an analogous technique. We calculate probability distributions of the flux-factors, given source spatial template size and SNR, by weighing the distribution of flux-factors found for given measured sizes, by the probability that the given measured size originates from the different possible injected sizes. Specifically, the correction ratio depends both on the injected and the measured size, as a larger ratio is needed to correct for a compact source that appears extended: \footnote{The only distribution which can be directly estimated through counting artificial sources is conditional to injected size. We cannot marginalize over those sizes without specifying a distribution of expected source sizes first.}
\begin{equation}
\begin{split}
p(\textrm{flux ratio}|\textrm{measured size, SNR})=\\
&\hspace*{-150pt} \sum_{injected} p(\textrm{flux ratio}|\textrm{measured, injected, SNR}) \times\\
&\hspace*{-150pt}\times p(\textrm{injected}|\textrm{measured, SNR}).\\
\end{split}
\end{equation}

These flux-factor distributions (Fig.~\ref{fig:flux_prob}) can be approximated by log-normal distributions, peaking near a factor of one, but with a tail to larger ratios to correct for the bias towards larger spatial size, which is introduced by including positive noise as part of the candidate source. We stress that this is just a convenient parametrization of the measured distributions.
We calculate the probability distribution of flux ratios for each measured spatial size in bins of SNR. Since the distributions are noisy due to the limited number of artificial sources, and since they do not strongly depend on SNR in the narrow 5--6 range of interest for our candidates, we consider the mean distribution over the $5<$SNR$<6$ range (Fig.~\ref{fig:flux_prob}).
 The fitted log-normal curves to these mean distributions, which provide a good interpolation to the noisy distribution estimates, will be utilized in the construction of the luminosity function. We can understand the general trend seen in these shapes as follows: the smallest template selects point sources. Therefore, the distribution peaks near one.  Slightly extended sources have a larger mean flux correction, reflecting the finding that they are most likely noise-smeared point sources and so their flux needs to be reduced.  Slightly larger (intermediate-size) sources then require less correction, because it becomes more likely that they  are somewhat extended in reality. Even larger sources show a long tail of larger flux-factors, because it is extremely unlikely that the real source is very extended.  Therefore their fluxes need to be significantly corrected (or rather, there is significant uncertainty as to their real flux, and we need to account for this in constructing the luminosity function). 

\subsubsection{Completeness}

\begin{figure}[htb]
\centering{
\includegraphics[width=.5\textwidth]{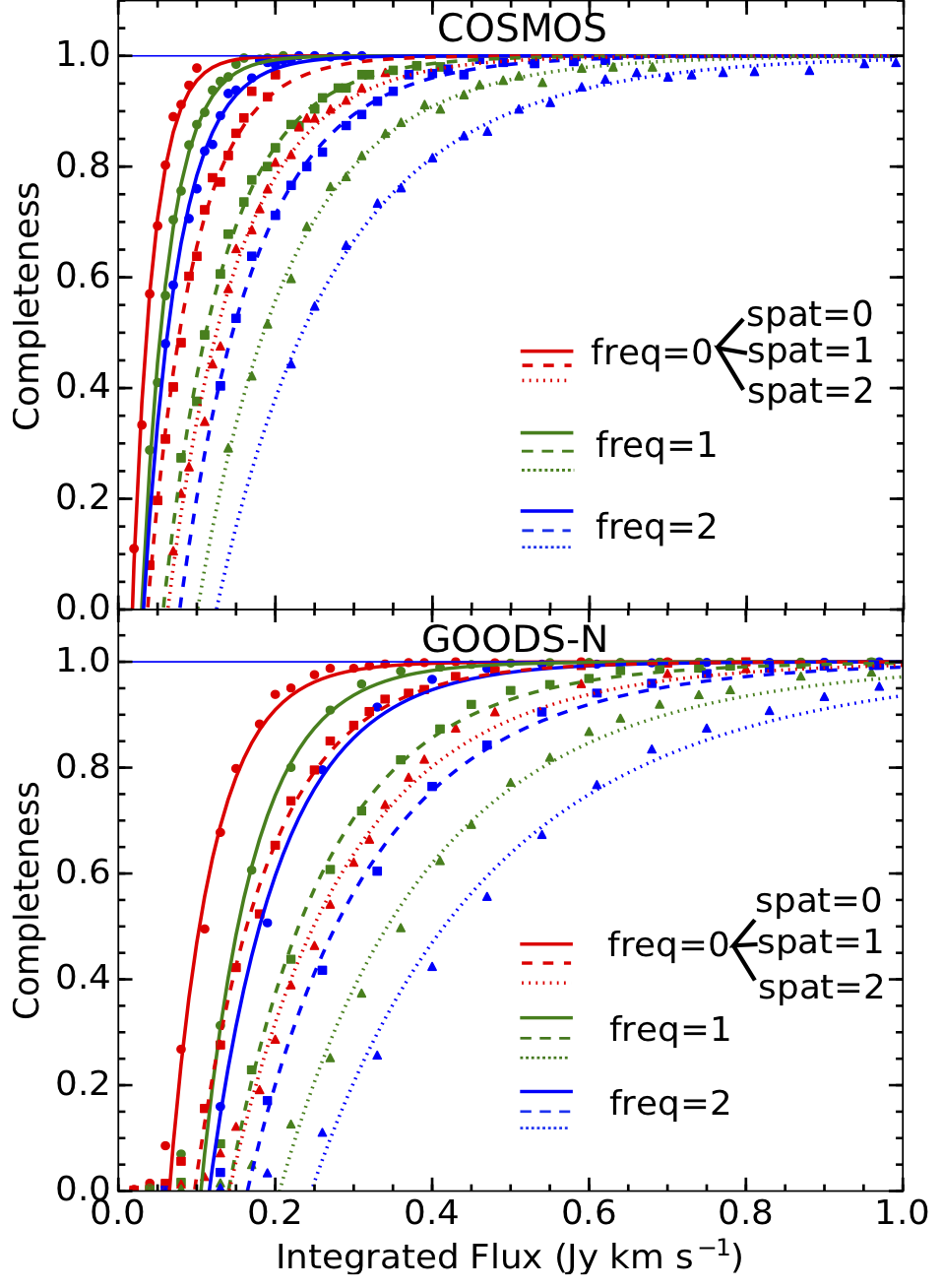}
}
\caption{Completeness corrections calculated as a function of injected flux of the artificial sources.
Top: Completeness in the COSMOS field. Bottom: Completeness in the shallower GOODS-N field. Colors distinguish the three different velocity widths of the artificial sources, and the line (marker)  style distinguishes different spatial sizes. Markers represent the measured completeness, in bins of line flux, and  lines represent the best-fitting interpolating functions.}
\label{fig:completeness}
\end{figure}

In order to estimate the completeness of our detection process, we utilize the artificial sources to measure the fraction of the injected sources that are detected.
%above a SNR threshold (we choose 5$\sigma$ but the precise threshold has a very small effect on the completeness values). %The detailed definition of completeness corrections is dependent on the use and as a starting point we consider how to best correct the number counts that enter the binned CO luminosity function estimation.
The objective of the completeness correction is to account for the fraction of the mosaic volume where a given line candidate would be detectable, and to account for the fraction of objects of given intrinsic line luminosity that would be missed by a fixed SNR threshold.

We assume that the fraction of detected lines (a proxy for the probability of detection)  only depends on the integrated line flux, on the spatial size, and on the velocity width. By injecting artificial sources that uniformly sample random positions within the edges of the mosaic, we derive completeness corrections that account for the effects of the spatial and frequency variation of the sensitivity, as previously adopted by \cite{ASPECS1}. 
%Lines with a fixed injected flux result in spatially varying signal-to-noise ratios and, assuming an isotropic distribution, uniform distribution is frequency space, and equal angular densities at different redshifts for the real sources (which is only approximate, angular distances at different redshifts correspond to slightly different physical distances, but the redshift range is narrow enough that the effect should be negligible compared to other sources of error) as with the artificial sources, the fraction of lines above the detection threshold allows correcting the observed numbers to the numbers inferred for the whole cosmic volume covered by the mosaic.

While completeness is a property of the overall number counts in a luminosity bin, we partially adopt the spirit of the $1/V_{max}$ method of calculating a ``per source" completeness only in so far as this depends on the line velocity width and especially spatial size of the detection. While this may potentially introduce a bias\footnote{Regions of parameter space that have very low completeness tend to be poorly sampled and hence cannot properly be accounted for in a completeness calculation that is weighted by the detected candidates} (see e.g., \citealt{Hogg10}), it saves us from  additional assumptions about the size distribution. We therefore point out the important caveat that our luminosity function estimate does not correct for missed objects due to poor sampling of the size distribution. The effect can be seen by noticing, for example,  that the completeness for extended objects at low flux values drops very quickly. This is reflected in the fact that all of our low-flux objects are point sources (therefore, the completeness correction in the lowest line luminosity bins misses the potential contribution from undetected extended sources).

The completeness is measured from the artificial sources as a function of ``injected" properties, i.e. injected integrated flux, and spatial size and frequency width (in three bins each, see Table~\ref{injected_sizes}), as the ratio of  injected sources recovered with SNR above a threshold value of 5$\sigma$ to the total number of injected sources (the precise choice of a threshold does not  change the result appreciably). These measured completeness values are shown in Figure~\ref{fig:completeness}, together with the interpolating functions that we use in deriving the luminosity function: the two-parameter ($I_0$ and $I_1$) family of functions $1-e^{-I/I_0}/(I+I_1)$, where I is the integrated line flux. While this chosen family of interpolating functions has no specific significance, we found it to provide an appropriate description of the measured completeness.

The optimal way to correct for the completeness of a luminosity function bin would require calculating the mean completeness within the bin, over the full ``internal" parameter space (in our case these are spatial and velocity line sizes, frequency, and precise luminosity within the bin), weighted by the model expectation for the distribution of sources within this space. We adopt an intermediate approach between this ``mean completeness" approach and a purely ``per-candidate" approach. We assume that the distribution of line luminosity within a bin and the frequency distribution are uniform and average over this subset of the internal space by randomly sampling it. On the other hand, we do not assume an intrinsic spatial size and velocity width distribution, in order to avoid biasing our result. We therefore adopt the sizes and line widths of the candidates to calculate the appropriate completeness, thereby letting the data determine the size and velocity width distributions.

In summary, for each detected line candidate, the size properties (i.e. the spatial size and frequency width) of the ``real" underlying source are estimated probabilistically  based on the measured size and width (from the peak template).  Then, probability-weighted completenesses are the factors that enter the evaluation of the luminosity function.
The dependence of these completenesses on the line flux are mean values evaluated for the full luminosity function bin, rather than depending on the candidate line flux measurement. This hybrid approach helps us mitigate the bias which would derive from a purely ``per-source" correction (e.g., \citealt{Hogg10}).

\subsection{Implementation of the statistical corrections}
In order to assemble the luminosity function, we use a variation on the  method used by \cite{Decarli14,ASPECS2}.  We weigh the contribution of each line candidate to the luminosity bin by its purity and inversely by its completeness, and use the total cosmic comoving volume covered within the edges of the mosaic. The completeness correction converts this volume to an effective $V_{max}$, for each galaxy, also accounting for the spatial variation in sensitivity.
In order to estimate the systematic uncertainty introduced by our assumptions, we evaluate the luminosity function with many random realizations of flux-factor, purity assignment and  luminosity bin widths and boundaries. 
One of the differences between our approach and the approach employed by \cite{Decarli14,ASPECS2} consists in including a larger number of candidates. 
We have also calculated the luminosity function using the same method employed by \cite{ASPECS2}, and the result is  consistent with our more extensive method. The advantage of our approach consists in relying less heavily on the properties of the  few moderate SNR, individual candidates that happen to be located near the top of the SNR list, but that still have a limited  probability of being real. Although a fraction of the moderate SNR candidates are expected to be real, it is not clear that those near the top (of the set of uncertain candidates) of the SNR list have a significantly greater likelihood of being real given the limited range in SNR considered ($\sim5-6$).
By utilizing a larger sample of candidates in deriving constraints to the luminosity function, down-weighted by appropriate purities, we do not introduce additional bias, but rather better explore the implications of the systematic uncertainties.
In particular, the statistical justification for a ``per-source" purity and completeness correction (which we cannot fully avoid) only holds for large enough samples. By better sampling the ``internal" space of possible candidate sizes and line widths, and by adopting average per-bin completenesses (i.e., not using the uncertain, measured fluxes), we aim to achieve a more accurate completeness correction and, evaluation of the systematic uncertainties introduced by these factors.

In detail, for each $L'_{\rm CO}$ bin we calculate a completeness factor appropriate for each of the nine bins in the spatial size-frequency width grid, by averaging over 1000 random realizations of values of $L'_{\rm CO}$ in each bin, using random redshifts to calculate the corresponding integrated flux, and hence the appropriate completeness correction for each.  We  average over this frequency distribution and precise $L'_{\rm CO}$ within each bin. We therefore enforce a uniform prior, and maintain the dependence on the spatial-frequency size information separate. We subsequently apply them, for each line candidate, as weighted by the candidate probability distribution for its spatial-frequency size assignment. In this way, we use the measured relative occurrence of different spatial and velocity sizes as weights for the appropriate completeness,  for each luminosity function bin.

In order to explore the range of luminosity function values allowed by our systematic uncertainty, we use 10,000 Monte Carlo realizations of the luminosity function calculation, for each bin width and shift, where we vary the purity assignment independently for each candidate, and the flux-factor to be applied. We therefore ``move candidates around" among adjacent luminosity bins, simulating the effect of the uncertainty in their intrinsic fluxes.
We separately implement the purity in one of the two ways that we previously described in Section F.1, either as normally or uniformly distributed. 
We also implement the flux correction as taking a random value drawn from the appropriate log-normal distribution, which was derived for each spatial size in the previous section. We also add a normal uncertainty of $20\%$ to the measured flux, to reproduce the uncertainty in our flux calibration.

Finally, in order to describe the range of values for the luminosity function (in log comoving volume density space) spanned by our 10,000 Monte Carlo realizations, we calculate the median value for each bin and a measure of the scatter around the median. %We evaluate the scatter conservatively by taking for the upper edge: the largest between the median plus standard deviation and the 84th percentile; and for the lower edge: the lowest between median minus standard deviation and the 16th percentile. 
%We evaluate the scatter conservatively by taking the largest value between the standard deviation and the percentile that corresponds to the standard deviation for Gaussian statistics (i.e. the 84th or 16th percentile for the upper or lower bound, respectively).
We evaluate the scatter conservatively by quoting luminosity function ranges that include 90\% of the probability.
We also evaluate the statistical Poisson uncertainty as appropriate for each bin as a relative uncertainty of $1/\sqrt N$, where {\it N} corresponds to the number of candidates in a $L'_{\rm CO}$ bin.

%We can check that our estimated luminosity function in COSMOS provides reasonable and conservative limits by appreciating that in the 10.5 to 11 bin we have 3 confirmed galaxies at least, and that gives a level for the bin (if 0.5 dex bin width) of 10$^{-3.5}$ Mpc$^{-3}$ dex$^{-1}$ (10$^{-4}$ Mpc$^{-3}$ dex$^{-1}$ would correspond to 1 candidate in a bin of 0.5 dex in $L'_{\rm CO}$).

\section{Additional stacking results of galaxies with optical redshifts }
In addition to the stacking of individual sets of galaxies described in the main text, we have attempted to stack three more sets of galaxies, based on optical redshift information. We do not detect significant  signal in these stacked spectra.

\begin{figure*}[htb]
\centering{
 \includegraphics[width=\textwidth]{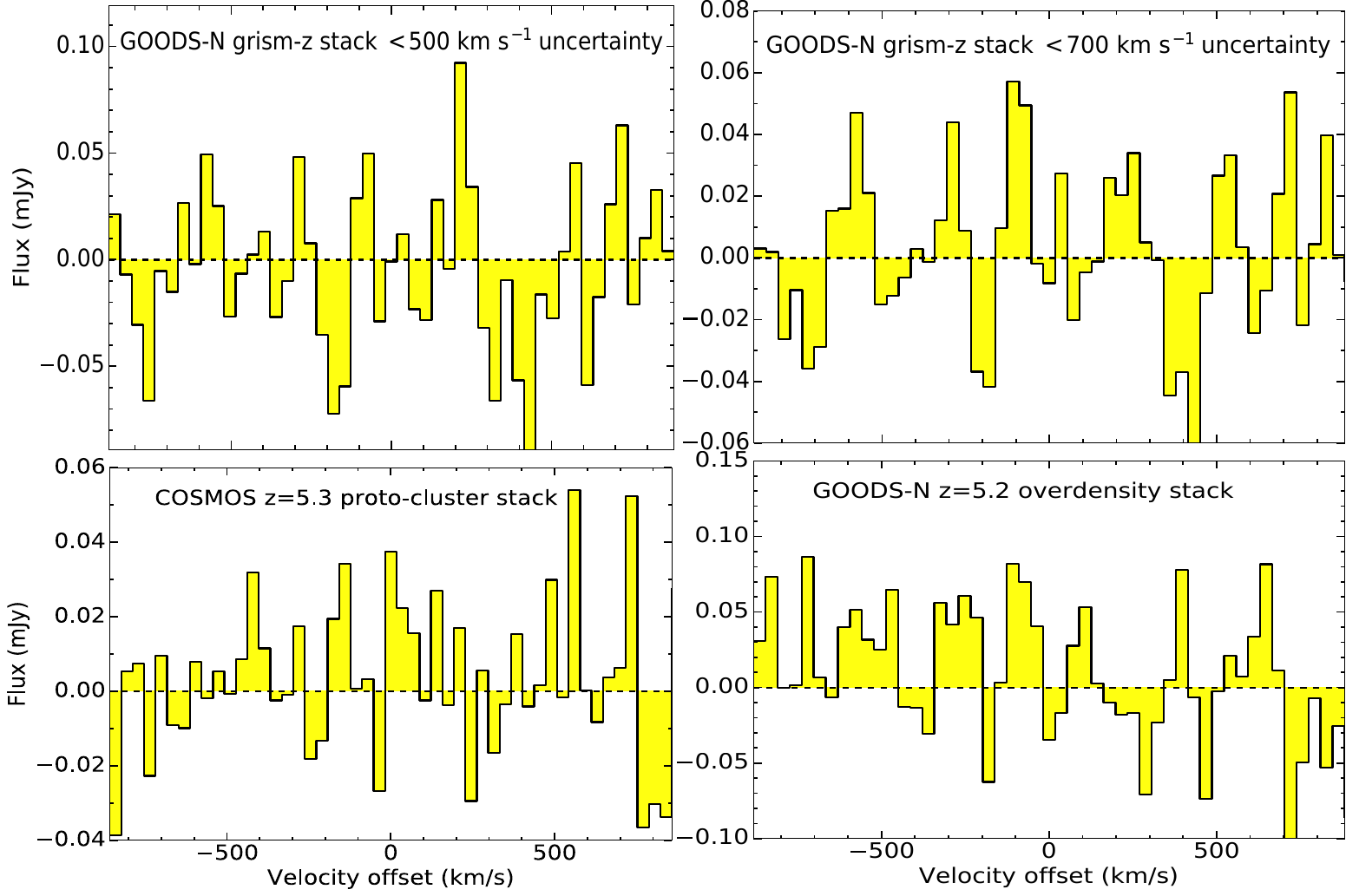}
}
\caption{Additional spectral stacks of sets of galaxies  with  spectroscopic or grism redshifts. Top: CO(1--0) stacks of subsets of galaxies with the highest quality grism spectra, with redshift uncertainty $<$500 (left) and $<$700 km s$^{-1}$ (right), respectively.  Bottom: CO(2--1) stacks for  samples of potential  $z>5$ galaxies  belonging to  previously identified over-densities in our fields, i.e. the AzTEC-3 proto-cluster at $z=$5.3 (left) and the overdensity around HDF850.1 (right; \citealt{Capak11,Walter12}). 
The spectral resolution is the same as in Fig.~\ref{fig:spectra_COS1}.}
\label{fig:stacked_spectra_appendix}
\end{figure*}

\subsection{COSMOS proto-cluster at $z=5.3$}
We search for CO(2--1)  from nine member galaxies of the AzTEC-3 proto-cluster, as identified through the Lyman Break (LBG) technique and color selection by \cite{Capak11}. While two of them show a  hint of a CO emission line signal ($\sim2.5\sigma$), all others (including LBG-1, the one with the strongest [C{\sc ii}] emission; \citealt{Riechers14} and in prep., \citealt{Capak15}), are consistent with noise. 
The positions of the LBGs with tentative CO detections are J2000 10:00:21.96 +02:36:08.5; and 10:00:20.13 +02:35:53.9. Assuming a line FWHM of 250 km s$^{-1}$ (i.e., the width of the [C{\sc ii}] line in LBG-1) we derive CO line fluxes of approximately $0.03\pm0.012$ Jy km s$^{-1}$, corresponding to $L'_{\rm CO21}=7\pm3\times 10^9$ K km s$^{-1}$ pc$^2$. We do not claim any detections due to the low significance and only quote these as approximate limits for reference.
We also stack all spectra at the positions of  LBGs (limited to the seven galaxies that are not contaminated by  emission from the bright CO(2--1)  line in AzTEC-3 at the resolution of our survey),  and do not detect a significant signal (Fig.~\ref{fig:stacked_spectra_appendix}). We therefore place a $3\sigma$ limit of $<$0.012 Jy km s$^{-1}$ on their average CO(2--1)  emission, corresponding to  $L'_{\rm CO21}<3\times 10^9$ K km s$^{-1}$ pc$^2$. The average stellar mass for the stacked LBGs, as reported by the COSMOS2015 catalog, is $4\times 10^9\,{\rm M}_\odot$. If we assume an $\alpha_{\rm CO}=3.6\, {\rm M}_\odot$ (K km s$^{-1}$ pc$^2$)$^{-1}$, we obtain a gas mass upper limit of $<1.1\times10^{10}\,{\rm M}_\odot$, and therefore a gas fraction of ${\rm M}_{\rm gas}/{\rm M}_*<3$ which is not strongly constraining. None of the LBGs are detected in  3 GHz radio continuum emission  \citep{Smolcic17}. This would place a limit of $<165 \,{\rm M}_\odot \, {\rm yr}^{-1}$ on their star formation rate if we adopted the redshift evolution of the radio-FIR correlation measured by \cite{Delhaize17}. This may be converted to an expected limit of $L'_{\rm CO}<2.2\times10^{10}$ K km s$^{-1}$ pc$^2$ by assuming the star formation law  \citep{Daddi10b,Genzel10}. This limit is  higher than what we derive from our CO non-detection, implying that our deep observations provides strong constraints on the CO luminosity of $z>5$ LBGs.

\subsection{Grism redshifts in the GOODS-N field}
The 3D-{\em HST} catalog \citep{Momcheva16}  contains 694 galaxies in GOODS-N with grism redshifts, for which the CO(1--0) line is covered by our data. Nevertheless, the majority of these grism spectra do not significantly improve the redshift determination over the photometric redshift, and are therefore not usable for stacking. We search for matches to our line search candidates (down to $4\sigma$) within a radius of $2^{\prime\prime}$ and  $\sim500$ km s$^{-1}$. We find 16 potential matches and assess the contamination by chance association by also matching our blind detection catalog to 694 random positions within the signal data cube. We find that the distribution due to random associations is well described by a Gaussian with mean 8 associations and a standard deviation of 3.5 associations. Therefore, the majority of our line associations are likely to be random, but some may be expected to be real. 
The measured line fluxes of these candidate counterparts would imply gas masses that are sometimes larger than the stellar masses. This is possible, but we consider it more likely that those cases may be random noise associations.

We also stack the spectra extracted at the positions of the galaxies with high quality grism redshift (Fig.~\ref{fig:stacked_spectra_appendix}). 37 of them have redshift uncertainties  less than 500 km s$^{-1}$ (based on the 95th percentile of the redshift probability distribution reported by \citealt{Momcheva16}), and 35 more have redshift uncertainties less than 700 km s$^{-1}$. At this level of uncertainty, it would be likely that a fraction of the line signal contributes to the stacked spectrum. Both of these stacks  show no detection. Assuming a line FWHM of 300 km s$^{-1}$, this implies $3\sigma$ upper limits of $<$0.011 and $<$0.008 Jy km s$^{-1}$ ; corresponding to $\lesssim$3 and 2 $\times 10^9$ K km s$^{-1}$ pc$^2$, respectively.

\subsection{HDF850.1 $z=5.2$ galaxy overdensity in GOODS-N}
We also search for potential CO(2--1)  emission from galaxies in the $z\sim5.2$ overdensity around the sub-millimeter galaxy HDF850.1 \citep{Walter12}, taking advantage of the abundance of available spectroscopic redshifts in this region. 24 of the 105 galaxies with spectroscopic redshifts presented in that work fall within our data, but none of them are individually detected. We  stack the  spectra to obtain an average spectrum (Fig.~\ref{fig:stacked_spectra_appendix}). No significant emission is found after stacking.   Assuming a line FWHM of 300 km s$^{-1}$ implies a $3\sigma$ upper limit of $<$0.015 Jy km s$^{-1}$, and $L'_{\rm CO21}<3.5\times 10^9$ K km s$^{-1}$ pc$^2$ at $z\sim5.2$. We match these 105 galaxies to galaxies within $1^{\prime\prime}$ in the photometric catalog by \cite{Skelton14}, finding 83 matches, but we do not adopt their stellar mass estimates because the redshifts of these galaxies were often greatly under-estimated.

%\begin{figure*}[tbh]
%\ContinuedFloat
%\includegraphics[scale=1.8]{GN7_fig.pdf} 
%\caption{GOODS-N 7 of 7. }
%\label{fig:spectra_GN7}
%\end{figure*}

%% The reference list follows the main body and any appendices.
%% Use LaTeX's thebibliography environment to mark up your reference list.
%% Note \begin{thebibliography} is followed by an empty set of
%% curly braces.  If you forget this, LaTeX will generate the error
%% "Perhaps a missing \item?".
%%
%% thebibliography produces citations in the text using \bibitem-\cite
%% cross-referencing. Each reference is preceded by a
%% \bibitem command that defines in curly braces the KEY that corresponds
%% to the KEY in the \cite commands (see the first section above).
%% Make sure that you provide a unique KEY for every \bibitem or else the
%% paper will not LaTeX. The square brackets should contain
%% the citation text that LaTeX will insert in
%% place of the \cite commands.

%% We have used macros to produce journal name abbreviations.
%% AASTeX provides a number of these for the more frequently-cited journals.
%% See the Author Guide for a list of them.

%% Note that the style of the \bibitem labels (in []) is slightly
%% different from previous examples.  The natbib system solves a host
%% of citation expression problems, but it is necessary to clearly
%% delimit the year from the author name used in the citation.
%% See the natbib documentation for more details and options.

%\begin{thebibliography}{}
   %\bibitem[daCunha et al.(2010)]{dC1} da Cunha, E., Eminian, C., Charlot, S., Blaizot, J. 2010, \mnras, 403, 4

%\end{thebibliography}

\end{document}